\documentclass[traditabstract]{aa}

\usepackage{graphicx}
\usepackage{txfonts}

\begin{document}

   \title{The nature of nuclear $H_{\alpha}$ emission in LINERs}
   
   \author{J. Masegosa\inst{1}, I. M\'arquez\inst{1} \and
           A. Ramirez\inst{1,2} \and
           O. Gonz\'alez-Mart\'in\inst{3,4}}
   \institute{Instituto de Astrof\' isica de Andaluc\' ia (CSIC), Apdo. 3004, 18080 Granada, Spain
      \thanks{\email{pepa@iaa.es}} \and
      Instituto de Astronom\' ia de la Universidad Nacional Aut\'onoma de M\'exico, Apdo. Postal 70-264 M\'exico D.F., M\'exico
\and
       IESL, Foundation for Research and Technology, 711 10, Heraklion, Crete, Greece
\and
Physics Department, University of Crete, P.O. Box 2208, Gr-710 03 Heraklion, Crete, Greece
}
   \date{}

  \abstract { To get insight in the nature of the ionized gas in the nuclear region of LINERs  we have performed 
a study of HST H$\alpha$ imaging of 32 LINERs. The main
    conclusion from this analysis is that for the large majority of
    LINERs (84\%) an unresolved nuclear source has been identified
    as well as extended emission with equivalent sizes ranging from
    few tens till about hundredths of parsecs. Their morphologies appear
    not to be homogeneous being basically grouped into three classes:
    nuclear outflow candidates (42\%), core-halo morphologies (25\%)
    and nuclear spiral disks (14\%). Clumpy
    structures reminiscent of young stellar clusters are not a common
    property on LINERs. The remaining 5 galaxies are too dusty to allow a clear 
    view of the ionized gas distribution.

    A size-luminosity relation has been found between the equivalent
    radius of the H$\alpha$ emission and the (2-10 keV) X-ray luminosities. 
    Both ionised gas morphologies and the size-luminosity relation are 
    indistinguishable from those of low
    luminosity Seyferts, suggesting the same origin for the NLR of
    LINERs and Seyferts. Also a relation between soft X-rays and
    ionized gas has been suggested for the first time in LINERs. 
     From multiwavelength 
    data, only 4 out of the 32 LINERs have no evidences on an AGN nature of their
    nuclear sources from multiwavelength data, but extremely obscured
    AGNs cannot be discarded out given the Compton thick signatures of
    their X-ray emission. For the confirmed AGN LINERs, their
    H$\alpha$ imaging favour core-halo and outflow morphologies (65\%
    of the cases). Finally, their calculated Eddington ratios
    show that our LINER sources radiate at sub-Eddington regime, with
    core-halo systems having on average larger Eddington ratios than outflow candidates.}


\titlerunning{$H_{\alpha}$ emission in LINERs}
\maketitle

\section{Introduction}

It has been suggested that { \emph{Low Ionization nuclear
emission-line regions} (LINERs, Heckman 1980) are placed at the low
luminosity end among the active galactic nuclei (AGN) family (Ho
2008). Although LINERs are found in a large population of nearby
galaxies (30\%; Ho et al  1997), a debate still does
exist on the nature of their energy source. 

Ho (2002, 2008) summarizes the main lines of evidence supporting the AGN 
nature of LINERs: host galaxies properties similar to Seyferts, most of  
the more massive black holes residing in LINERs, incidence of broad line 
regions (hereinafter BLR), compact nuclei at both radio and X-ray frequencies. He 
atributes the large progress made during the last two decades to
multifrequency analyses and HST high spatial resolution studies.

The extensive work made by Nagar et al. (2000, 2002, 2005) have shown that 
radio cores are found in 44\% of LINERs, a percentage similar to that 
observed in Seyferts (47\%). Also when radio data at different frequencies 
exist their spectra tend to be flat as it is expected when non-thermal 
processes take place. 

At X-ray frequencies large progress has also been made thanks to the 
large X-ray facilities Chandra and XMM-Newton. X-ray observations can 
be considered of paramount importance, constituting one of the best tools
to identify AGN. From the different studies carried out in the last 
decade (Ho et al. 2001, Eracleous et al. 2002, Dudik et al. 2005, 
Gonz\'alez-Mart\'in et al. 2006 and 2009a), it has been proved that an AGN 
is present in at least 60\% of the LINERs. Moreover when multifrequency 
information is taken into account (basically the incidence of broad lines and 
the properties at radio frequencies) the percentage of AGNs rises 
up to 90\% (Gonz\'alez-Mart\'in et al. 2006 and 2009a).  

On their hand, HST observations have provided a large advance in the
physics of LINERS. The pioneering UV imaging surveys by Maoz et al. (1995) 
and Barth et al. (1998) concluded that 25\% of the observed
LINERs had an UV compact source in their nuclei. But of course one of the
most outstanding results during the last decade has been the
discovery that sources with detected radiocores show variability at
UV frequencies on month scales (Maoz 2007). Four of their 13
sources (namely M81, NGC 3998, NGC 4203 and NGC 4579) have been
confirmed to be variable also at X-ray frequencies (Pian et al. 2010).
Recently Gonz\'alez-Mart\'in et al. (2010) have also detected X-ray variability 
for the LINER NGC 4102.

HST optical works (Pogge et al. 2000; Sim{\~o}es Lopes et al. 2007;
 Gonz\'alez-Delgado et al. 2008; Gonz\'alez-Mart\'in et al. 2009a) have 
confirmed that almost all the observed LINERs show a nuclear source on 
top of an irregular distribution of circumnuclear dust. Dust obscuration 
can explain the existence of dark-UV LINERs.
The importance of an obscuring environment, maybe linked to the acretion physiscs, 
has been recognized in our X-ray aproach to the nature of LINERs (Gonz\'alez-
Mart\'in et al. 2009b). We found that a large percentage of them (50\%) 
show clear signs to be  Compton thick. This fraction is even larger than 
that reported for Seyferts (30\%) (Gonz\'alez-Mart\'in et al. 2009b, Panessa 
et al. 2006) and so the location and nature of their obscuring matter needs to 
be further investigated. 
Until new high resolution X-ray images become available, only indirect 
information can be obtained on the nature of LINERs by looking for correlations
between X-ray properties taken at lower resolution and optical/NIR properties 
taken at much larger spatial resolution. In this vein, it is worthwhile to 
search for the properties of the ionized gas and its relation to 
the X-ray results.

Previous works have concluded that the H$\alpha$ morphology of LINERs
mainly consists on a point source embeded in an extended
structure sometimes clumpy, filamentary and in some particular cases
with clear indications of nuclear obscuration, but mostly
indistinguishable from what is observed in low luminosity Seyfert
galaxies (Pogge et al. 2000, Chiaberge et al.  2005 and Dai \& Wang
2008). Based on STIS spectroscopic observations of 13 LINERs, Walsh et
al. (2008) clearly demonstrate that at scales of tens of parsecs
their energy source is consistent with photoionization by the central
nuclear source, but with a NLR kinematics dominated by
outflows. Following Barth's (2002) considerations, by analogy with
Seyfert unification models, it is natural to wonder whether the
various types of low luminosity AGN (LLAGN), which LINERs could belong to,
 are different manifestations of the same underlying phenomenon, with observed
differences being only orientation or obscuration. The main goal of this
paper is twofold: (1) to evaluate if the ionized gas in the central
regions of LINERs shows characteristics indicative of ionized emission
from the AGN (a NLR), and also (2) to investigate their relation to the
Seyfert population.

In this paper, we present an update of the properties of the Narrow
Line Region for a large sample of 32 LINERs. 
Archival \emph{HST} narrow imaging data have been
used (WFPC2 and ACS). In Section 2 the sample and the 
\emph{HST} image processing are described.  In Section 3 we present
the results and discussion. Section 4 summarizes our main conclusions.

\section{{\bf Sample and data reduction}}

\begin{table*}
\tiny
\begin{center}
\caption{Archival {\emph HST} data for the LINER galaxies.}
\label{tab-datos}
\begin{tabular}{l l l l l  l l l}
\hline
Galaxy &Righ Ascension & Declination & Instrument & Proposal & Prop.& Filter & Exposure\\
       & (2000)        & (2000)      &            & number   & PI name&      & time (sec)\\
  (1)    & (2)        & (3)      & (4)           & (5)   & (6)&   (7)   & (8)\\
\hline
IC1459   &22:57:10.607  &-36:27:44.00 &  WFPC2    &  6537 &De Zeew      &F631N        &2300(3) \\
       &              &             &             &       &             &FR680P15     &2000(3)  \\
NGC0315  &00:57:48.883  &+30:21:08.81 &  WFPC2    &  6673 &Baum         &F555W        &460(2)\\        
       &              &             &    	  &	     &      	&F814W        &460(2)\\
       &              &             & 		  &	     &          &FR680N       &3300(3)\\
NGC2639  &08:43:38.078  &+50:12:20.01 &  WFPC2    &  7278 &Falcke       &F547M        &320(2)\\
       &              &             &             &  7278 &Falcke       &F814W        &140(2)\\
       &              &             &             &  7278 &Falcke       &FR680P15     &1200(2)\\
NGC2681  &08:53:32.730  &+51:18:49.30 &  ACS      &  9788 &Ho           &F658N,F814W  &720(2),120(1)\\
NGC2787  &09:19:18.560  &+69:12:12.00 &  WFPC2    &  6785 &Malkan       &F658N,F702W  &2800(4),800(2)\\
NGC2841  &09:22:02.634  &+50:58:35.47 &  ACS      & 10402 &Chandar      &F658N,F814W  &1400(2),750(2)\\
NGC3226  &10:23:27.008  &+19:53:54.68 &  ACS      &  9293 &Ford         &F658N,F814W  &1400(2),700(2)\\
NGC3245  &10:27:18.392  &+28:30:26.56 &  WFPC2    &  7403 &Filippenko   &F658N,F702W  &2200(2),140(1)\\
NGC3379  &10:47:49.600  &+12:34:53.90 &  WFPC2    &  6731 &Ciardullo    &F502N,F547M  &9785(5),600(5)\\
NGC3607  &11:16:54.660  &+18:03:06.50 &  ACS      &  9788 &Ho           &F658N,F814W  &700(2),120(1)\\
NGC3623  &11:18:55.960  &+13:05:32.00 &  WFPC2    &  8591 &Richstone    &F547M,F658N  &1600(4),1600(4)\\
NGC3627  &11:20:15.028  &+12:59:29.58 &  WFPC2    &  8591 &Richstone    &F658N        &1600(4)\\
         &              &             &  WFPC2    &  8597 &Regan        &F606W        &560(2)\\
NGC3998  &11:57:56.133  &+55:27:12.91 &  WFPC2    &  5924 &Dressel      &F658N,F791W  &1660(3),100(2)\\
NGC4036  &12:01:26.753  &+61:53:44.81 &  WFPC2    &  5419 &Sargent      &F547M        &300(1)\\
       &              &             &             &  6785 &Malkan       &F658N        &2800(4)\\
NGC4111  &12:07:03.130  &+43:03:55.40 &  WFPC2    &  6785 &Malkan       &F658N,F702W  &1200(2),600(1)\\
NGC4278  &12:20:06.826  &+29:16:50.71 &  WFPC2    &  6731 &Ciardullo    &F502N,F547M  &12700(5),600(5)\\
NGC4314  &12:22:31.990  &+29:53:43.30 &  WFPC2    &  8597 &Regan        &F606W        &560(2)\\
       &              &             &             &  6265 &Jefferys     &F658N        &600(2)\\
NGC4374  &12:25:03.743  &+12:53:13.14 &  WFPC2    &  6094 &Bower        &F547M,F658N  &1200(2),2600(2)\\
       &              &             &             &       &             &F814W        &520(2)\\
NGC4438  &12:27:45.594  &+13:00:31.77 &  WFPC2    &  6791 &Kenney       &F656N,F675W  &5200(4),1450(4)\\
NGC4486  &12:30:49.423  &+12:23:28.04 &  WFPC2    &  5122 &Ford         &F547M,F658N  &800(2),2700(2)\\
NGC4552  &12:35:39.807  &+12:33:22.83 &  WFPC2    &  6099 &Faber        &F555W,F814W  &2400(4),1500(3)\\
       &              &             &             &  8686 &Goudfrooij   &F658N        &2300(2)\\
NGC4579  &12:37:43.522  &+11:49:05.50 &  WFPC2    &  6436 &Maoz         &F502N,F547M  &3200(3),726(5)\\
       &              &             &             &       &             &F658N,F791W  &1400(2),726(5)\\
NGC4594  &12:39:59.432  &-11:37:22.99 &  WFPC2    &  5924 &Dressel      &F658N,F791W  &1600(2),100(1)\\
NGC4636  &12:42:49.870  &+02:41:16.00 &  WFPC2    &  8686 &Goudfrooij   &F547M        &1000(4)\\
       &              &             &             &       &             &F814W        &400(4)\\
       &              &             &             &       &             &FR680P15     &2300(2)\\
NGC4676A&12:46:10.110  &+30:43:54.90 &  WFPC2     &  8669 &Van derMarel &F555W,F814W  &320(2),320(2)\\
       &              &             &             &       &             &FR680N       &1200(2)\\
NGC4676B&12:46:11.243  &+30:43:21.87 &  WFPC2     &  8669 &Van derMarel &F555W,F814W  &320(2),320(2)\\
       &              &             &             &       &             &FR680N       &1200(2)\\
NGC4696  &12:48:49.276  &-41:18:40.04 &  WFPC2    &  5956 &Sparks       &F702W,FR680N &320(2),1200(2)\\
NGC4736  &12:50:53.061  &+41:07:13.65 &  WFPC2    &  5741 &Westphal     &F555W        &296(3)\\
       &              &             &             &  8591 &Richstone    &F656N        &1700(5)\\
NGC5005  &13:10:56.231  &+37:03:33.14 &  WFPC2    &  6436 &Maoz         &F502N,F658N  &2400(2),1400(2)\\
       &              &             &             &       &             &F791W        &720(4)\\
NGC5055  &13:15:49.330  &+42:01:45.40 &  WFPC2    &  8591 &Richstone    &F547M,F656N  &1400(4),1700(5)\\
NGC5846  &15:06:29.286  &+01:36:20.24 &  WFPC2    &  6357 &Jaffe        &F702W        &1000(2)\\
       &              &             &             &  6785 &Malkan       &F658N        &2200(2)\\
NGC5866  &15:06:29.499  &+55:45:47.57 &  ACS      &  9788 &Ho           &F658N,F814W  &700(2),120(1)\\
\hline                                                                                           
\end{tabular}
\end{center}
\end{table*}

We have searched for archival \emph{HST} data for the 82 LINERs in our
sample (Gonz\'alez-Mart\'in et al. 2009a) in the \emph{Hubble Legacy
  Archive} (HLA hereinafter) web page\footnote{http://hla.stsci.edu/hlaview.html}. 
HLA data are fully processed (reduced, co-added, cosmic-ray cleaned etc.) images ready 
for scientific analysis. 
All the files for narrow band observations centered either in H$\alpha$ or
[O III] emission lines (at the redshift of the galaxy) and their
corresponding continum have been retrieved. For thirty two galaxies,
this kind of narrow-band imaging data are available\footnote{Also
  available are the narrow band images of NGC6240 and NGC6241, which
  are not considered in this paper since the NLR physical sizes for
  these two galaxies cannot be resolved even with HST data due to
  their much { larger} distances.}. HLA data products are available for
all of them, so we have
retrieved the fits files corresponding to averaged, processed
data. Table \ref{tab-datos} provides the
galaxy names (Col. 1), coordinates as provided by HLA (Cols. 2 and 3),
instrument (Col. 4; { most of the data comes from WFPC2, only 5 galaxies coming from  ACS}),  
{ proposal number and principal investigator's name (Cols. 5 and 6), the filters  used in this analisys (Col. 7) and the total exposure time for such filters (Col. 8). The number of images used for each filter is shown in brackets in column 8. When only a single image was available, a cosmic ray extraction was applied by using the LACOSMIC{\footnote{http://www.astro.yale.edu/dokkum/lacosmic/}} routine (van Dokkum 2001).} 
 
\emph{HST} absolute astrometry does not guarantee the centering of two images at the level 
of its spatial resolution. For that reason, when needed, 
the narrow- and wide-band images have been aligned according to the center 
of the galaxy (maximum peak in brightness) and with the stars present
in the field.

In order to maximize the chances to get a reliable estimation of the continuum level 
at the wavelength corresponding to the emission line,
we use a common procedure
to get continuum subtracted images, based in a relative
calibration as follows. The fluxes in the narrow ($I(F_{narrow})$) 
and broad band ($I(F_{wide})$) filters are: 

\begin{eqnarray}
I(F_{narrow}) = I_{narrow}(cont) + I_{narrow}(line) \\
I(F_{wide}) = I_{wide}(cont)
\end{eqnarray}

\noindent where $I_{narrow}(line)$, $I_{narrow}(cont)$, and $I_{wide}(cont)$ 
are the intensities measured in the line itself, the continuum under the line, and 
the continuum in the wide filter, respectively.

To assume that line emission is less extended than the  continuum
 is equivalent to say that, far enough from the center, for
flux calibrated images one should have 
$I_{narrow}(cont) = I_{wide}(cont)$. Previous to any calculation, the background 
of the two images, narrow and wide-band, has been set to zero.

   \begin{figure}
    \centering
      \includegraphics[width=0.7\columnwidth,angle=-90]{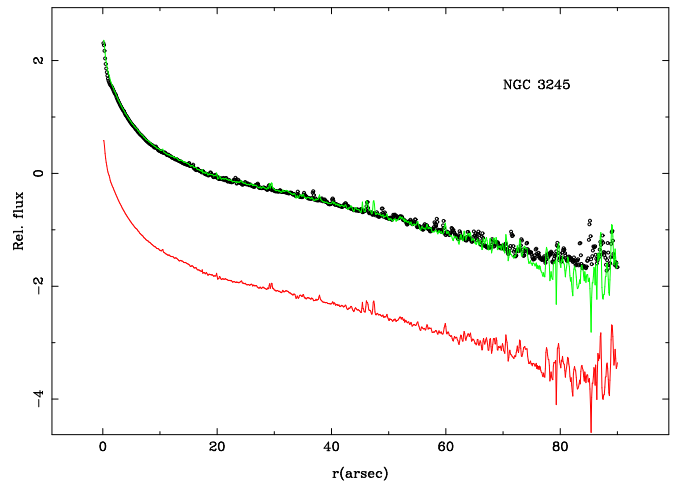}
\setcounter{figure}{0}
   \caption{Surface brightness profiles for NGC 3245. The broad-band profile, $I(F_{wide})$, is plotted 
in black (circles points). The narrow-band profile, $I(F_{narrow})$, is ploted in red. The green line
is the narrow-band profile scaled to that of the broad-band.}
           \label{perfiles}
    \end{figure}

Then, we have to calculate the
  factor required to put the two images at the same level. To do so,
  we obtain the surface brightness profiles (with {\sl ellipse}
within IRAF\footnote{IRAF is distributed by the National
  Optical Astronomy Observatories, which are operated by the
  Association of Universities for Research in Astronomy (AURA), Inc.,
  under contract with the National Science Foundation.}) for the {narrow- and broad-band} images. The comparison
of the two profiles allows us both to estimate the region where the
fluxes should be fixed at the same level, and then the
corresponding factor to be applied. 
An example of the two profiles obtained for the galaxy 
NGC 3245 is shown in Fig.~\ref{perfiles}}. The resulting emission-line image
is calculated as:

\begin{equation}
I(emission) =  I(F_{narrow}) - factor \times I(F_{wide})
\end{equation}

The emission line images coming from equation 3 are not flux calibrated,
but allow to recover emission-line morphologies and sizes, what is
our main purspose, as it will be explained below. 

Fig.~\ref{fig-images} shows the resulting continuum subtracted images
for all the galaxies listed in Tab.~\ref{tab-datos}. Both H$\alpha$ and [O III]
images are shown when available (for NGC 4579 and NGC 3379). Overlaid to the
grey scale images a number of contours are shown, the faintest
corresponding to 3 times the dispersion of the background, $\sigma$. 

Fig.~\ref{fig-images} also shows the  sharp divided
continuum images (SD hereinafter).  Briefly, the sharp-dividing method 
consists on dividing the original image by another one that results from 
applying a filter to the original image, with a box size several times that of the 
PSF FWHM. In our case, a median box of 30 pixels has been used.
SD images provide an enhancement of the small structures and thus will be in many cases 
a good tracer of the dust structures (see Marquez et al. 2003 and references therein), 
as it is the case, for instance, for NGC\,4374 (see Fig.~\ref{fig-images}).

Appendix A gives a description on the morphology for each object together 
with additional relevant information.

For estimating the sizes of the emission-line regions, we consider a
3$\sigma$ level above the background and measure the area of the region inside the
corresponding contour. The size is parametrized as the equivalent
radius of such an area, i.e.: $R_{eq} = (Area / \pi)^{1/2}$. The
$\sigma$ of the background level for each image has been measured
in several regions around the galaxy, so that the final $R_{eq}$ is the 
median of these values, and its accuracy is provided by the dispersion of
the various $R_{eq}$ around the median. The results are presented 
in Table~\ref{tab-req}. No radius has been determined neither for
NGC 3379 due to low S/N of the images, nor for 
NGC 3627 and NGC 5866
due to the large amount of dust wich difficults their determination.  Col. 1 shows 
the galaxy name,  Col. 2 the filter used for continuum substraction, Col. 3 the 
distance as taken from Gonz\'alez-Mart\'in et al. (2009a),
Cols. 4 and 5 show the X-ray soft and hard luminosities taken from 
 Gonzalez-Mart\'in et al. (2009a)\footnote{The L$_X$ (2-10 keV) are Compton-thick corrected.}
 and Cols. 6 and 7 the equivalent radius and its dispersion, $\sigma_{Req}$. Two
estimates of the equivalent radius have been obtained when two
continuum filters were available. The deepest resulting image has been chosen 
to estimate the final equivalent radius.  In these cases, the corresponding filter  
is flagged with 
an asterisk (in Col. 2). This estimation can be compared with sizes from 
other analysis based on 3$\sigma$ detection limits for the extended emission
(see for instance Schmitt et al. 2003)

Nevertheless, since the data are inhomogeneous, a S/N threshold
  does not have a well defined physical meaning, which complicates the
  interpretation of equivalent radius. Therefore, we have used 
  the flux calibration corresponding to the images taken with narrow
  band filters, $I(F_{narrow})$ (done in the standard way, using the
  information available on the image headers), for the resulting emission-line 
image, once the continuum is rescaled and subtracted, $I(emission)$. 
Due to the uncertainties in the flux calibration for ramp filters, the images 
obtained with such filters have not been used. 
 For the flux calibrated images, we have also calculated R$^*$$_{eq}$ (Col. 8
in Table 2) 
  as the isophotal equivalent radius at the isophotal level of
  2.9$\times$ 10$^{-9}$ erg s$^{-1}$cm$^{-2}$ arcsec$^{-2}$. This rather
arbitrary surface brightness was chosen to optimize the measure for all the 
available data. This
  radius allows a measure of a physical characteristic size of the
  regions independently on the individual S/N ratios of the
  images. This has been done for the 22 objects with flux calibrated
  images.

\begin{tiny}
\begin{table*}
\begin{center}
\setcounter{table}{1}
\caption{X-ray luminosities$^a$  and H$\alpha$ equivalent radii$^b$.}
\label{tab-req}
\begin{tabular}{l l r l l r r l }
\hline
Galaxy &Cont. &  D$^a$  & log(L$^b_{soft}$)      & log(L$^b_{hard})$      & R$_{eq}^c$ & $\sigma_{Req}$ &
R$^*$$_{eq}^c$ \\
       &Filter&     (Mpc)         &  (erg/s)      &  (erg/s)      & (pc)    & (pc)    & (pc)\\
(1)& (2) &  (3)  & (4)    & (5)  & (6) & (7) & (8)\\
\hline
IC 1459   &FR680P15&  29.24&     40.6&      40.5&      245.52&   0.17    &         \\   
NGC315    &F555W   &  68.11&     42.0 &     41.8&      398.66&   0.02    &         \\
          &F814W*&     68.11&     42.0 &     41.8&      528.32&  0.22    &         \\    
NGC2639   &F547M*&     45.45&     42.2&      40.1&      212.37&  0.18    &         \\    
          &F814W&     45.45&     42.2&      40.1&      185.79&   0.12    &         \\    
NGC2681   &F814W&     17.22&     38.6&      41.0&      209.75&   0.07    & 48.27   \\
NGC2787   &F702W&      7.48&     38.9&      38.8&      43.50 &   0.03    & 29.11   \\
NGC2841   &F435W*&     11.97&     39.4&      39.2&      167.91&  0.02    &         \\
          &F814W&     11.97&     39.4&      39.2&      142.97&   0.25    & 16.36$^d$ \\    
NGC3226   &F814W&     23.55&     40.7&      40.8&      100.49&   0.02    &  29.53$^d$ \\
NGC3245   &F702W&     20.89&     38.8&      40.8&      117.78&   0.21    &  158.90 \\    
NGC3379   &F547M&     10.57&     38.0&      39.9&            &           &         \\    
NGC3607   &F814W&     22.80&     38.6&      40.5&      283.65&   0.31    & 27.17$^d$\\    
NGC3623   &F547M&      7.28&     39.1&      39.4&      86.33 &   0.08    &   35.52  \\
NGC3627   &F606W&     10.28&     39.2&      41.2&            &   0.08    & 109.45   \\
NGC3998   &F791W&     21.98&     42.7&      40.6&      237.93&   0.08    &   233.01 \\
NGC4036   &F547M&     24.55&     39.0&      40.9&      199.21&   0.06    &   116.29 \\
NGC4111   &F702W&     15.00&     40.9&      40.4&      132.19&   0.03    &   308.69 \\
NGC4278   &F814W&     16.07&     39.6&      41.0&       67.0 &   0.08    &   118.34 \\
NGC4314   &F606W&      9.68&     39.6&      39.1&      165.01&   0.04    &          \\
NGC4374   &F547M&     18.37&     39.5&      41.3&      401.75&   0.03    &          \\
          &F814W*&    18.37&     39.5&      41.3&      340.93&   0.06    &   197.53 \\
NGC4438   &F675W&     16.83&     40.1&      40.8&      255.57&   0.02    &   153.53 \\
NGC4486   &F547M&     16.07&     40.9&      40.8&      224.74&   0.08    &   81.99  \\
NGC4552   &F814W&     15.35&     39.5&      39.3&      216.95&   0.15    &   217.22 \\
NGC4579   &F791W&     16.83&     40.9&      41.2&      145.98&   0.03    &   174.22 \\
          &F547M&     16.83&     40.9&      41.2&      45.09 &   0.01    &          \\
NGC4594   &F791W&      9.77&     39.6&      39.9&      93.35 &   0.09    &   107.34 \\
NGC4636   &F814W&     14.66&     39.0&      40.9&      54.75 &           &          \\
NGC4676A &F555W*&     88.00&     39.7&      39.9&      598.38&   0.10    &          \\
          &F814W&     88.00&     39.7&      39.9&      769.16&   0.06    &          \\
NGC4676B &F555W*&     88.00&     40.0&      40.1&      1001.37&  0.06    &          \\
          &F814W&     88.00&     40.0&      40.1&      843.25&   0.05    &          \\
NGC4696   &F702W&     35.48&     41.6&      40.0&      205.77&   0.15    &          \\
NGC4736   &F555W*&     5.20&     38.8&      38.6&      146.49&   0.03    &          \\
          &F814W&      5.20&     38.8&      38.6&      252.39&   0.01    &   226.27 \\
NGC5005   &F791W&     21.28&     40.7&      41.6&      248.99&   0.03    &   393.56 \\
NGC5055   &F547M&      7.14&     38.6&      39.6&      76.04 &   0.06    &   62.92  \\
NGC5846   &F702W&     24.89&     40.2&      40.8&      156.31&   0.08    &   88.26  \\
NGC5866   &F814W&     15.35&     40.1&      38.1&            &            &         \\
\hline                                              
\end{tabular}
\end{center}
{$^a$ Distances have been taken from table 1 in Gonz\'alez-Mart\'in et al. (2009a)}\\
$^b$ Note that these luminosities have been corrected from intrinsic absorption; 
L$_X$ (2-10 keV) has been also corrected for Compton-thickness. L$_{hard}$ and L$_{soft}$ hold for the 
logarithm of the (2-10) kev (from Gonzalez-Mart\'in et al. 2009b) and (0.3-2) keV 
(from Gonzalez-Mart\'in et al. 2009a) \\
{ $^c$ R$_{eq}$ is the equivalent radius corresponding to a level 3 times higher that the dispersion of the background. R$^*$$_{eq}$ corresponds to the isophotal level at 2.9$\times$ 10$^{-9}$  erg s$^{-1}$cm$^{-2}$ arcsec$^{-2}$.}\\
$^d$ The resulting radius is smaller than 2 pixels.
\end{table*}
\end{tiny}



\section{Results and discussion}

\subsection{H$\alpha$ emission as a tracer of the morphology of the NLR in LINERs}

The first result from our analysis is that for most LINERs the
H$\alpha$ emission is composed of a nuclear source and extended
emission, revealing a complex structure, with a large range of
different morphologies. The exceptions are NGC 2639, NGC 3379, NGC
3627, NGC 4036 and NGC 5005, for wich an unresolved nuclear source has
not been identified. We have grouped our sample galaxies into 4 types of objects
according to the morphology of the extended H$\alpha$ emission in the
central 1-2 arcseconds.  The objects belonging to each sub-cathegory
are shown in Table \ref{clasif-Halpha}.

\begin{enumerate}
\item {\sl Core-halo:} When a clear unresolved nuclear source,
  surrounded by diffuse  emission, has been
  identified.  Nine out of the 32 objects belong to this
  class. In most cases the putative nucleus is sitting in a 
  linear elongated structure. 
    In five cases (IC\,1459,
  NGC\,315, NGC\,2639, NGC\,3623, and NGC\,5055; see individual
  comments in Appendix A) the extended emission appears to be
  sitting in the disk of the galaxy and the elongation of the
  emission follows the major axis of the galaxy (taken from the NED 
database\footnote{The NASA/IPAC Extragalactic Database (NED) is operated 
by the Jet Propulsion Laboratory, California Institute of Technology, 
under contract with the National Aeronautics and Space Administration.}). In three of
  them (NGC\,2787, NGC\,3998, and NGC\,4111), the nuclear disk axis
  seems to be perpendicular to the galaxy major axis. NGC 2681 does not 
 show any elongation (see Fig.~\ref{fig-images}).  

\item {\sl Outflows:} Eleven galaxies show morphological evidences 
to have  nuclear outflows (Veilleux et al 2005). 
Some of them present debris/filamentary extension (NGC\,4486,
NGC\,4676A and B, NGC\,4696, NGC\,5005, and NGC\,5846), biconical
structures (NGC\,4036 and NGC\,5005) and also buble-like structures
(NGC\,3245 and NGC\,4438) coming out from the nucleus.  The
  high spatial and spectral resolution spectroscopic data (\emph{HST-STIS})
  for NGC\,3245, NGC\,4036 and NGC\,4579, reported by Walsh
  et al. (2008), indeed evidence outflow kinematics 
  strengthening our suggestion. For the remaining objects, such a
  kinematical confirmation has to await until similar spectroscopic
  data are available.

 \item {\sl Disky:} Seven galaxies present face-on structures that can
   be associated to H$\alpha$ emission along the spiral arms
   (NGC\,2681, NGC\,2841 and NGC\,4736), diffuse emission along the
   disk (NGC\,3379, NGC\,4552 and NGC\,4636), nuclear plus star
   formation rings (NGC\,4314). NGC\,4594 has also been included in
   this class because, although it is not seen face on, it appears
   that its H$\alpha$ emission is concentrated in the nuclear region
   and their spiral arms.

\item {\sl Dusty:} Those where clear dust lanes obscure the 
underlying H$\alpha$ structure. This  prevents us from getting information on 
the morphology of these inner regions. Five objects have been 
classified as such (NGC3226, NGC3607, NGC3627, NGC4374 and NGC5866).
Different structures can be identified depending on the dust distribution 
along the galaxy, but mostly nuclear sources surrounded by an 
inhomogeneous dusty disk are found.

\end{enumerate}

\begin{table}
\begin{center}
\caption{Morphological classification of H$\alpha$ nuclear emission.}
\label{clasif-Halpha}
\begin{tabular}{l l l l l }
\hline
core-halo & outflow& dusty & disky \\
\hline
IC 1459  &   NGC 3245 &  NGC 3226  & NGC 2681\\
NGC 315  &   NGC 4036 &  NGC 3607  & NGC 2841\\
NGC 2639 &   NGC 4438 &  NGC 3627  & NGC 3379\\   
NGC 2787 &   NGC 4486 &  NGC 4374  & NGC 4314\\
NGC 3623 &   NGC 4579 &  NGC 5866  & NGC 4552\\
NGC 3998 &   NGC 4636 &            & NGC 4594\\
NGC 4111 &   NGC 4676A&           ~& NGC 4736\\  
NGC 4278 &   NGC 4676B&           ~& ~        \\
NGC 5055 &   NGC 4696 &           ~&        ~\\
        ~&   NGC 5005 &           ~&        ~\\
        ~&   NGC 5846&            ~&        ~\\
\hline                                              
\end{tabular}
\end{center}
\end{table}

Our main concern here is to understand whether the detected H$\alpha$
  nuclear regions correspond to the expected NLR for AGN.  
Pogge et al. (2000) made an extensive \emph{HST} investigation on
the NLR of 14 LINERs,  and concluded that at {\emph HST}
  resolution the NLRs are resolved showing complex morphologies,
  different from galaxy to galaxy, that come from a combination of
  knots, filaments and diffuse gas. Dai \& Wang (2008) concluded similarly
  with an extension of Pogge's sample up to 19 LINERs.  

  Among our 32 sample galaxies, 17 LINERs are studied in this paper
  for the first time. Pogge et al. (2000) already analized 7 of the
  LINERs in our sample (namely NGC 3998, NGC 4036, NGC 4374, NGC 4486,
  NGC 4579, NGC 4594 and NGC 5005) and Dai \& Wang (2008) studied
  another 4 (namely NGC 404, NGC 2768, NGC 3718 and NGC 4192 ) not included
in the sample because of our X-ray selection.  

All together, including 
the new 17 LINERs from
our work plus the 19 ones from Dai \& Wang (2007) (we have 15 objects in
common with them), they conform a rather homogeneous set of data for 36
LINERs, which seems to be the larger sample homogeneously analyzed so far.
It is worth noticing that the 4 objects from Dai \& 
  Wang's paper not included in our sample can be fit into the outflow-like group. 
Thus from the total sample of 36 LINERs, 42 \%  would be outflow candidates, 
25\% core-halo systems, 19\% disk-like systems and 14\% dusty LINERs.
These results stress the intereresting possibility of shock heating as an extra 
contribution to the ionization in addition to nuclear 
photoionization. This scenario needs to be explored at length with high S/N 
spectroscopy for the outflow candidates to investigate if at least for
these LINERs the long standing problem
of ionizing-photon deficit  
can be solved (see Eracleous et al. 2010b for a full discussion).

The question then to be answered is whether the origin of the outflow can
be circumnuclear star formation or it is a nuclear outflow predicted
by the unified AGN models (Elvis 2000). From { the STIS spectroscopic analysis 
by Gonz\'alez Delgado et al. (2004) it is found that recent star forming processes 
(with ages lower than 10$^7$ years) 
are almost absent in LINERs}, being the dominant stellar population that of old stars 
with, in some particular cases, 
some contribution from intermediate age (10$^8$ years) stars.  
The H$\alpha$ identified
structures appear to be consistent with such a picture.  Indeed, at 
the HST resolution of few tens of parsecs, a
knotty appearance should be expected when young star clusters are present,
which is not observed in most of the images. Their inspection 
appears to indicate that such knotty structures are only present 
in the Mice system. 
In disk-like systems, star formation can be distinguished in their 
disks (e.g. see the star formation ring on NGC\,4314 at $\sim$200 parsecs 
from the nucleus, Fig.~\ref{fig-images}). The structure of core-halo galaxies is more likely 
originated from the gas ionized by the nucleus. For dusty galaxies, although a
faint nuclear source is visible in most of them, the dust distribution
prevents us from drawing any conclusion on the extended
ionized gas.

\subsection{Quantification of the H$\alpha$ emission: equivalent radius}

To further investigate the origin of the extended H$\alpha$ emission,
and considering its irregularity,  we
have calculated a characteristic radius for estimating the
  size of the ionized region: the equivalent radius, 
$R_{eq}$, {and R$^*$$_{eq}$} provided in 
Table~\ref{tab-req}\footnote{Radii smaller than 2 pixels, 
identified with  $^c$ in Column 7, are not considered.} 
(see Section 2, for a detailed explanation of the methodology). 


   \begin{figure}
    \centering
      \includegraphics[width=0.7\columnwidth,angle=-90]{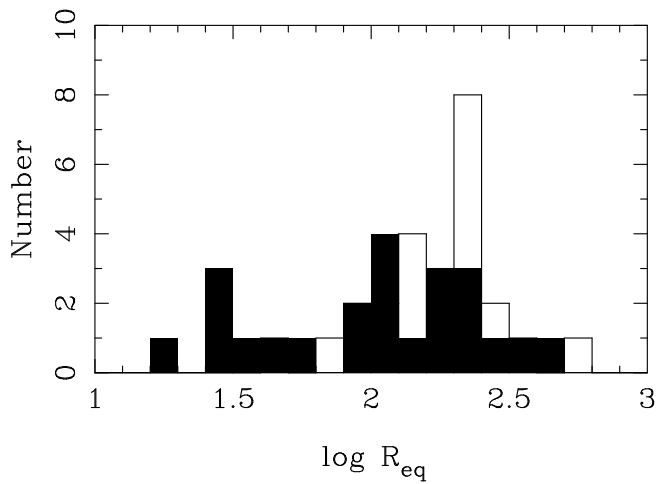}
\setcounter{figure}{2}
   \caption{Histograms of equivalent radii $R_{eq}$ (in parsecs),in our LINER sample. The 
black filled area shows the R$^*$$_{eq}$ distribution.}
           \label{histoReq}
    \end{figure}

We searched { for a distance dependence that could bias our result}, so we plotted
equivalent radii in arcseconds {\sl versus} distance and did not found
any correlation between the two quantities. In Fig.~\ref{histoReq}
the distribution of $R_{eq}$ in parsecs is shown { as the empty histogram and 
the corresponding R$^*$$_{eq}$ distribution as the black filled area}. 
A range of values between 43 and 528 pc 
with a median
value of 200 pc has been obtained { for $R_{eq}$ and between 16 and 469 pc, with a median value 
of 116 pc for R$^*$$_{eq}$.  Comparing both estimations, it is found that 
with the exception of NGC 4111, NGC 4314 and NGC 5005 for which it is found
the largest values of R$^*$$_{eq}$ and much smaller values from the $R_{eq}$ 
estimation , for the remaining cases  R$^*$$_{eq}$ is equal or lower than 
$R_{eq}$. Thus we can conclude that the currently more extended used size estimation
at 3$\sigma$ detection limits tend to overestimate the true physical size 
of the nebula. Finally it is worth to} note that no
significative difference is found on the size among the different
morphologies.  

This range of values is similar to that reported by Dai \& Wang
(2008).  For the { 14} galaxies with measured radii in common in both
works, our estimations for NGC\,2787, NGC\,4111, and\,NGC 4594
are smaller; large discrepancies are found for three objects (for
  NGC\,4314 and NGC\,4736 Dai \& Wang measured very small 
    values and for NGC\,4374 a rather large value was measured
    compared to ours); for the remaining ten objects our estimations
    are larger than theirs.  
    We stress that the method used by Dai \&
    Wang (2008) relies on the estimation of the annulus 
    at which the 3-$\sigma$ level above the continuum is reached 
(see also Bennert et al. 2002). 
The general irregularity of the isophotes makes this method rather uncertain, what 
has motivated us to use $R_{eq}$, that we consider a 
more realistic estimation of the size of the emitting regions.  

Our sizes cover the lower end of the distribution of values for
the major axis obtained, with [OIII]-HST imaging (Schmitt et al. 2003)
for the NLR of Seyfert galaxies. For the 10 Seyfert galaxies with HST
data from Schmitt et al.'s sample included in the X-ray catalog
CAIXA\footnote{Catalog of AGB in the XMM-Newton archive, Bianchi et
  al. (2009).}, we have recalculated the sizes using our definition
of $R_{eq}$ and obtained a range of values between 56 and 314
pc with a median of 169 pc, very much the same than the value for
LINERs. Althought the comparison is not straightforward since
for Seyferts most of the data comes from the [OIII] line, it is
however very suggestive that their NLR morphologies and sizes are not
very different from those of LINERs.

\subsection{Luminosity - size relation}

The luminosity - size relation can also be used to get insight onto
the nature of the ionized emission. This has been raised as an important
relation for AGN since Peterson et al. (2002) found that it can be
defined for the BLR of Seyferts. { Greene et al (2010) have revisited 
such a dependence and found that it is consistent with R$_{BLR}$ $\alpha$ $\sqrt {L}$
based both in Balmer lines and hard (2-10 keV) luminosities. This is the
expected dependence when 
the BLR density is independent on luminosity.
Their data also suggest a steeper relation for the narrow line luminosities,
R$_{BLR}$ $\alpha$ L$^{0.6}$.}

{ Bennert et al. (2002) and Schmitt et al. (2003) searched for
such a relation for the NLR of Seyferts and Dai \& Wang
(2008) extended the work to LINERs. They concluded} that LINERs 
follow the same relation than Seyferts and
QSOs. In this work we present this relation, but { for the first
time} using the X-ray luminosity instead of 
that of  H$\alpha$, which is a more robust tracer of the power of 
the AGN (Maiolino et al. 2002). { H$\alpha$ is expected to be 
more contaminated by other processes as recent star formation events.}

The X-ray luminosity can be used as a measure of the
bolometric intrinsic luminosity of an AGN (Gonzalez-Mart\'in et al
2006, 2009a and b). Therefore, it is worthwhile to investigate whether it is
  related to the size of the NLR.  In Fig. \ref{ReqLX} { the
 hard (2-10 keV) X-ray luminosity {\sl versus} the two determinations
of the equivalent radius is presented}. The different H$\alpha$ morphologies described in
  Section 3.2 are plotted with different symbols. The following { three
galaxies have been excluded from the plot:  NGC\,3379, NGC\,3627 and NGC\,5866.  
NGC\,3379 was excluded because the low count rates of its
narrow line image} impede the determination of a reliable equivalent
  radius; NGC\,3627 and NGC\,5866 were excluded because large
amounts of dust hampers the detection of their NLR. The two
galaxies conforming the Mice system (NGC 4676A and NGC4676B) 
show a large knotty
extension of star formation regions together with typical structures of
outflowing material, 
leading to a rather 
large value of equivalent
radius exceding the hypothetical NLR. Therefore, despite their inclusion 
in the plots, they won't be used for any correlation below.

A first attemp to look for a correlation between X-ray
  luminosities and $R_{eq}$ is based on a least square linear fit, that 
  results in the values reported in Tab. \ref{req-Lx}, and not plotted
    in Fig. \ref{ReqLX} for clarity. The correlations are quite bad,
    with all the galaxies classified as disky (but NGC\,4594) showing larger sizes than 
those expected from their luminosities for the remaining galaxies.
This is not unexpected, since 
$H\alpha$ emission in disky galaxies also comes from the contribution of 
ionized regions in their disk. Therefore we tried again a
linear fit, but this time excluding disky galaxies. The result is the
full line in Fig. \ref{ReqLX}. The resulting coefficients imply 
better correlations in this case (see Tab. \ref{req-Lx}). Finally,
we fitted just the core-halo systems (dashed line in
Fig. \ref{ReqLX}), resulting in the best correlation (see Tab. \ref{req-Lx}).  

   \begin{figure}
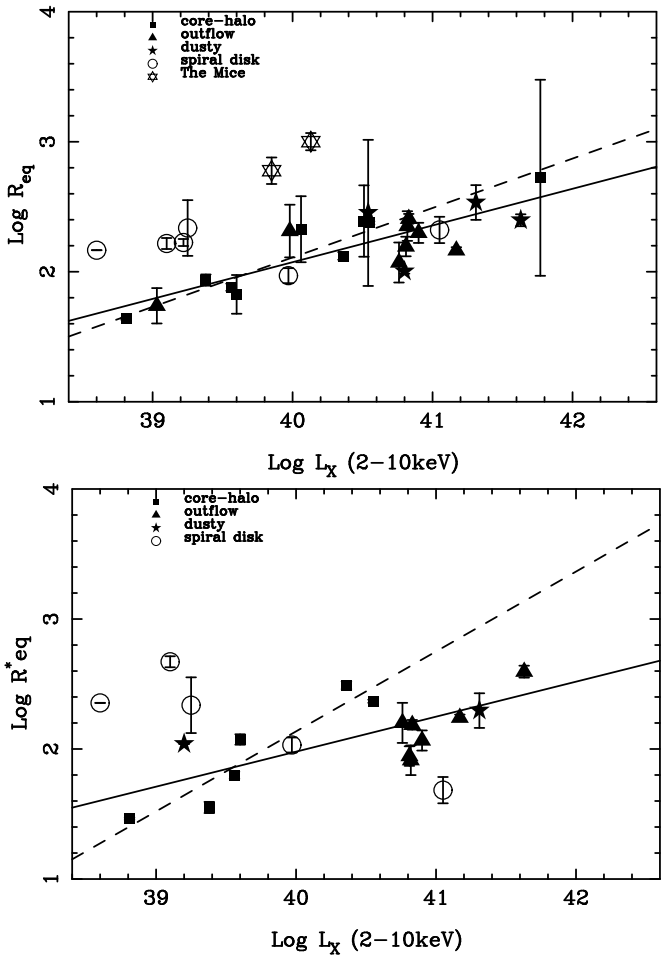

    \centering
       \includegraphics[width=0.7\columnwidth,angle=-90]{Req.vs.LX.ps}
       \includegraphics[width=0.7\columnwidth,angle=-90]{Reqcal.vs.LX.ps}
   \caption{Top: (2-10) keV band absorption corrected luminosity versus the equivalent radius to the contour 
corresponding to 3$\sigma$ times the background, $R_{eq}$. Bottom: {The same for the equivalent radius 
of the level corresponding to 2.9$\times$ 10$^{-9}$  erg s$^{-1}$cm$^{-2}$ arcsec$^{-2}$, R$^*$$_{eq}$.}
The equi\-va\-lent radii are  derived thought narrow band \emph{HST} images. The full lines show the best linear fit to all the galaxies excluding disky systems. The dashed lines show the best linear fit to the core-halo systems.}
           \label{ReqLX}
    \end{figure}

Therefore, two main results appear from Tab. \ref{req-Lx}:
(1) The  luminosity-size relation is better when using X-ray luminosities 
at harder energies and (2) a better correlation is found 
when disky systems are excluded  from the fitting,  
the best fit resulting when only core-halo systems are taken into account.

This later result can be due to the fact that core-halo systems appear
to be less dusty and therefore provide a clear insight on the NLR. 
Since the hard X-ray luminosities cannot be produced by
stars, this relation support the nature of the emission regions as the
result of the ionization by the AGN.  In that respect it is very
suggestive of the similarity of LINERs with higher power AGN,
that the slope for core-halo systems ({ 0.38}) has the same value than
that reported for Seyfert galaxies by Schmitt et al. (2003) by using
the  [O III] luminosity as a proxy of the AGN power.

{ The resulting correlations for the subset of calibrated data
are shown in Fig. \ref{ReqLX} { (bottom)} and in Tab. \ref{req-Lx}. It is very interesting to 
notice that a significant correlation remains only for the core-halo systems. 
Both dusty and outflow galaxies appear to have lower equivalent radii 
for their X-ray luminosities. For the dusty systems it is obvious that 
the presence of large amounts of dust obscures the inner regions and thus lowers 
the measured size of the H$\alpha$ emission. The explanation for the 
outflow candidates is not so straightforward. 
It appears that they cover a narrow range on X-ray luminosities.
This result may suggest a different origin 
for the emission mechanism in these systems but needs to be further investigated}.   

\begin{table*}
\begin{center}
\caption{Fitting parameters for the correlations between the equivalent radius and X-ray luminosity.}
\label{req-Lx}
\begin{tabular}{l | l |l l | l l  |}
       &  &R$_{eq}$        &               & R$^*$$_{eq}$ & \\  
\hline
       & energies& slope& Correl. coeff.& slope& Correl. coeff. \\
\hline
full sample & (2-10keV) & $0.179 \pm0.044$ &0.630&0.069$\pm$0.082&0.193\\
            & (0.5-2 keV)&$0.056 \pm0.040$ &0.266&0.114$\pm$0.064&0.389\\
\hline
non-disklike & (2-10keV) & $0.282 \pm0.054$ &0.814&0.269 $\pm$0.070&0.727\\
             & (0.5-2 keV)&$0.074 \pm0.049$ &0.324&0.144 $\pm$0.064&0.530\\
\hline
core-halo & (2-10keV) &  $0.380 \pm0.047$   &0.949&0.615$\pm$0.116&0.936\\
          & (0.5-2 keV)& $0.173 \pm0.056$   &0.757&0.221$\pm$0.073&0.834\\ 
\hline                                              
\end{tabular}
\end{center}
\end{table*}

\subsection{Soft X-rays vs NLR morphologies}

For a collection of 8 Seyferts, Bianchi et al. (2006) have reported a
spatial correlation between the soft X-ray emission and the NLR as
reflected by the [OIII] emission, taking this result as an important
evidence on the photoionized nature of soft X-rays. Given the
morphological similarity between the NLR of Seyferts and LINERs 
(Schmitt et al 2003, Pogge et al 2000), it will be
interesting to explore if such a relation does exist also in LINERs.

In Fig.~\ref{softXvsHST} the soft X-ray isocontours are overploted
(in black) over the H$\alpha$ images.  Only the { 28} galaxies with available 
Chandra imaging have been included. The remaining { 4} galaxies have only 
XMM-Newton X-ray data, as indicated in Table~\ref{multiwav} with an asterisk 
in Col. (6). Althought a very detailed 
comparison cannot be made due to the different resolutions at both
wavelenghts (around 1" and 0.1'' for Chandra and HST data,
  respectively) it is remarkable that soft X-rays and H$\alpha$ data
show a rough coincidence in their shapes, the soft X-ray
  contours following the structures identified with HST. This is
  not the case for the hard X-rays (red contours in
  Fig.~\ref{softX-halpha}).  Few galaxies depart
from the general behaviour, { NGC 3226, NGC4486, NGC 4676A and B, NGC 5846 and NGC 5866.  
NGC 3226 show a compact structure both at soft and hard X-rays, whereas the H$\alpha$ 
distribution seems to suggest an outflow coming out from that compact nucleus}.  For NGC
4486, both soft and hard X-rays follow the radio jet also visible in
the continuum images, being the H$\alpha$ outflow perpendicular to
it. 
NGC 4676A and B  and NGC 5846  shows in
  H$\alpha$ a  structure non-coincident with either hard or
  soft X-rays, suggesting that the emitting gas
has a different origin. No conclusion can be obtained for NGC 5866: 
its H$\alpha$ emission appear to be very obscured by large amounts of dust 
and  soft X-rays show a spatial distribution which appears to be out of the plane 
of the galaxy.

Unfortunatelly there does not exist yet a sample of good RGS XMM-Newton
data for LINERs to allow the modelling of the soft X-ray
emission. However the data reported by Starling et al. (2005) on the
LINER galaxy NGC 7213 and those collected for 53 LINERs
by Gonz\'alez-Mart\'in et al. (2010, in preparation) seem to suggest that their
soft emission comes from photoionized gas, in good agreement
with the conclusions obtained with the systematic work on Seyfert 2
galaxies by Bianchi and colaborators (Bianchi et al. 2006, Bianchi et
al. 2010).

\subsection{Multiwavelength properties}

Different authors (Ho et al. 1999; Maoz 2007; Eracleous et al. 2009, 2010a; 
Gonz\'alez-Mart\'in et al. 2009a)
 have recognized the importance of the multiwavelenght information to get a 
clear picture of the enegy source in LINERs. Table \ref{multiwav} shows relevant 
information for the LINERs in this sample. The information in Cols. from 
(2) to (10)  has been extracted from Tab. 12 in 
Gonz\'alez-Mart\'in et al. (2009a), with 
Col. (1) providing their number code for each galaxy. In Col. (6) the word CT 
has been added when a LINER shows
Compton-thick (CT hereinafter) characteristics as defined in
Gonz\'alez-Mart\'in et al. (2009b). Col. (7) has been updated with the corresponding 
references for three objects. Col. (9) gives the final
classification considering the multiwavelength information from
Cols. (3) to (8). Col. (10) provides the HST morphological class
as defined in this paper.  In Col. (11) we present
the Eddington ratio, R$_{Edd}$, calculated using the formulation given in Eracleous et al. (2010a):\\ 

            R$_{Edd}$ = 7.7 x 10$^{-7}$ L$_{40}$ M$_8$$^{-1}$\\

where L$_{40}$ is the bolometric luminosity in units of 10$^{40}$ erg s$^{-1}$ 
and M$_8$ is the black hole mass in units of 10$^8$ M\sun. {The SED obtained
by Eracleous et al (2010) for LINERs leads to a bolometric luminosity 50
times L$_{2-10keV}$. Both L$_{2-10keV}$ and M$_8$ values have been taken 
from Gonzalez-Mart\'in et al (2009b).}

\begin{table*}
\begin{center}
\caption{Multiwavelength properties of LINERs. }                       
\label{multiwav}
\begin{tabular}{l l l c c l l c l l l c  l c} 
\hline\hline                 
Num& 	      Name	   &UV	  &X-ray &UV	 &X-ray &Radio	&Broad           &Final     &\emph{HST}& R$_{Edd}$ \\
   & 			   &Var.  & Var. &Comp.  &Class.&Comp.	&H$\sf{\alpha}$  &Class     & Class.   &         \\ 
(1)& 	      (2)	   & (3)  &(4)	 &(5)   &  (6) &(7)    & (8)             & (9)      &      (10) &(11)	    \\ 
\hline 
80 &		IC\,1459   &  & & &AGN        & &Y& Y&core-halo &5.0x10$^{-5}$  \\
1  &	       NGC\,0315   &  & & &AGN        &Y&Y& Y&core-halo &1.0x10$^{-3}$   \\ 
9  &	       NGC\,2639   &  & & &Non-AGN* CT&Y&Y& Y&core-halo &7.9x10$^{-3}$   \\
11 &	       NGC\,2681   &  & &Y&AGN CT     &Y&N& Y&core-halo &3.2x10$^{-5}$   \\
15 &	       NGC\,2787   &  & &N&AGN        &Y&Y& Y&core-halo &2.5x10$^{-6}$   \\
16 &	       NGC\,2841   &  & & &AGN        & & & Y&disky     &6.3x10$^{-6}$  \\
19 &	       NGC\,3226   &  & & &AGN        &Y&Y& Y&dusty     &1.3x10$^{-4}$  \\
20 &	       NGC\,3245   &  & & &AGN CT     &Y$^a$&N&Y&outflow   &1.0x10$^{-6}$  \\
21 &	       NGC\,3379   &  & & &Non-AGN CT & &N& N&disky     &2.0x10$^{-7}$  \\
24 &	       NGC\,3607   &  & & &Non-AGN CT & &Y& Y&dusty     &5.0x10$^{-7}$  \\
26 &	       NGC\,3623   &  & & &Non-AGN*   & &N& N*&core-halo &1.2x10$^{-5}$  \\
27 &	       NGC\,3627   &  & & &Non-AGN* CT& &Y& Y&dusty     &7.9x10$^{-6}$  \\
32 &	       NGC\,3998   & Y&Y&Y&AGN        &Y&Y& Y&core-halo &2.5x10$^{-3}$   \\
33 &	       NGC\,4036   &  & & &AGN CT     &Y&N& Y&outflow   &3.2x10$^{-6}$  \\
34 &	       NGC\,4111   &  & &N&AGN        & &N& Y&core-halo &7.9x10$^{-4}$   \\
39 &	       NGC\,4314   &  & &N&Non-AGN CT & &N& N&disky     &1.0x10$^{-4}$   \\
39 &	       NGC\,4278   &  & &N&Non-AGN CT & &N& N&disky     &1.2x10$^{-4}$   \\
41 &	       NGC\,4374   &  & & &AGN CT     & &Y& Y&dusty     &1.6x10$^{-6}$  \\
43 &	       NGC\,4438   &  & &N&AGN CT     &Y&N& Y&outflow   &6.3x10$^{-5}$   \\
46 &	       NGC\,4486   & Y&Y&Y&AGN        &Y&Y& Y&outflow   &2.5x10$^{-5}$   \\
48 &	       NGC\,4552   & Y&Y&Y&AGN        & &Y& Y&disky     &3.9x10$^{-6}$   \\
50 &	       NGC\,4579   & Y&Y&Y&AGN        &Y&Y& Y&outflow   &2.5x10$^{-4}$   \\
52 &	       NGC\,4594   & Y& &Y&AGN        & &Y& Y&disky     &3.9x10$^{-6}$   \\
53 &	       NGC\,4636   &  & & &Non-AGN CT &Y&Y& Y&outflow   &2.1x10$^{-6}$   \\
54 &	      NGC\,4676A   &  & & &Non-AGN    & & & N&outflow   &...$^b$             \\
55 &	      NGC\,4676B   &  & & &AGN        & & & Y&outflow   &6.3x10$^{-6}$  \\
57 &	       NGC\,4696   &  & & &Non-AGN    &Y$^c$&Y& Y&outflow   &3.9x10$^{-4}$  \\
58 &	       NGC\,4736   & Y& &Y&AGN        & &Y& Y&disky     &1.0x10$^{-5}$   \\
59 &	       NGC\,5005   &  &Y& &AGN* CT?   &Y&N& Y&outflow   &1.0x10$^{-5}$  \\
60 &	       NGC\,5055   &  & &Y&AGN  CT     & &N& Y&core-halo &1.0x10$^{-5}$   \\
70 &	       NGC\,5846   &  & & &Non-AGN CT &Y$^c$&Y& Y&outflow   &2.0x10$^{-5}$  \\
71 &	       NGC\,5866   &  & & &Non-AGN CT & &Y& Y&dusty     &5.0x10$^{-5}$  \\
\hline  
\end{tabular}
\end{center}
$^a$ From Wu \& Cao (2005)\\
$^b$ No BH mass is available for this source\\
$^c$ From Dunn et al. (2010)
\end{table*}

As we did in Sect. 3.2., the two components of the Mice system will
not be included in the discussion; their rather complex nature
resulting from the merger-like interaction between NGC 4676A and B, is
unique among our sample galaxies and may contaminate our results. Our
discussion will therefore be dealing with the remaining 30 galaxies.

From the 4 LINERs with a final classification as non-AGNs according to
Col. (9), NGC 3623 has an uncertain X-ray classification since it is
based in XMM-Newton data. The other 3 (namely NGC 3379, NGC 4314 and
NGC 4278) show an H$\alpha$ classification as disk-like systems. Their
X-ray data show evidences to be Compton-thick (Gonzalez-Mart\'in et al
2009b), what suggests that they could well host extremely obscured AGN activity.

For 3 out of the 26 confirmed AGN LINERs (namely NGC 3607, NGC 3627 and NGC
5866), their classification is only based on the detection of a broad
H$\alpha$ line, being classified at X-rays as non-AGNs.  They all have an
HST classification as dusty objects and appear to
be Compton thick at X-ray frequencies, so in these three cases a hint of a 
relationship 
between the obscuring materials could be claimed. 

In addition to these, 10 more AGN LINERs show evidences of CT nature. In 5
of them (namely NGC 2639, NGC 4374, NGC 4552, N4636 and NGC 5846) 
BLRs have been detected. Only a dusty environment is seen with the
HST data for NGC 4374. For the other 4 galaxies the obscuring material
seems to be most probably sitting in the innermost regions.  Seven out
of the 13 CT LINERs show their BLR, among which 4 of them have  dusty
H$\alpha$ morphologies.  Therefore, for the remaining 3 out of the 7 CT
LINERs there is no obscuring material at HST resolution that could be
invoked as the origin of its CT nature.  A similar result has been
found for Seyferts 1 (Malizia et al. 2009, Panessa et al. 2008),
questioning the dichotomy type 1/type 2 AGNs in the current
unification models (Urry and Padovani 2000). Finally for the remaining
CT narrow-line LINERs (NGC 2681, NGC 3245, NGC 4036, NGC 4438, NGC
5005 and NGC 5055) the obscuration cannot be atributed to important
dust lanes obscuring the nuclei. Summarizing the results on CT LINERs,
a large incidence is found in the dusty systems, since 4 out of the 5
dusty systems are CT; the remaining are distributed among the
different types.

Althought the number statistics are rather low, it is very interesting
to notice that among secured X-ray AGN-classed LINERs, based on
Chandra observations, ({ 19 out of 24 galaxies}, see Col. (6) in
Tab.~\ref{multiwav}) outflow and core-halo morphologies prevail (6
outflow systems, { 7} core-halo, 4 disk-like and { 2} dusty) amounting to 
{ 68\%}.
Taking the 26 galaxies AGN-classed based on
multifrequency data, 8 have been classified as
core-halo, 9 as outflow, 4 as disky and 5 as dusty. Therefore outflows and
core-halo represent 65\% of the AGNs.
   
Considering the Eddington ratios, the large range obtained (from
10$^{-7}$ to 10$^{-2}$) overlaps with the values found for Seyfert
galaxies (Panessa et al. 2006, Gonzalez-Mart\'in et al 2009a),
suggesting that LINERs are not always the low accretion cousins of
Seyferts.  We have found a slight trend for the Eddington ratios to
decrease when moving from core-halo to outflow and disky systems (see
Fig. \ref{Redd}). 
Dusty galaxies are not considered in the general trend since in the absence of dust 
they should fit in one of the other 3 classes. 
Different authors have claimed that strong radio jets are responsible for the bulk
of the radio emission observed in LINERs (Nagar et al. 2005, Filho et al. 2002)  and
that the radio loudness parameter (see Maoz 2007) can be related to the Eddington ratios
in the sense that Eddington ratios are larger for lower radio loudness ratios. Maoz (2007)
speculate that, in order to explain high Eddington ratios in low luminosity AGNs, mechanisms
preventing gas to reach the inner parts of the accretion disk would be at work; they suggest radio 
loudness at low
luminosities as such a solution, with the gas joining a jet or an outflow. Our data seem to support such a
hypothesis since 1) radio loud systems are found in core-halo and outflow systems and, even more
important, 2) all the outflow systems appear to be radio-loud.


   \begin{figure}
    \centering
\setcounter{figure}{4}
       \includegraphics[width=0.7\columnwidth,angle=-90]{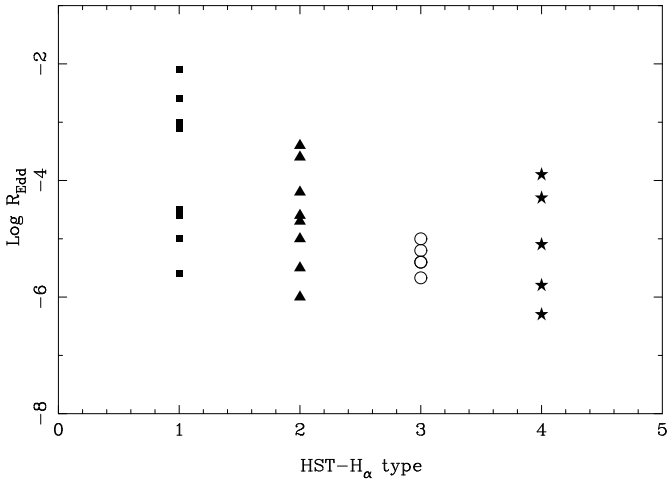} 
   \caption{Eddington ratios as a function of the HST-H$\alpha$
     classification for the AGN LINERs in our sample. Symbols are the
     same as in Fig.~\ref{ReqLX}.}
           \label{Redd}
    \end{figure}

\section{Summary and conclusions }

We have presented HST-H$\alpha$ imaging of 32 LINERs, 
selected from the X-ray sample studied in our previous works 
(Gonz\'alez- Mart\'in et al. 2009a and b).
A full description of the extraction and reduction process is given and the 
resulting emission-line images are also presented together with the { sharp divided}
continuum images 
for each galaxy. The description of the most relevant properties for each 
individual galaxy is also given. 

The main conclusion from this analysis is that, for the large majority
of LINERs, an unresolved nuclear source has been identified,
together with extended emission with equivalent sizes ranging from few
tens of parsecs till about 500 pc. Adding up additional 4 LINERs from
the literature to our sample, we conclude that their emission-line morphologies
appear not to be homogeneous, being basically grouped into three
classes: nuclear outflow candidates (42\%), core-halo morphologies
(25\%) and nuclear spiral disks (14\%), being the remaining 5 galaxies
too dusty to allow a clear view of the ionized distribution. Except
maybe for the only case of a merger-like interaction (the two galaxies
in the Mice system), no signatures of clumpy structures reminiscent of
star clusters have been identified, in agreement with { results
from stellar population analisis (Gonz\'alez-Delgado et al. 2004 and Sarzi et al. 2006).}

A size-luminosity relation has been found between the equivalent radius 
of the H$\alpha$ emission and the hard X-ray luminosity. This correlation ressembles that reported 
for the NLR of Seyferts galaxies based in the [OIII] 
luminosity (Schmitt et al. 2003). This relation is another piece of 
evidence confirming  
the AGN-NLR nature of the ionized gas 
in LINERs (Pogge et al. 2000, Walsh et al. 2008).

Indications of a relationship between soft X-rays and H$\alpha$ emission in LINERs are also 
reported for the first time. This spatial correlation looks
similar to the one reported by Bianchi et al (2006) for Seyferts, 
evidencing the photoionized nature of the soft X-rays.

For the only 4 LINERs with no evidences for AGN nature of their nuclear emission, 
a CT AGN cannot  be discarded out given the properties of their X-ray emission. For
the confirmed AGN-LINERs, their H$\alpha$ morphologies favour
core-halo and outflow systems (65\% of the cases). Finally, Eddington
ratios have been calculated showing that LINER nuclei radiate in the
sub-Eddington regime, in agreement with previous data (Maoz 2007, Ho 2008,
Eracleous et al. 2010a). However core-halo systems tend to have
larger Eddington ratios than outflow candidates on average. These result
may be consistent with the suggestion by Maoz (2007) of 
radio-loud outflow related systems showing smaller Eddington ratios. 

\begin{acknowledgements}
J.M. and I.M. acknowledge financial support from the Spanish grant
AYA2007-62190 and Junta de Andaluc\'{\i}a TIC114 and the Excellence Project P08-TIC-03531.
 O.G-M acknowledges finantial support by the EU FP7-REGPOT 206469 and ToK 39965 grants.
{ A.R. acknowledge finantial support from CONACyT grant number +081535. We acknowledge 
the valuable feedback from an anonymous referee}.
Based on observations made with the NASA/ESA Hubble
  Space Telescope, and obtained from the Hubble Legacy Archive (HLA
  hereinafter), which is a collaboration between the Space Telescope
  Science Institute (STScI/NASA), the Space Telescope European
  Coordinating Facility (ST-ECF/ESA) and the Canadian Astronomy Data
  Centre (CADC/NRC/CSA). { This research has made use of the NASA/IPAC Extragalactic Database (NED) which is operated by the Jet Propulsion Laboratory, California Institute of Technology, under contract with the National Aeronautics and Space Administration.}
\end{acknowledgements}

\begin{appendix}
\section{Comments on individual objects}

{\bf IC 1459.} 
At HST resolution Lauer et al. (2005) classified this
galaxy as an starting dusty nuclear ring and Verdoes Kleijn et al.
(2000), based in H$\alpha$+[NII] WFPC2 images, identify an ionized gas disk
with PA 37$^o$ and inclination  60$^o$ following what is found
by Goudfrooij et al. (1990) at larger scales (100 arcsec). 
Here we report a residual H$\alpha$
emission with a central source and a rather  biconical
distribution extended $\sim$ 500 pcs at PA 37$^o$ (Fig.~\ref{fig-images}). 
At soft 
 X-ray energies (0.3-2 keV), it extends along the same direction than that in H$\alpha$
HST data (Fig.~\ref{softXvsHST}).\\

\noindent {\bf NGC 315.} 
A compact unresolved source of 
ionized gas on top of a dusty 
disk, together with  an extension 
of the disk of 200 pc  at PA 49$^o$ is detected (see Fig.~\ref{fig-images}; see also Verdoes-Keijn et al. 1999). The high spatial resolution 
provided by Chandra imaging allowed the detection of the X-ray jets 
(Donato et al., 2004; Worrall et al., 2003, 2007; Gonzalez-Martin et al. 
2009). Soft X-rays are extended along the axes of the jet and the host galaxy (Fig.~\ref{softXvsHST}). 
\\

\noindent {\bf NGC 2639.} 
At HST resolution, its H$\alpha$ shows an elongated asymetric extended
structure at PA -29$^o$, but not nuclear compact source is
identifiable. It also shows extended filaments being more prominent
towards the NW. The SE region is maybe obscured by dust. The SD images
and the broad band data show a rather
dusty morphology (see Fig.~\ref{fig-images} and Sim{\~o}es Lopez et al. 2007).\\

\noindent {\bf NGC 2681.}
The H$\alpha$ emission image show a central source with extended emission along 
the central spiral structure with major axis at PA 40$^o$ and a radial 
extension of 4 arcsecs ($\sim$ 440 pcs). Spiral dust lanes are clearly detected
in the SD image (Fig.~\ref{fig-images}). Broad agreement is found in the elongation of soft X-ray and H$\alpha$ emissions (Fig.~\ref{softXvsHST}).\\

\noindent {\bf NGC 2787.}
On the HST SD images, a near-nuclear dust-lane 
is clearly resolved into a spectacular set of concentric, elliptical dust 
rings,  covering a radial range of 5–10\arcsec (see also Shields et al. 2007, Sim{\~o}es-Lopes et al.
2008, Gonzalez Delgado et al. 2008). In H$\alpha$,
 a nuclear component has been 
detected, in good agreement with Dai \& Wang (2009). An elongation 
at PA 49$^o$ can also be identified, which is perpendicular to the major axis of the galaxy (Fig.~\ref{fig-images}). Soft X-rays roughly follow the H$\alpha$ emission (Fig.~\ref{softXvsHST}).\\

\noindent {\bf NGC 2841.}
At HST resolutions
it shows a rather face-on ring-like structure and a clearly well identified 
unresolved nuclear source. A small NLR can be identified  at PA of 90$^o$. Dust morphology becomes 
apparent from the SD image (Fig.~\ref{fig-images}). 
Soft X-rays extend along two main axes, one following the hard X-ray emission (PA about 10) and the other one close to that of the H$\alpha$ emission (Fig.~\ref{softXvsHST}).\\

\noindent {\bf NGC 3226.} 
This galaxy shows a bright nucleus with some evidences of 
dusty environment clearly seen in the SD image (see also Gonzalez Delgado et al. 2008) . 
In $H_{\alpha}$ it shows an extended morphology quite similar 
to that observed in continuum (Fig.~\ref{fig-images}). {At X-ray frequencies 
it shows a compact structure both at soft and hard energies. The  H$\alpha$ however 
seems to suggest an outflow-like morphology coming out from that compact nucleus 
(Fig.~\ref{softXvsHST}).}\\


\noindent {\bf NGC 3245.} 
The $H_{\alpha}$ image 
shows a kidney-like structure slightly brighter 
to the North, with a nuclear unresolved source
(see also (Gonzalez Delgado et al. 2008, Walsh et al. 2008). Kinematical data from
Walsh et al. (2008) support our outflow classification. 
The western dust structure is clearly appreciated in the SD image. (Fig.~\ref{fig-images}). One of the two axes shown by the soft X-ray contours follows the H$\alpha$ emission (Fig.~\ref{softXvsHST}).\\

\noindent {\bf NGC 3379.}
An extended structure 
emerging from the nucleus can be apreciated
althought the S/N ratio on the H$\alpha$ image is low. A tiny dust-lane crosses the nuclear regions at PA -50 in SD (Fig.~\ref{fig-images}). At HST resolution Lauer et al. (2005), based in the F555W filter,
classified this galaxy without a clear nuclei but with a dusty nuclear
ring morphology.  Shapiro et al. (2006) reported a well defined disk of
emission at H$\alpha$ with PA 118. The morphology of the soft X-ray contours is quite complex, but a rough agreement with the extended H$\alpha$ emission is found (Fig.~\ref{softXvsHST}).\\ 

\noindent {\bf NGC 3607.}
The H$\alpha$
image shows a clearly nuclear unresolved source and diffuse emission
following what it appears to be an inclined disk. The strong dust lanes visible in the SD images 
obscure the H$\alpha$ emission (Fig.~\ref{fig-images}). 
Lauer et al. (2008) suggested that it
contains a dusty outer disk dynamically old which appears to
transition rapidly but smoothly at the center to a second gas disk
that is perpendicular to the first and is seen nearly edge-on.  This
inclined disk seems to be settling onto a nuclear ring. Excepting the outermost contours, the soft X-ray emission elongates along the H$\alpha$ emission (Fig.~\ref{softXvsHST}).\\
 
\noindent {\bf NGC 3623.} 
H$\alpha$ emission has been detected, extending 
$\sim$ 130 pc at PA -10$^o$. 
Inside the more extended 
structure an inner disk is appreciated extending 30 parsecs along 
PA 53$^o$. Large scale dust-lanes clarly appear in the SD image (Fig.~\ref{fig-images}).\\

\noindent {\bf NGC 3627.}
The $H_{\alpha}$ data (Fig.~\ref{fig-images}) do not show a well defined nuclear source, most probably due to the dust lane crossing the nuclear region in the direction NS and obscuring
the SE-NW elongated extended emission (see the SD image). Gonzalez-Delgado et al. 2008 reported from HST data that  
chaotic dust lanes and  several compact sources are identified at the center.\\

\noindent {\bf NGC 3998.}
The H$\alpha$
image (Fig.~\ref{fig-images}) 
shows a 100 pc extended structure surrounding a compact nucleus. 
The major axis of this extension is oriented along a PA=0  (see also 
Pogge et al. 2000). The SD image shows 
little indication of dust in the nuclear region, in good agreement
with Gonzalez Delgado et al. (2008). Soft X-rays are elongated along the same direction as the H$\alpha$ emission (Fig.~\ref{softXvsHST}).\\

\noindent {\bf NGC 4036.}
The HST H$\alpha$ image (Fig.~\ref{fig-images})
shows, on top of a well identified nuclei, the existence of a
complicated filamentary and clumpy structure, with an extension of 390
pcs at PA 63$^o$, already reported by Pogge et al. (2000) and Dai and
Wang (2009) (see also the SD image).  Walsh et al. (2009) have shown
the presence of a gas velocity gradient of $\sim$ 300 km s$^{-1}$
across the inner 0.2", compatible with the outflow-like structure apparent in the ionised gas. 
The soft X-ray emission appears to follow the H$\alpha$
emission (Fig.~\ref{softXvsHST}).\\

\noindent {\bf NGC 4111.}
A rather knotty morphology surrounding a clear nuclear source is
observed, embebed in a diffuse halo. This morphology is interpreted as
a core-halo structure detected at HST resolution both with
medium size filters (Sim{\~o}es Lopes et al. 2007) and narrow band
H$\alpha$ data (Dai and Wang, 2009).  A crossing dust structure is
seeing perpendicular to the disk main plane (see SD image). Soft X-ray contours are elongated along the same PA as the H$\alpha$ emission (Fig.~\ref{softXvsHST}).\\

\noindent {\bf NGC 4278.}
%
{ A clear core-halo morphology is shown by its  H$\alpha$ emission on the top of a very faint continuum (Fig.~\ref{fig-images}). This emission seems to follow what it is observed in the soft X-ray emission (Fig.~\ref{softXvsHST})}.\\

\noindent {\bf NGC 4314.}
The H$\alpha$ image (Fig.~\ref{fig-images}) shows both an 
unresolved nucleus and a number of 
HII regions tracing the star formation ring.
The same features are well traced by the SD image, where the spiral dust lanes associated with the ring are conspicuous (see also Gonzalez Delgado et al. 2008). 
At soft energies, its emission follow the
star forming regions observed in H$\alpha$ (Fig.~\ref{softXvsHST}).\\

\noindent {\bf NGC 4374.}
H$\alpha$ image (Fig.~\ref{fig-images}) shows an inclined gas disk surrounding the nucleus.
This emission gas structure takes the form of filaments that extend 
roughly east-west and north-south (see also (Pogge et al. 2000).
The dust structure clearly appears in the SD image, where the nucleus is seen in the center of the dust-lane to the South. The soft X-ray contours are roughly aligned with the ionised gas (Fig.~\ref{softXvsHST}).\\

\noindent {\bf NGC 4438.}
Gonzalez Delgado et al(2008) defined it as galaxy with 
very perturbed central morphology and strong dust lanes cross 
the center along PA 0 obscuring the eastern side of the galaxy (see the SD image). The H$\alpha$ image (Fig.~\ref{fig-images}) shows a ring-like structure 
where a clear knot is seen in the south-east region coincident with the 
continuum nucleus. The other side would remain invisible due to obscuration 
by dust.  Two plumes can be 
apreciated to the north and south west extending about 150 pcs in both 
directions. This is one of the clearest examples to be a candidate of nuclear 
outflow, buble structures, as defined in Veilleux and Brandt (2007). Soft X-rays are aligned with the H$\alpha$ emission (Fig.~\ref{softXvsHST}).\\
 
\noindent {\bf NGC 4486.}
The H$\alpha$ image (Fig.~\ref{fig-images}) shows a compact source with filaments which
resemble an outflow from the nucleus (see also Pogge et al. 2000; Dai
and Wang 2009). As already noticed by Pogge et al. (2000), the
conpiscous jet clearly visible in the continuum images (see SD)
dissapears in the H$\alpha$ continuum substrated map. Soft X-rays are missaligned with respect to H$\alpha$ emission, the former following the jet axis (Fig.~\ref{softXvsHST}).\\

\noindent {\bf NGC 4552.}
At HST resolution the H$\alpha$ data show a compact unresolved nuclear
source located at the center of a symmetric extended emission in a disk-like structure.
No trace of dust-lanes is seen in the SD image (Fig.~\ref{fig-images}). Soft X-rays roughly follow the H$\alpha$ emission (Fig.~\ref{softXvsHST}).\\

\noindent {\bf NGC 4579.}
The H$\alpha$ emision (Fig.~\ref{fig-images}) traces a bright, nuclear point source surrounded 
by complex clumpy and filamentary emission (see also Pogge et al. 2000).
The higher ionization gas traced by [OIII] (Fig.~\ref{fig-images}) is composed of a compact
source and a filamentary, jet-like structure towards the NE.
Walsh et al. (2008) have shown that the gas is not in regular rotation, displaying two
kinematical components with a velocity separation of 450 km s$^{-1}$,
being consistent with an outflow from the nucleus. The dust-lanes
seen in the SD image conform a mainly chaotic structure together with a 
much stronger, offset, linear feature that goes at
PA$\approx$45 in the West side. Soft X-ray contorus follow the H$\alpha$ emission at large scales. There is a hint of an extension of hard X-rays along the PA of the [OIII] jet-like feature (Fig.~\ref{softXvsHST}).\\

\noindent {\bf NGC 4594.}
The H$\alpha$ image (Fig.~\ref{fig-images}) shows a compact nuclear source together with
fainter emission extending along the E-W direction in a bar-like
morphology, with two spiral arms emerging from it, with a total
extension of 300 pc. The kinematical data by Walsh et al. (2008) show
organized motion consistent with rotation but with significant
irregularities in the nucleus.  A strong velocity gradient 
and decoupled kinematics between gas and stars were found
by Emsellem \& Ferruit (2000). An overall extension of soft X-rays is seen along the same axis as the extended ionised gas (Fig.~\ref{softXvsHST}).\\

\noindent {\bf NGC 4636.} 
The H$\alpha$ data (Fig.~\ref{fig-images}) show a
central compact source and a very faint ring like structure more clearly visible
in the southern region of the galaxy (see also Sim{\~o}es Lopes et
al. 2007 and Dai and Wang 2009). 
{ Towards the north, a more prominent H$\alpha$ emission is seeing with a clear outflow-like morphology. This morphology seems to follow the soft X-ray data (Fig.~\ref{softXvsHST})}.\\

\noindent {\bf NGC 4676A and B. The Mice.} 
The H$\alpha$
images (Fig.~\ref{fig-images}) 
show in both galaxies a very clumpy and irregular structure. In 
galaxy B a more conspicuous knotty structure is visible (one of the knots coincides with 
the nucleus). In galaxy A however a more diffuse emission is seen.
Central dust-lanes are much stronger in galaxy B, as can be appreciated in the SD images (see also Laine et al. 2003). The soft X-ray contours are unrelated to the extended H$\alpha$ emission in both galaxes (Fig.~\ref{softXvsHST}).\\

\noindent {\bf NGC 4696.}
H$\alpha$ imaging (Fig.~\ref{fig-images}) shows a clear nuclear source with elongated extended emission
along PA 47$^o$, and  larger filamentary structures towards the west 
maybe resembling outflows out of the nucleus.Crawford et al (2005) report also 
a filamentary structure shared by the H$\alpha$ and soft X-ray emission (see also our Fig.~\ref{softXvsHST}).

\noindent {\bf NGC 4736.}
The H$\alpha$ image 
shows a circumnuclear spiral structure of extension 200 pc. Dust lanes in 
spiral arms are traced in the SD image (Fig.~\ref{fig-images}).
Gonzalez Delgado et al. (2008)
suggest the presence of spiral dust lanes down the nucleus and a compact 
nuclear stellar cluster. Despite the complexity of the soft X-ray emission, a rough agreement is found in the overall shape of both images (Fig.~\ref{softXvsHST}).\\

\noindent {\bf NGC 5005.}
H$\alpha$ data (Fig.~\ref{fig-images}) show a very asymmetric emission with a wide-angle cone-like structure extending to the SE (see Pogge et al. 2000 and Dai $\&$ Wang 2009), 
perpendicular to the major axis of the galaxy.
A strong dust lane crosses the galaxy from East to West, offset from the nucleus (see the SD image and Gonzalez Delgado et al. 2008).\\

\noindent {\bf NGC 5055.}
Its H$\alpha$ image (Fig.~\ref{fig-images}) shows 
a nuclear source and extended emission along 
PA 110$^o$. 
A floculent spiral structure, stronger to the Souht, is visible in the SD image (see also Gonzalez Delgado et al. 2008). Soft X-rays and H$\alpha$ emission are extended along roughly the same direction (Fig.~\ref{softXvsHST}).\\

\noindent {\bf NGC 5846.} 
The H$\alpha$ image (Fig.~\ref{fig-images}) shows a compact nucleus and diffuse emission
resembling a very wide outflow extending up to 2" in the W
direction. This strong asymmetry cannot be explained by dust absorption
(see the SD image). No correaltion is seen between H$\alpha$ and soft X-ray emission (Fig.~\ref{softXvsHST}).\\

\noindent {\bf NGC 5866.}
The H$\alpha$ image shows an extremely faint nucleus on top of a very
dusty structure along PA -45$^o$ strongly obscuring the nucleus ( see
also the SD image) what hampers either any classification of the emission (Fig.~\ref{fig-images}) or any comparison with the soft X-ray emission (Fig.~\ref{softXvsHST}).\\

\end{appendix}

   \begin{figure*}
    \centering
       \includegraphics[]{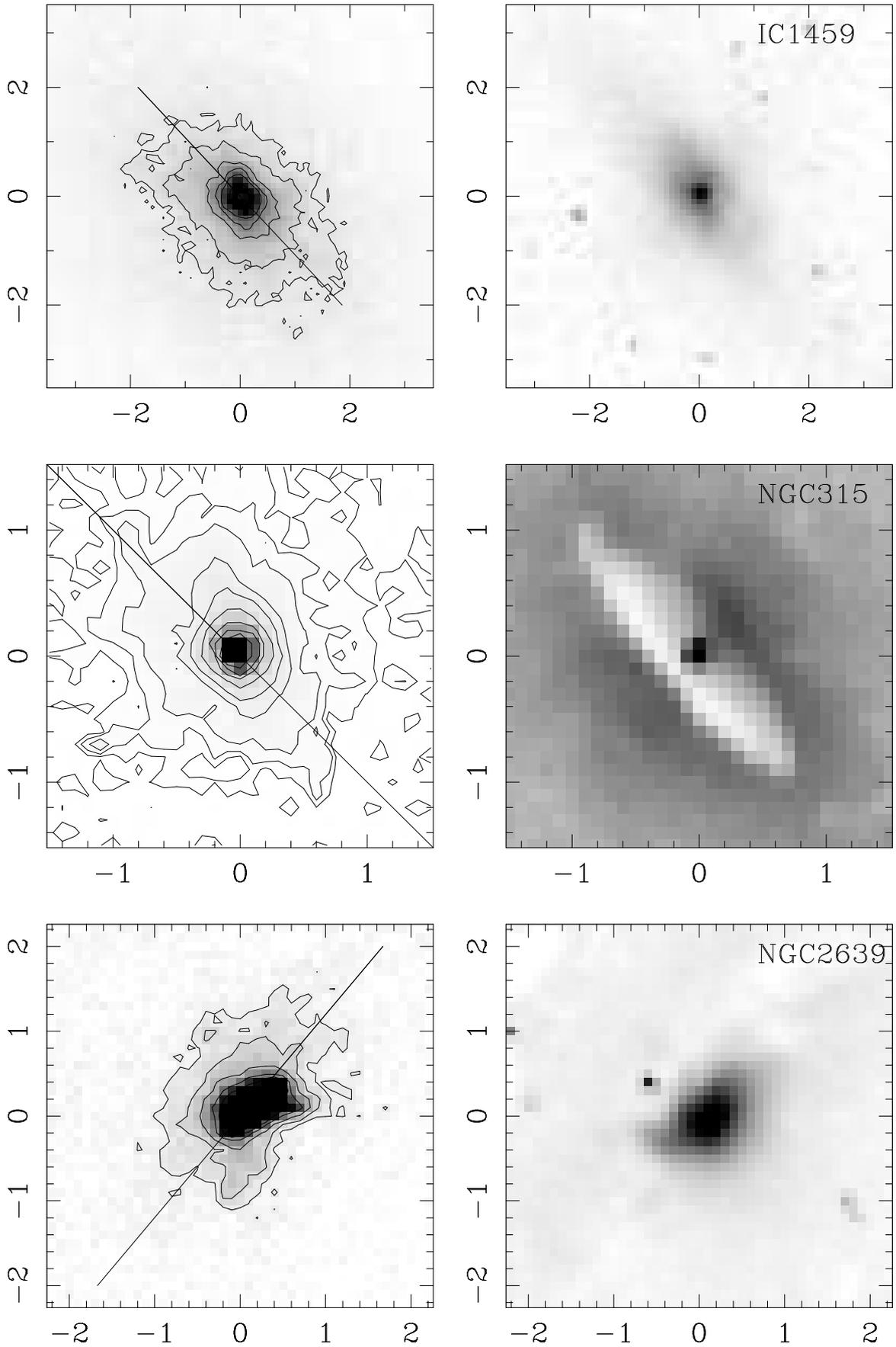}
\setcounter{figure}{1}
   \caption{Images of H$\alpha$ (left) and SD (right). Top is north and east is left. {The units of the plots are arcseconds. 
For clarity contours above  3$\sigma$ levels  have been  plotted in the H$\alpha$ images. For the outflow candidates the contour for which R$^*$$_{eq}$ has been estimated is 
also plotted in black or white thick line. The position angle of the host major axis has been taken from the ned database and it is shown as a solid line.}}
           \label{fig-images}
    \end{figure*}

   \begin{figure*}
    \centering
       \includegraphics[]{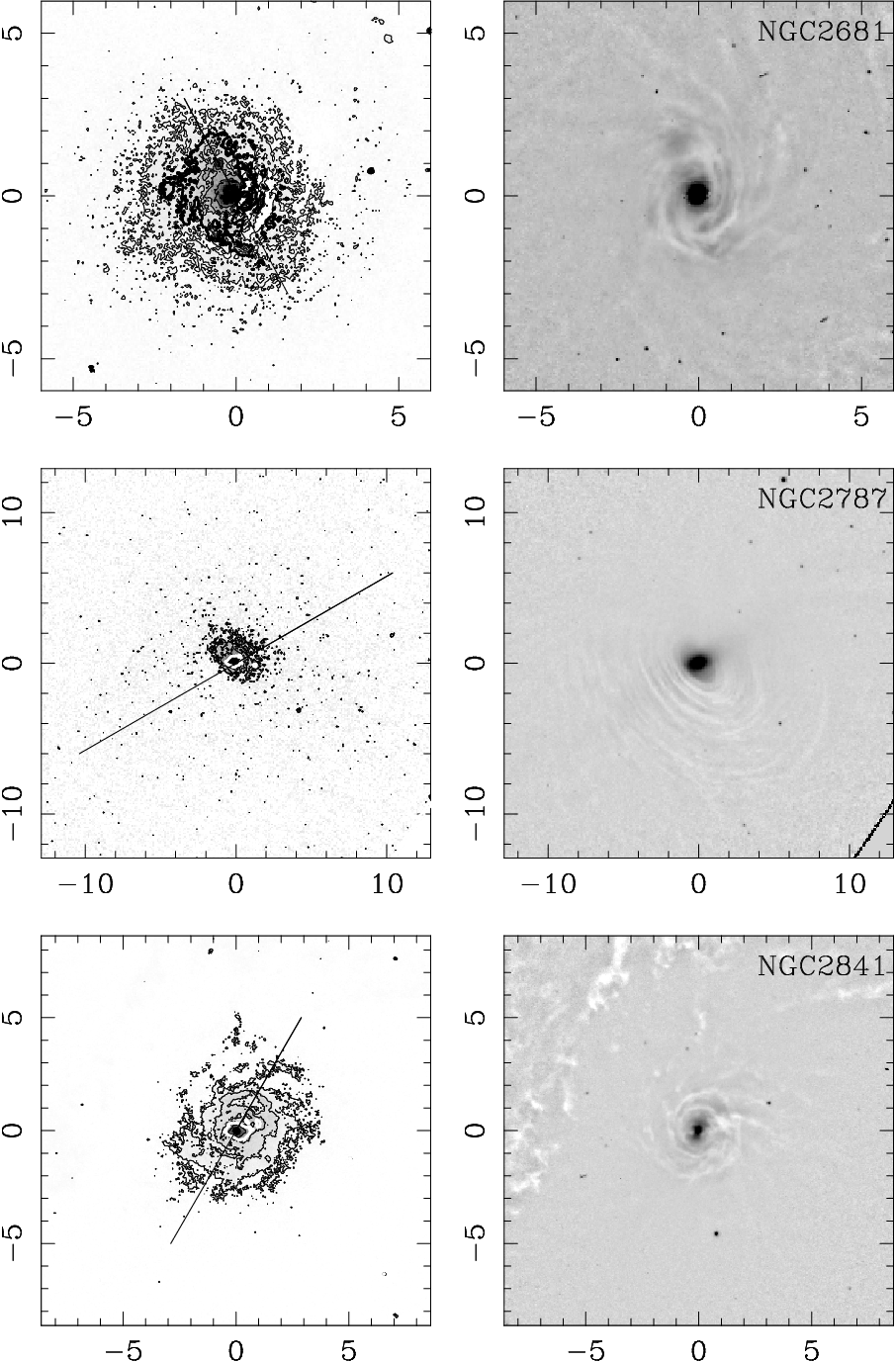}
\setcounter{figure}{1}
   \caption{Continued.}
           \label{}
    \end{figure*}

   \begin{figure*}
    \centering
      \includegraphics[]{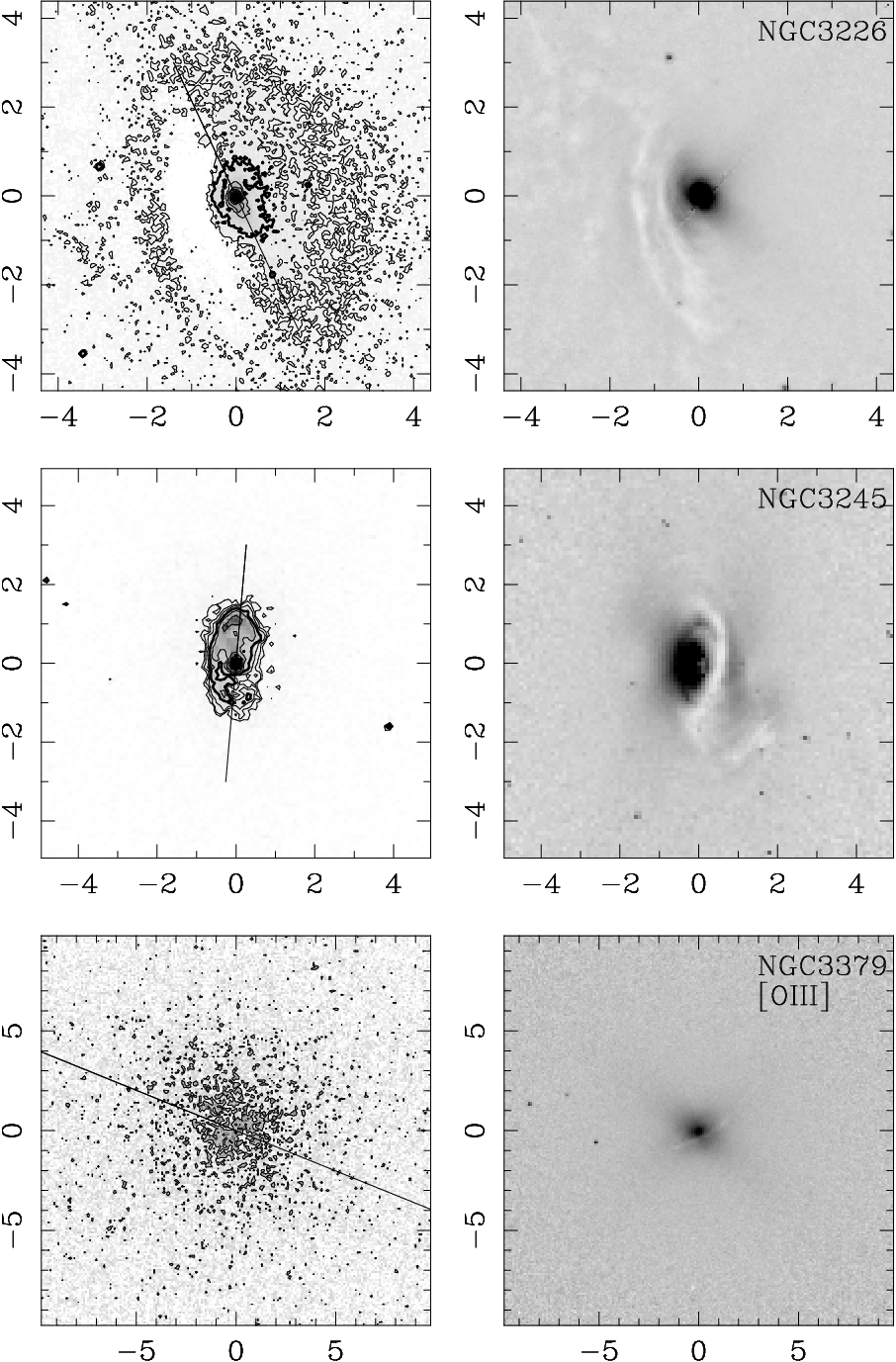}
\setcounter{figure}{1}
   \caption{Continued.}
           \label{}
    \end{figure*}

   \begin{figure*}
    \centering
       \includegraphics[]{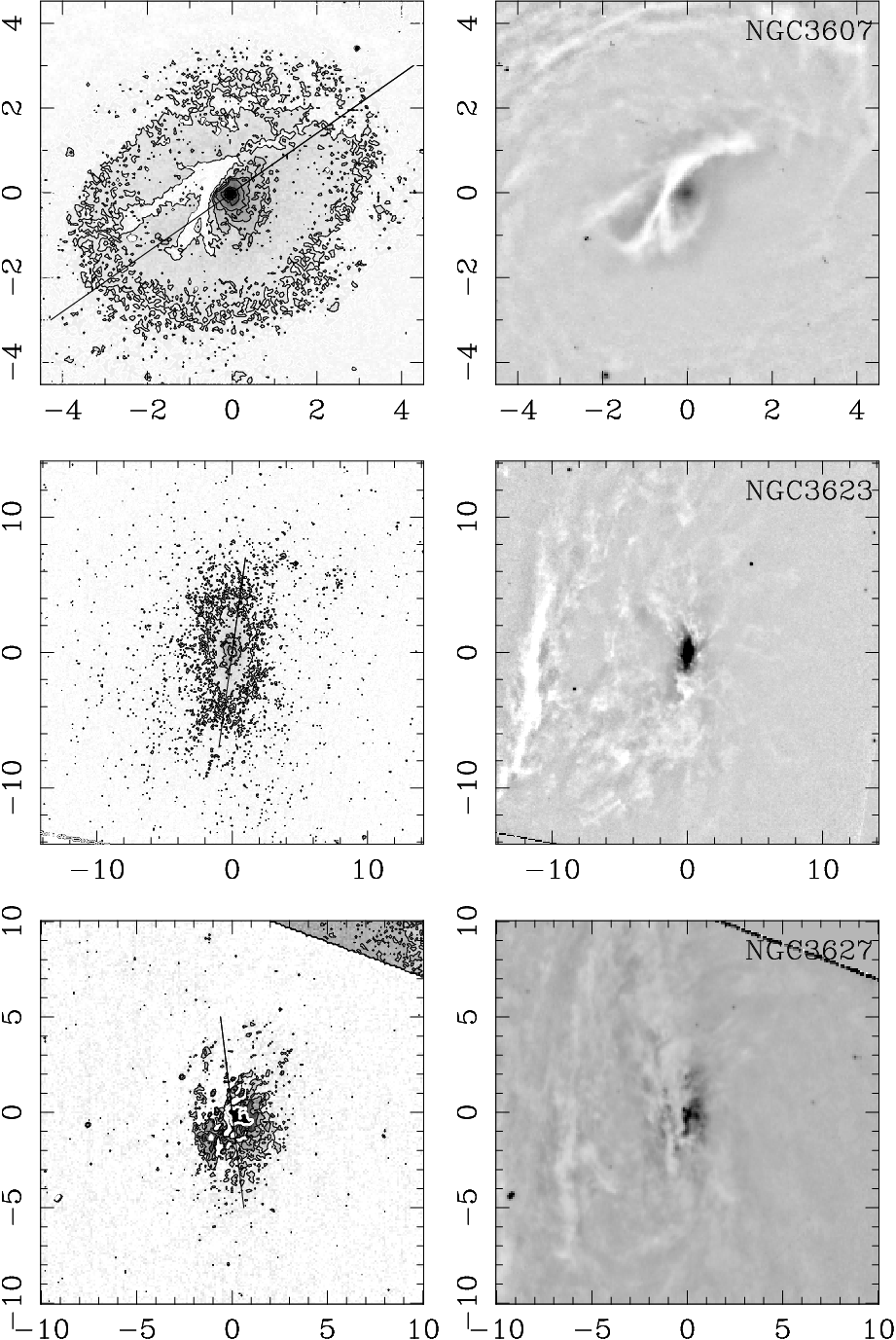}
\setcounter{figure}{1}
   \caption{Continued.}
           \label{}
    \end{figure*}

   \begin{figure*}
    \centering
       \includegraphics[]{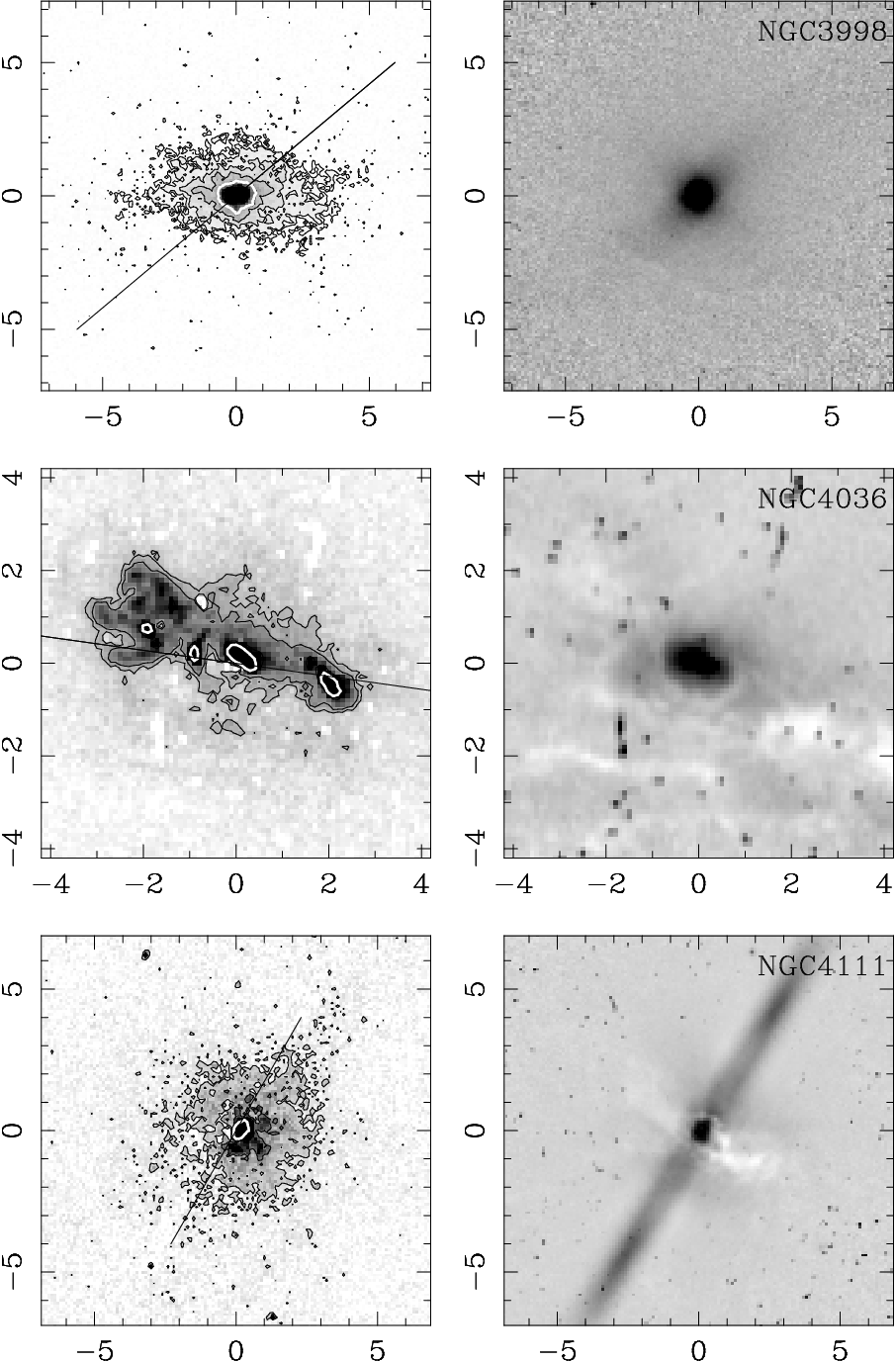}
\setcounter{figure}{1}
   \caption{Continued.}
           \label{}
    \end{figure*}

   \begin{figure*}
    \centering
       \includegraphics[]{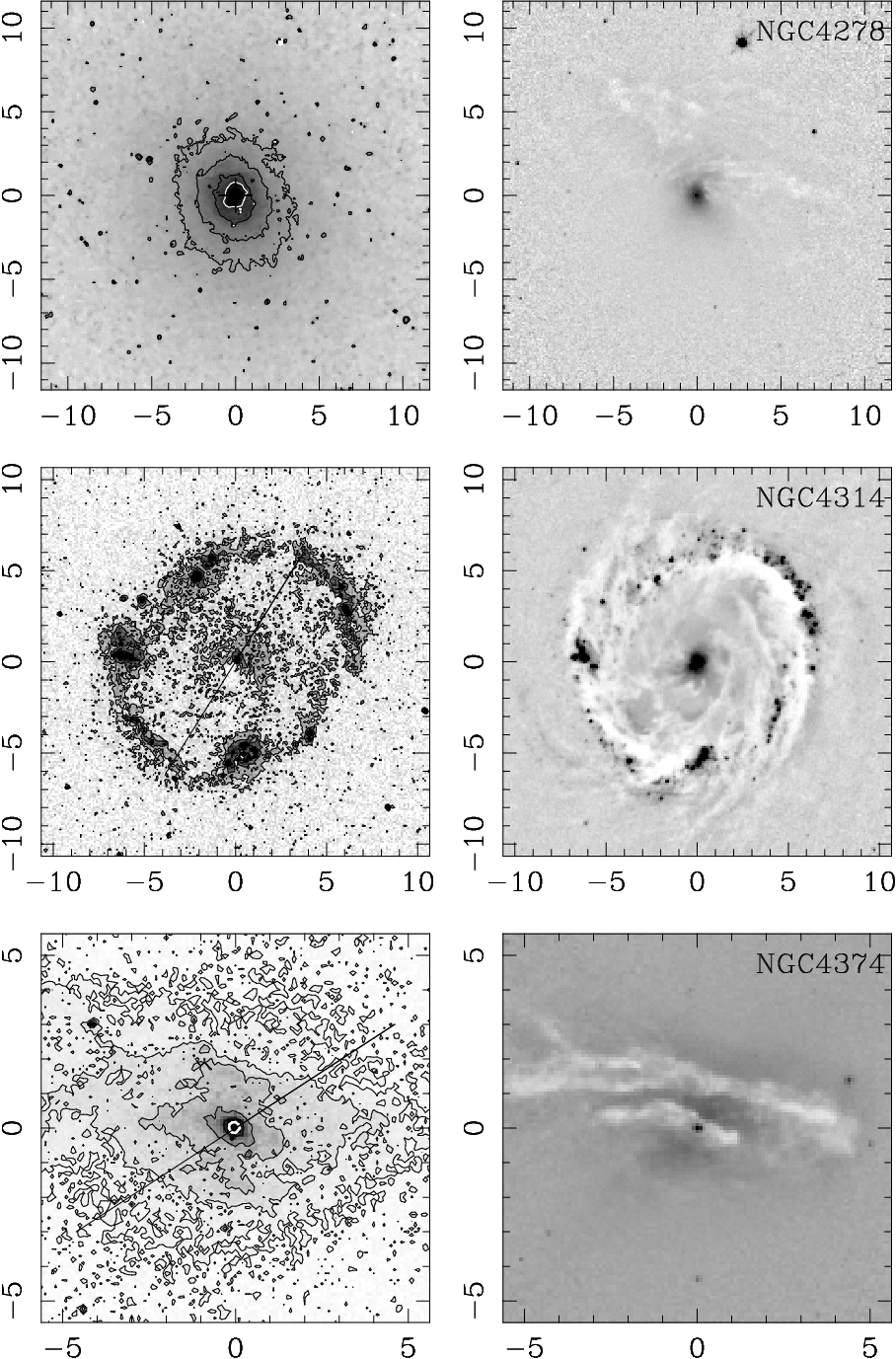}

\setcounter{figure}{1}
   \caption{Continued.}
           \label{}
    \end{figure*}

   \begin{figure*}
    \centering
       \includegraphics[]{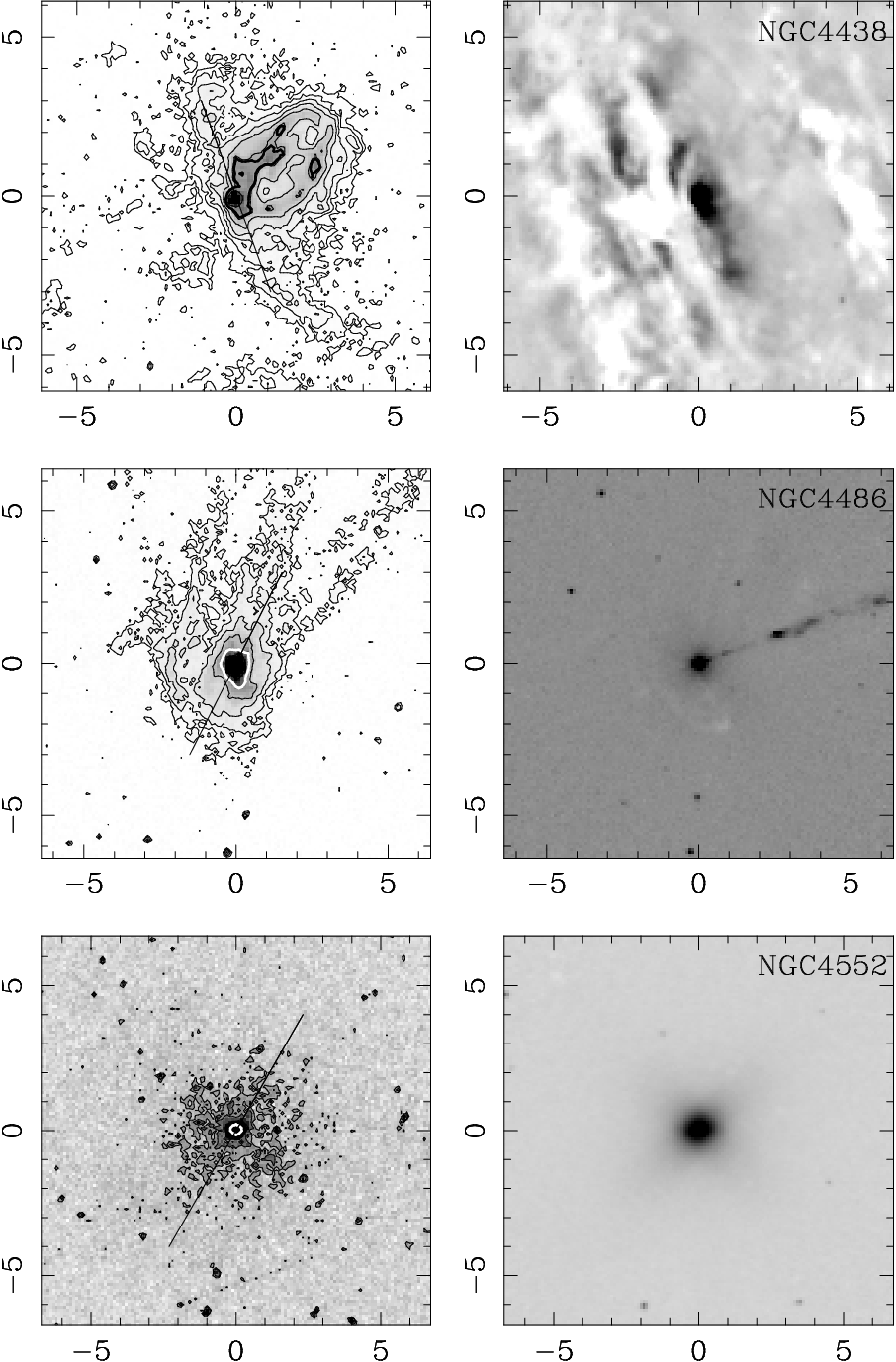}
\setcounter{figure}{1}
   \caption{Continued.}
           \label{}
    \end{figure*}

   \begin{figure*}
    \centering
       \includegraphics[]{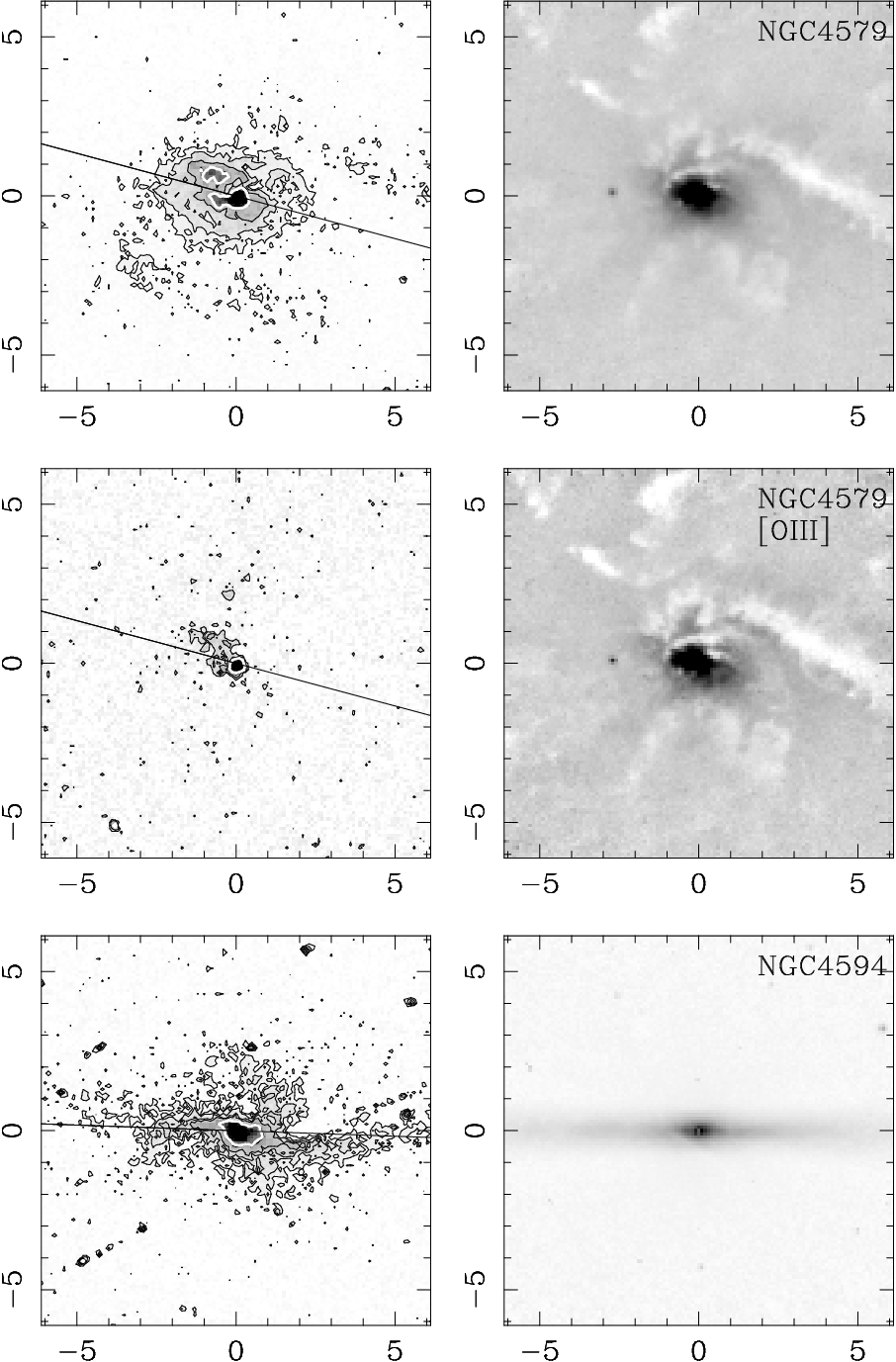}
\setcounter{figure}{1}
   \caption{Continued.}
           \label{}
    \end{figure*}

   \begin{figure*}
    \centering
       \includegraphics[]{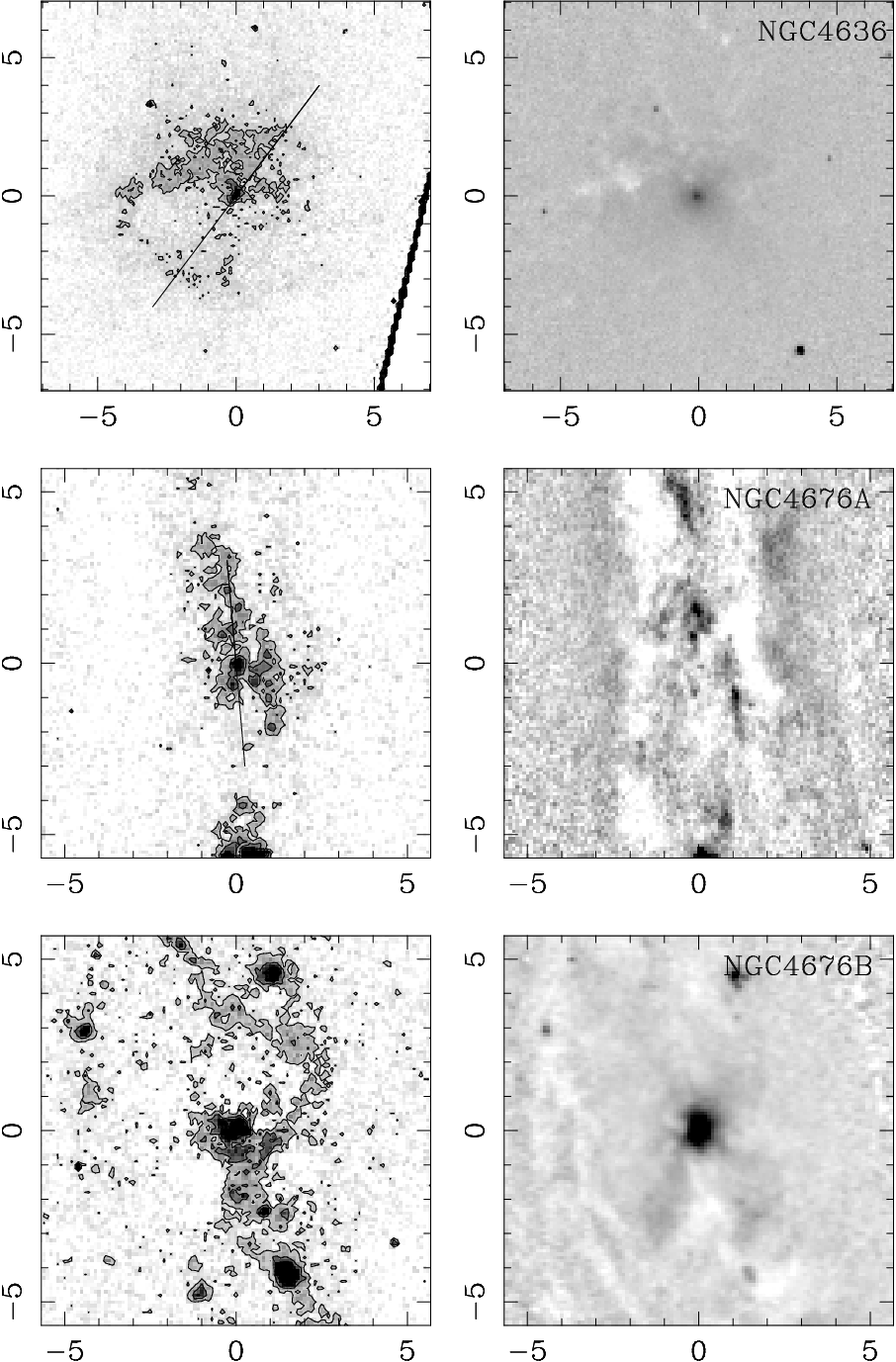}
\setcounter{figure}{1}
   \caption{Continued.}
           \label{}
    \end{figure*}

   \begin{figure*}
    \centering
       \includegraphics[]{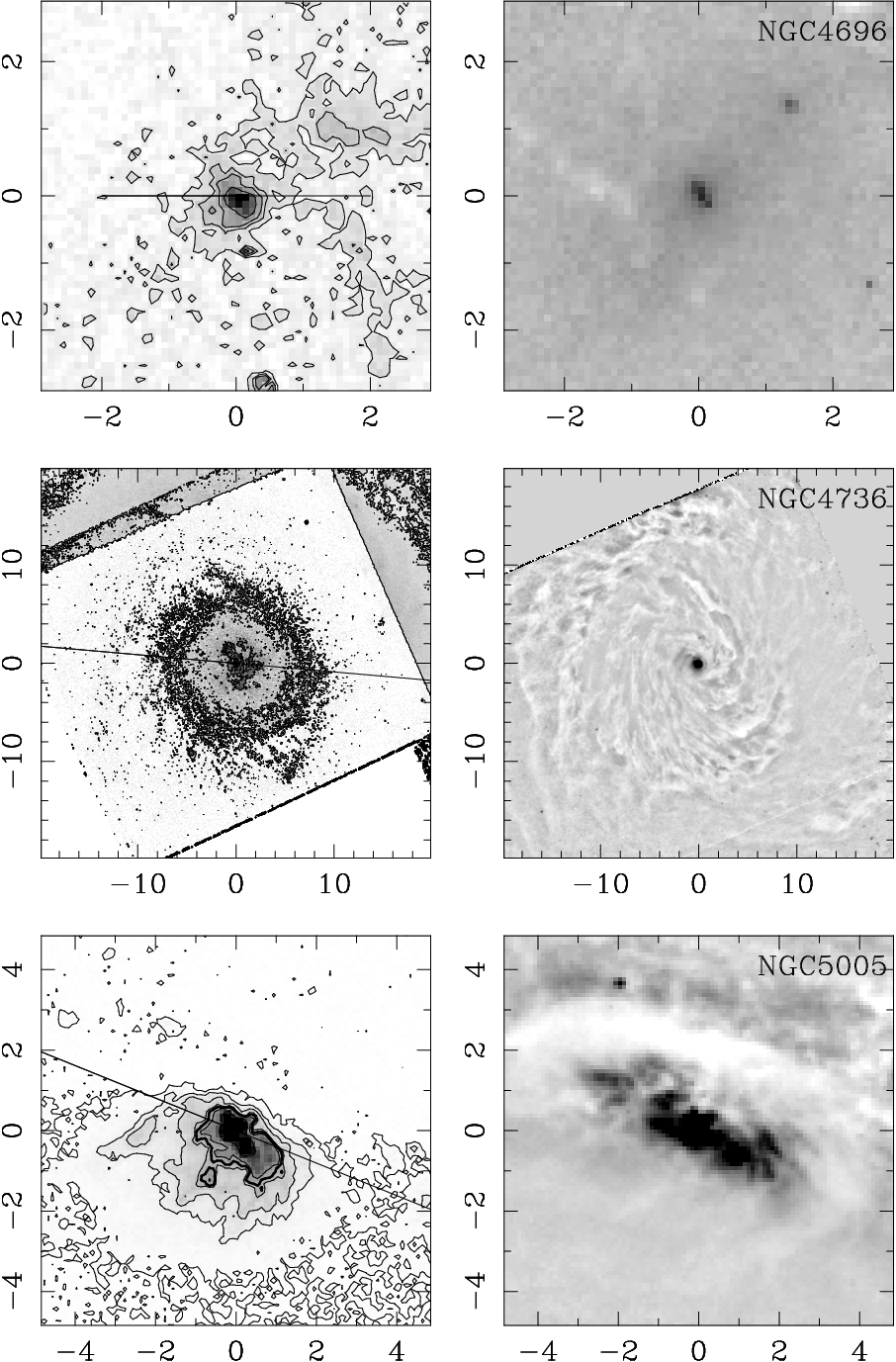}
\setcounter{figure}{1}
   \caption{Continued. }
           \label{}
    \end{figure*}

   \begin{figure*}
    \centering
       \includegraphics[]{fig-images-11.ps}
\setcounter{figure}{1}
   \caption{Continued.}
           \label{}
    \end{figure*}

\begin{figure*}[b]
  \centering
\includegraphics[width=0.90\columnwidth]{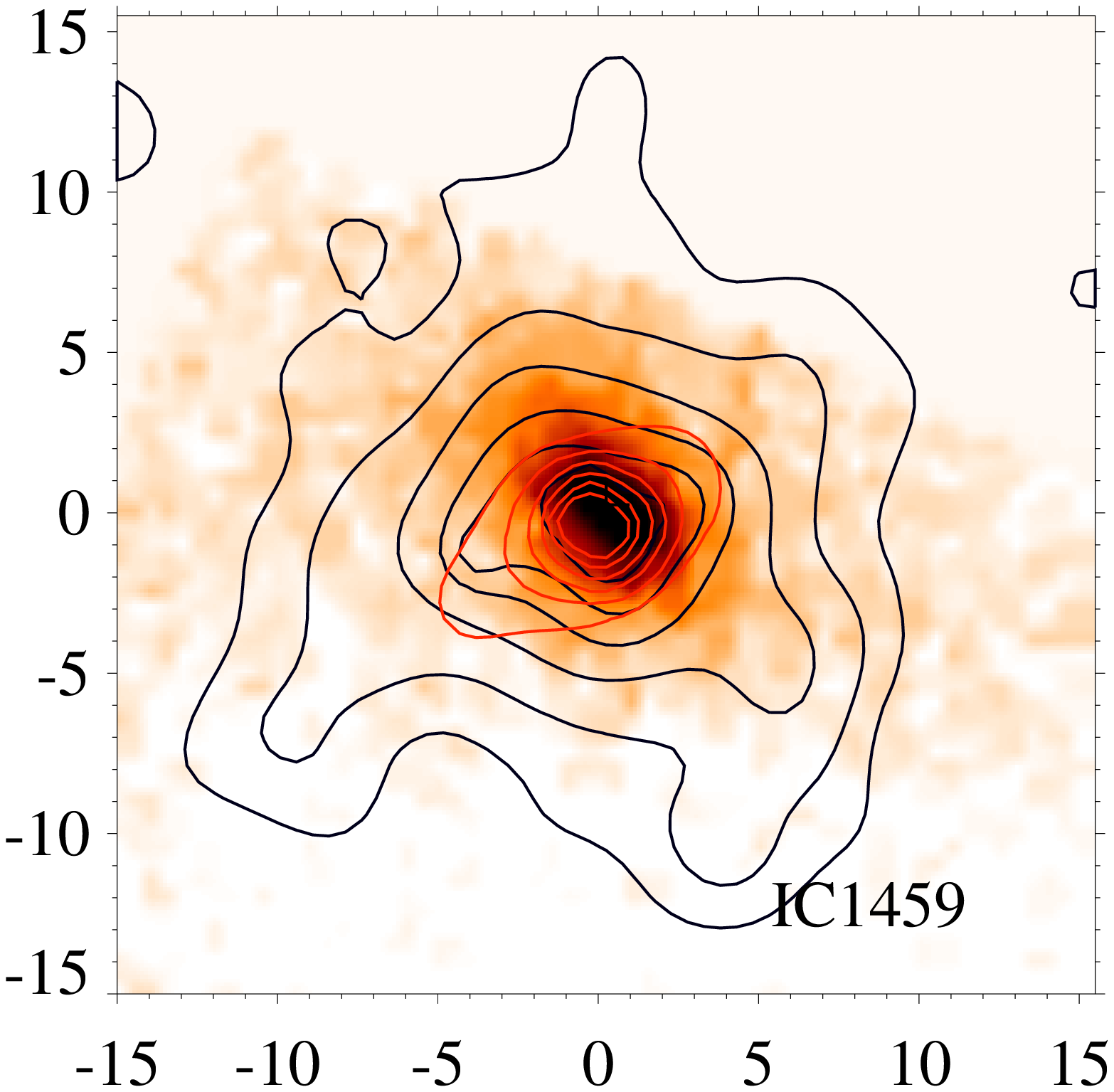}
\includegraphics[width=0.90\columnwidth]{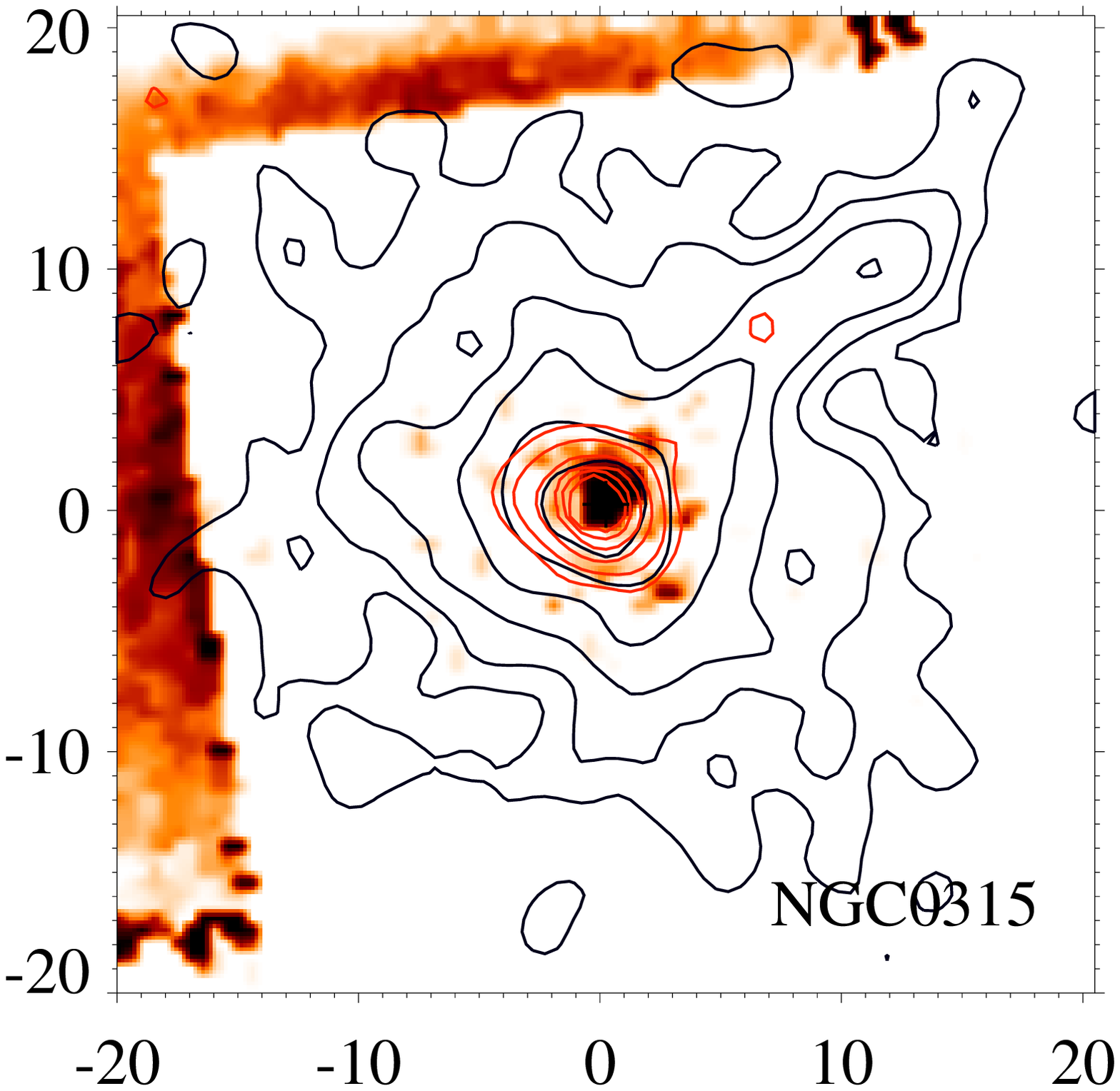}
\includegraphics[width=0.90\columnwidth]{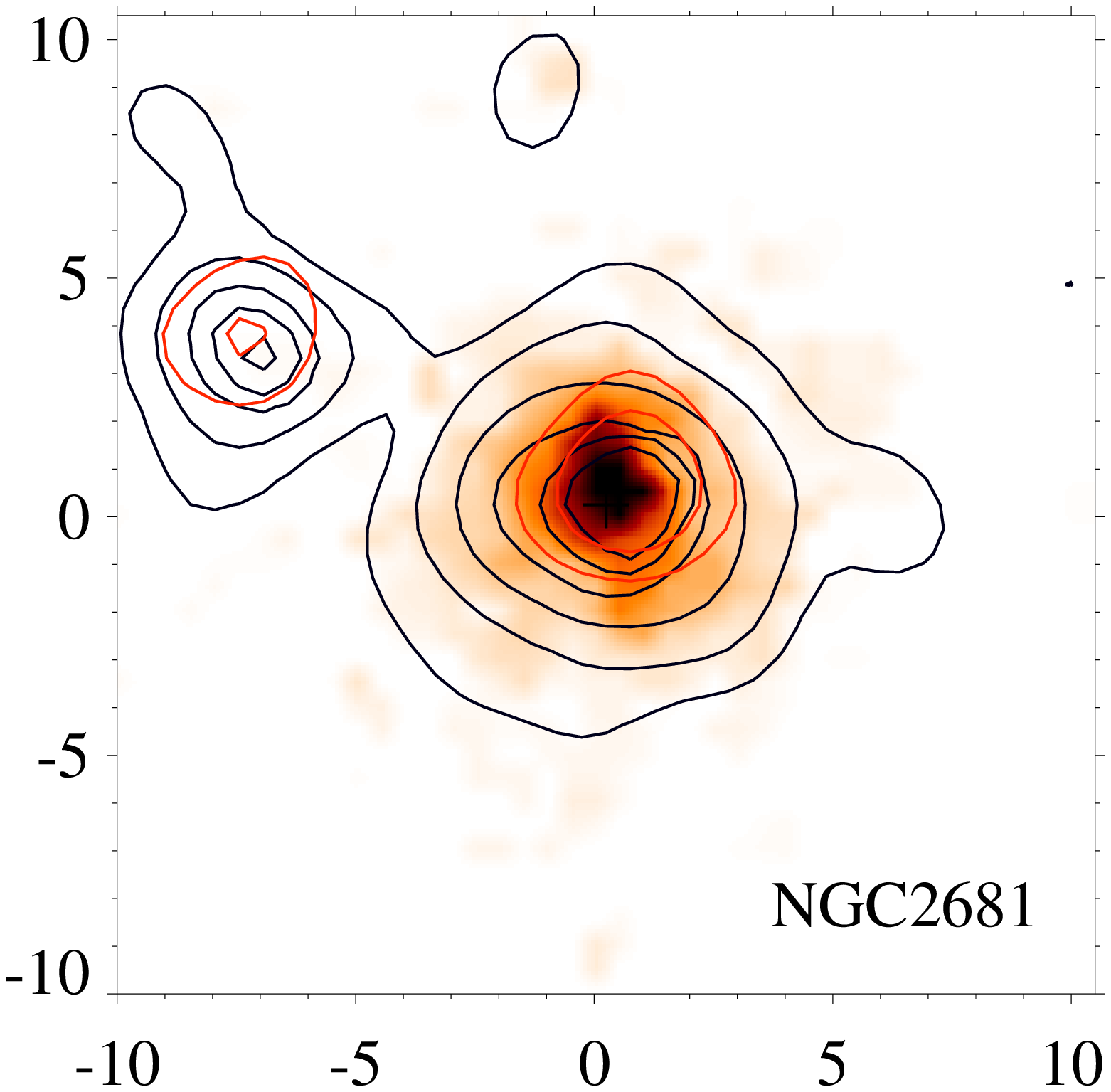}
\includegraphics[width=0.90\columnwidth]{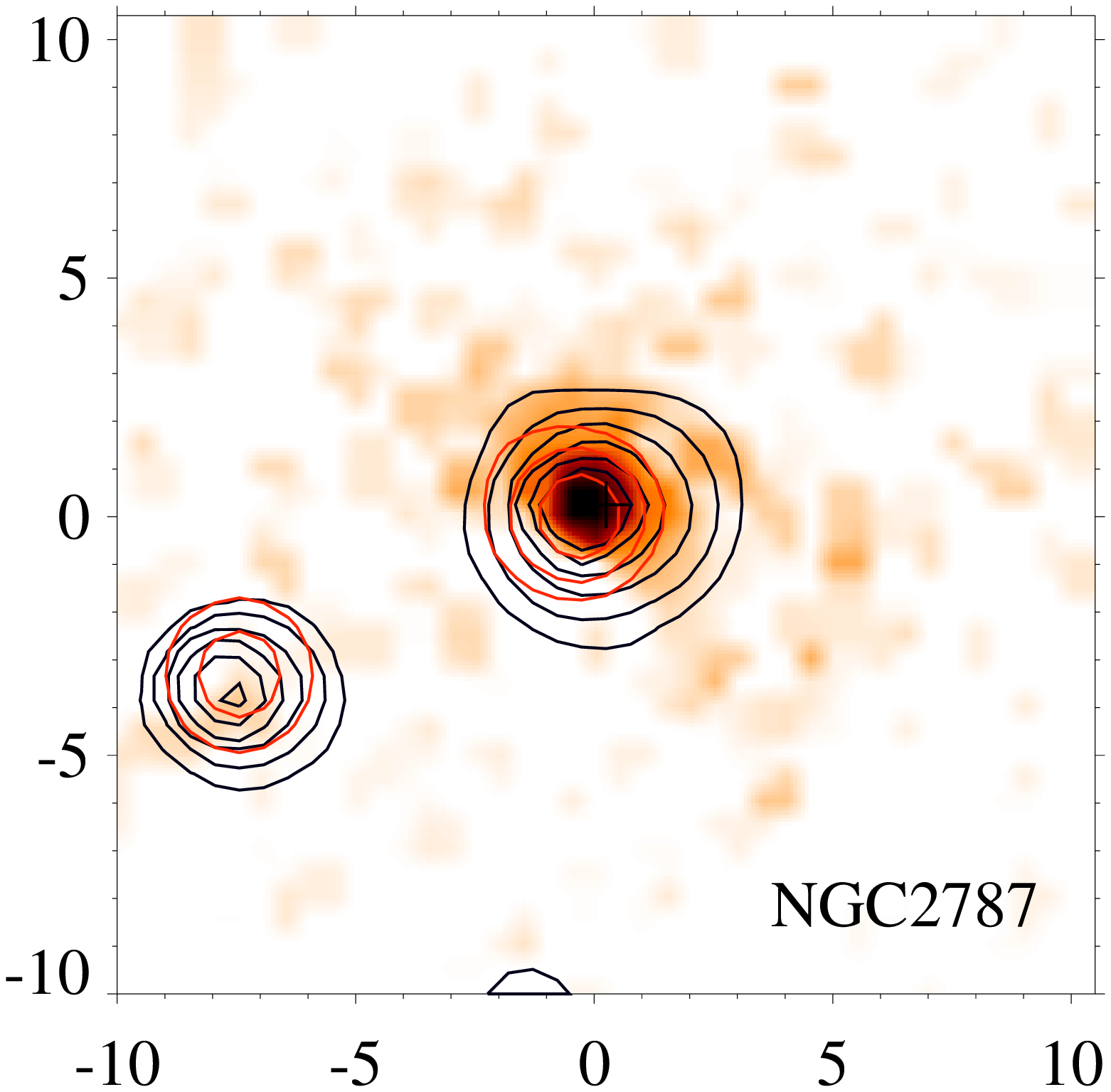}
\includegraphics[width=0.90\columnwidth]{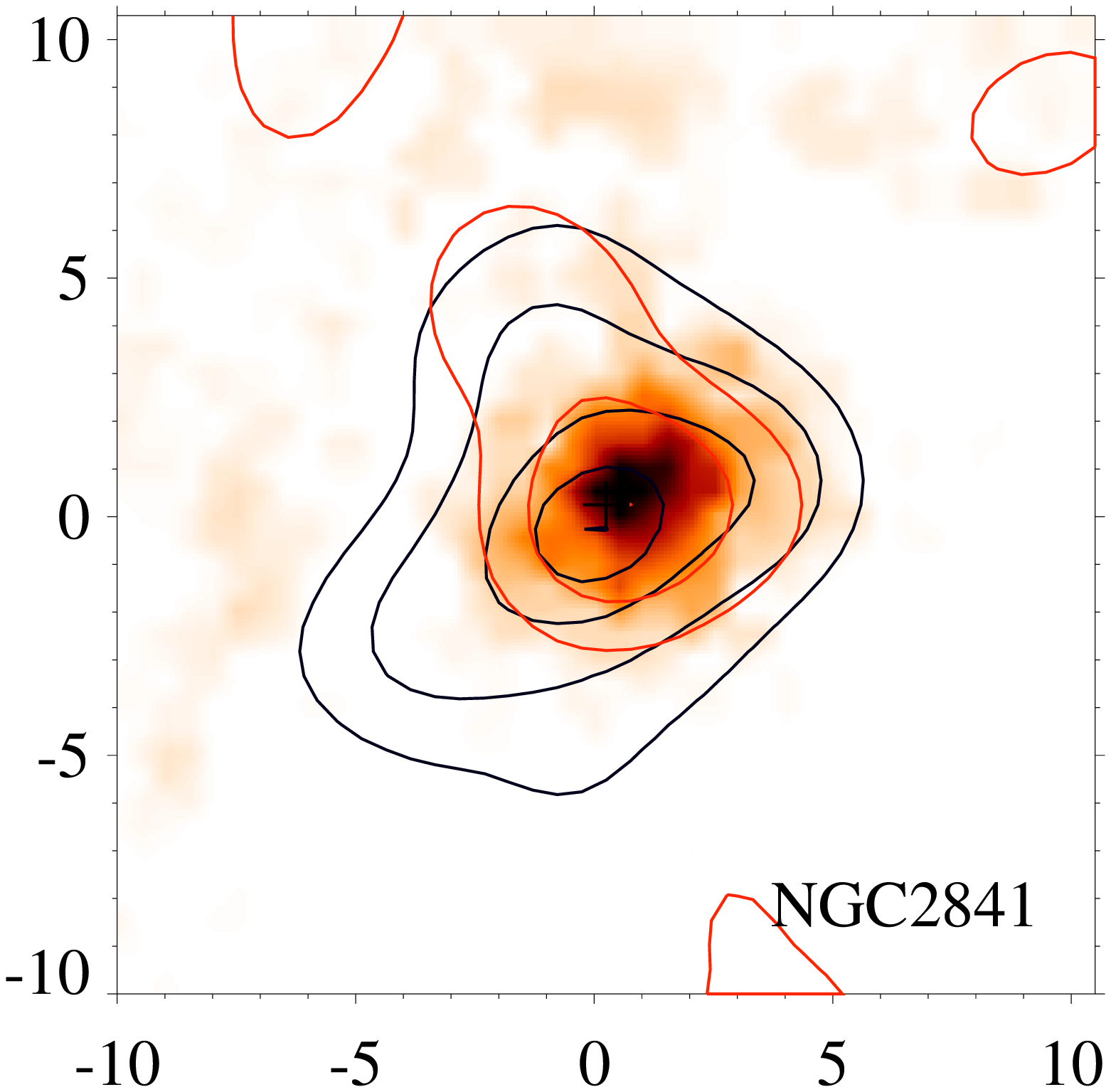}
\includegraphics[width=0.90\columnwidth]{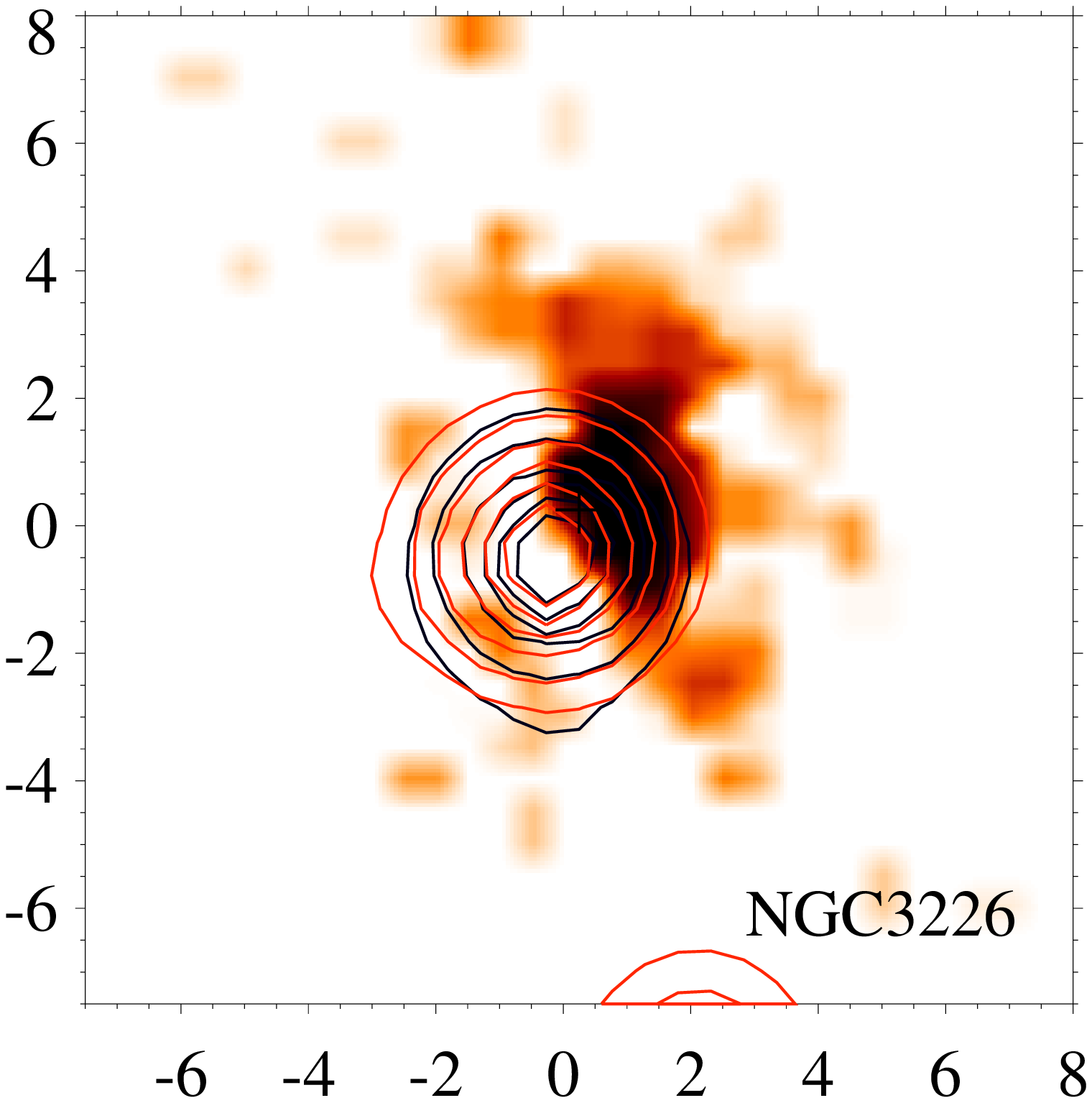}
\addtocounter{figure}{3}
 \caption{X-ray contours are overplotted onto the H$\alpha$ images. { Top is north and east is left.
The units of the plots are arcseconds. }
Soft (0.6-0.9 keV) X-rays { contours} are plotted in black and hard (4.5-8 keV) X-rays in red}
         \label{softX-halpha}
  \end{figure*}

 \begin{figure*}[b]
  \centering
\includegraphics[width=0.90\columnwidth]{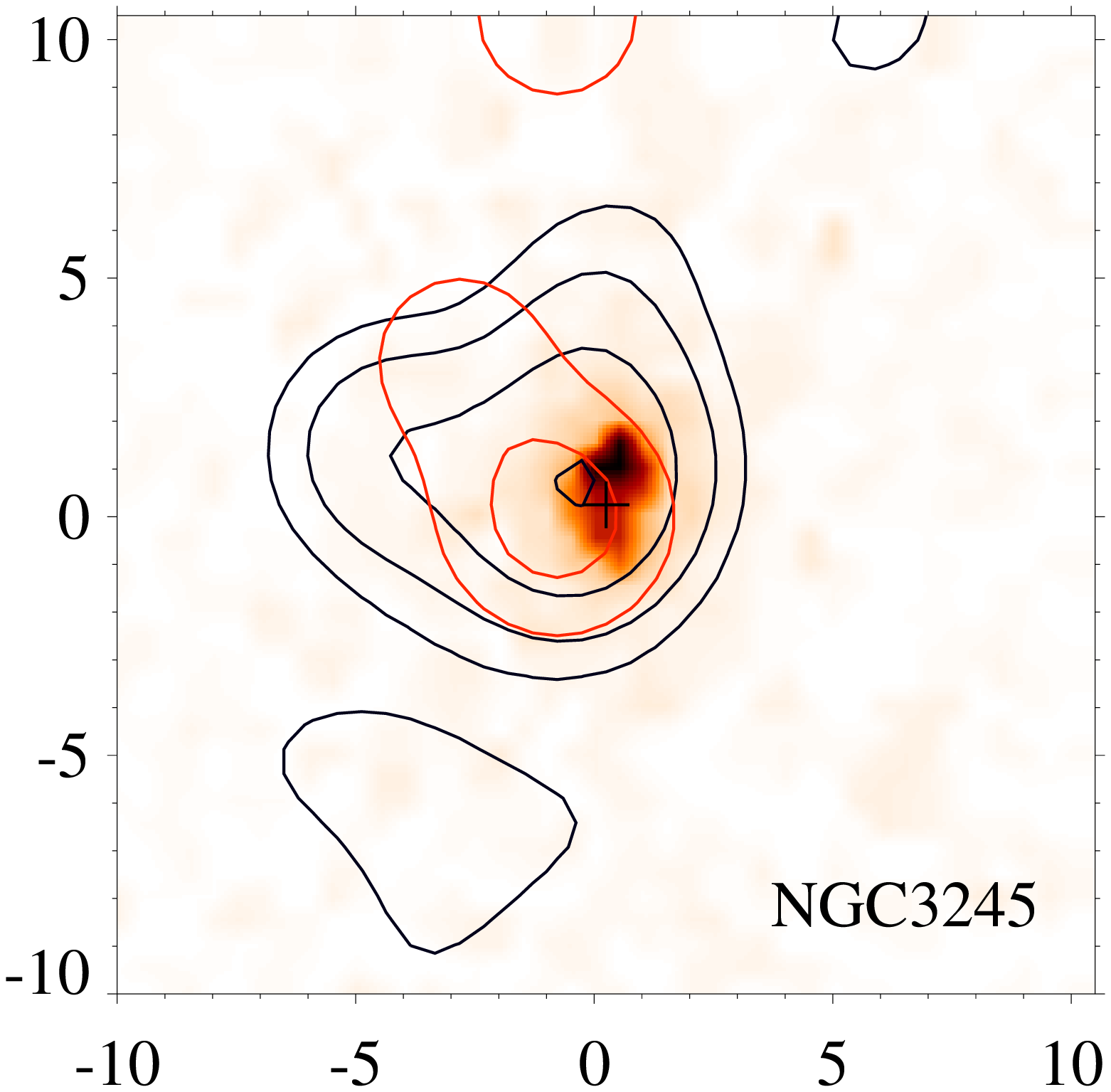}
\includegraphics[width=0.90\columnwidth]{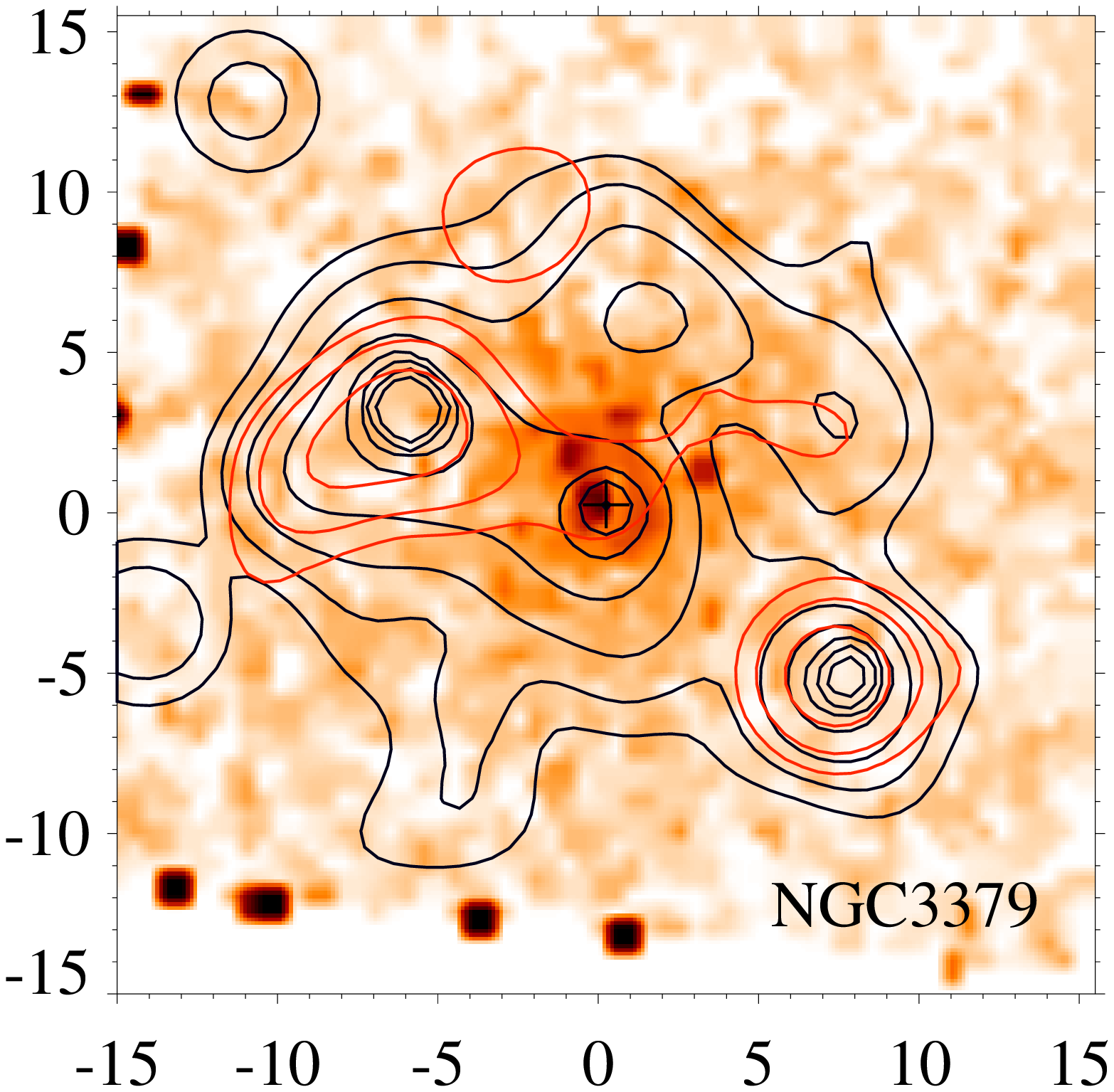}
\includegraphics[width=0.90\columnwidth]{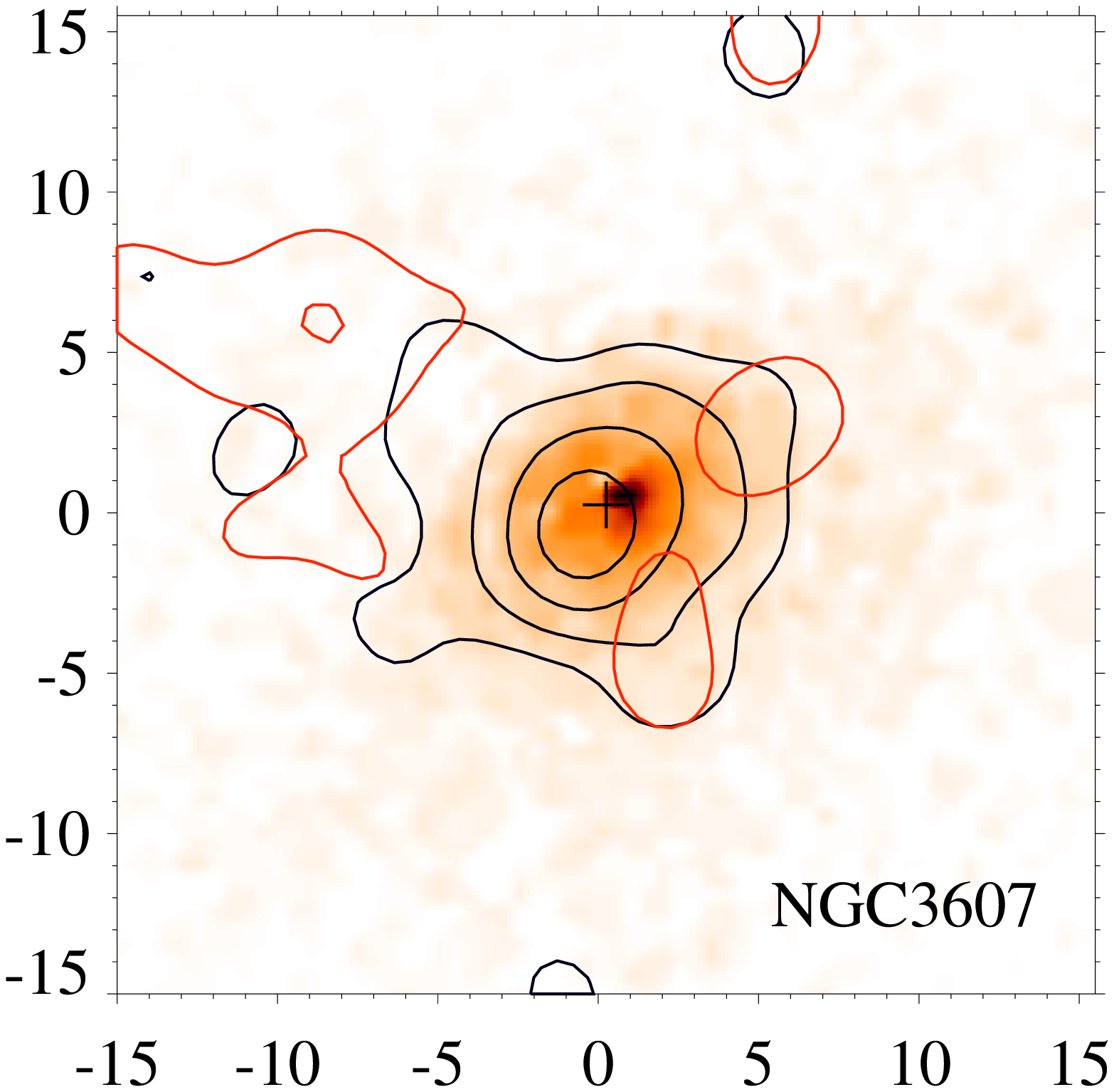}
\includegraphics[width=0.90\columnwidth]{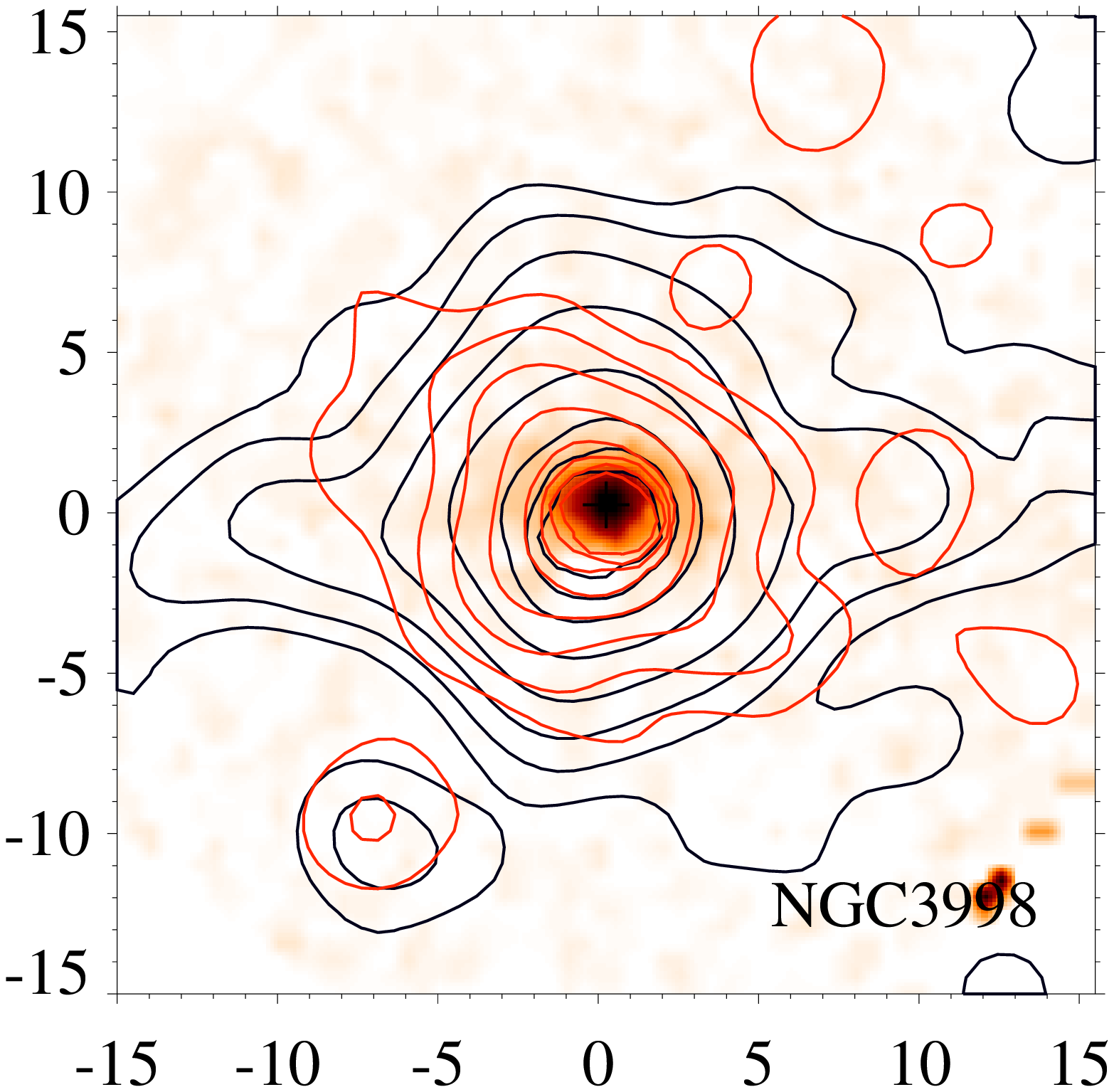}
\includegraphics[width=0.90\columnwidth]{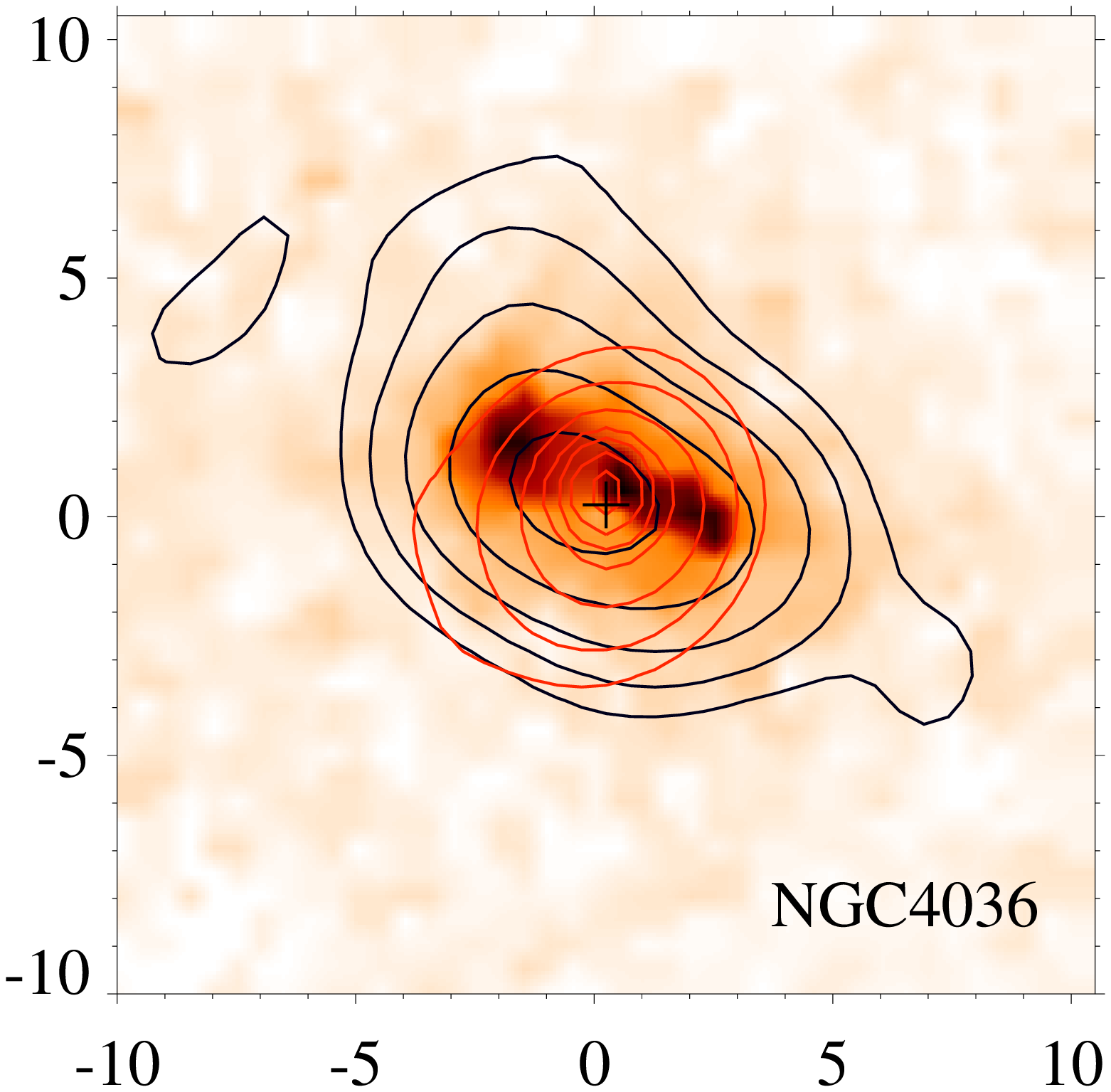}
\includegraphics[width=0.90\columnwidth]{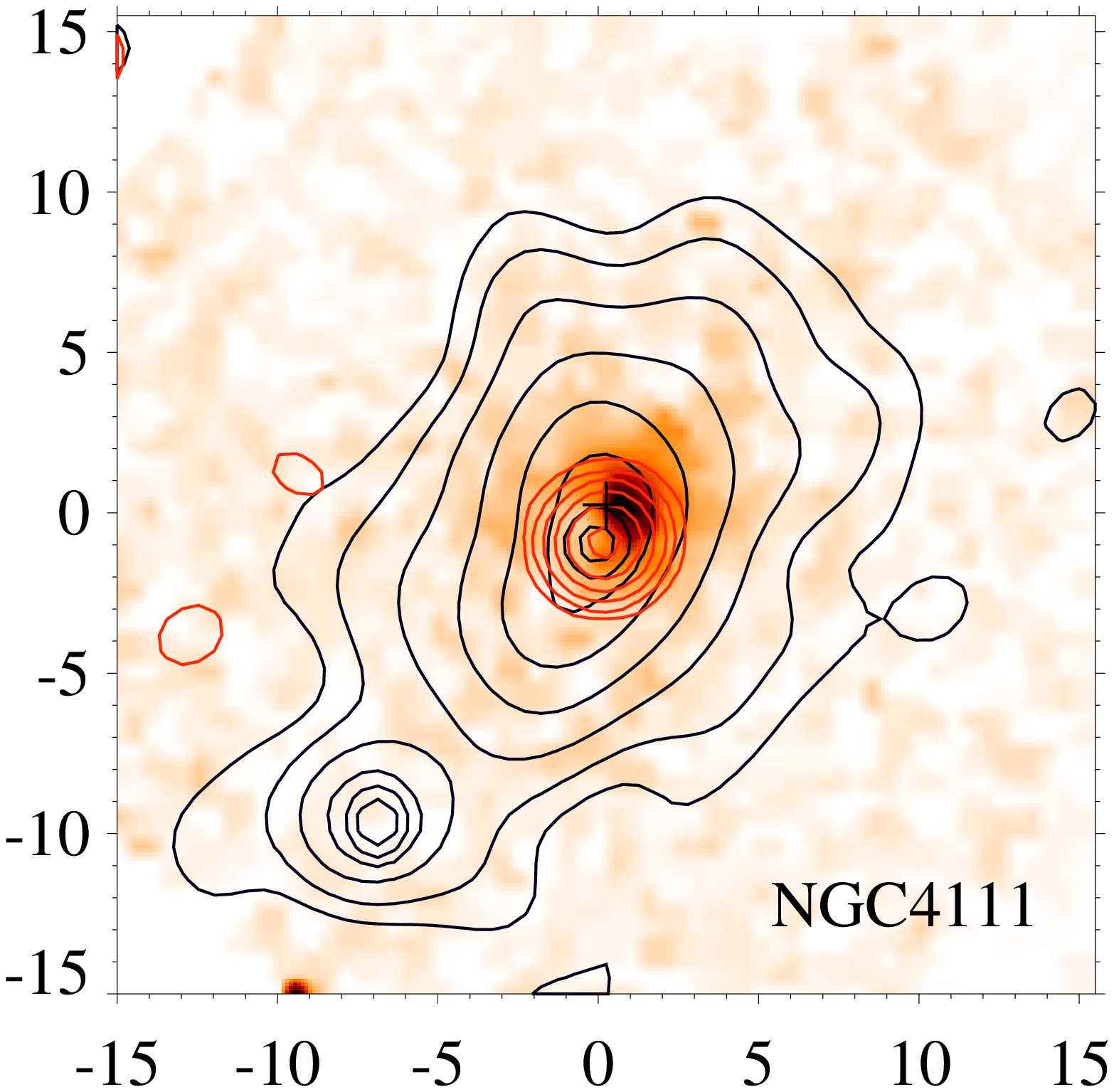}
\addtocounter{figure}{-1}
 \caption{Continued.}
         \label{}
  \end{figure*}

 \begin{figure*}
  \centering
\includegraphics[width=0.90\columnwidth]{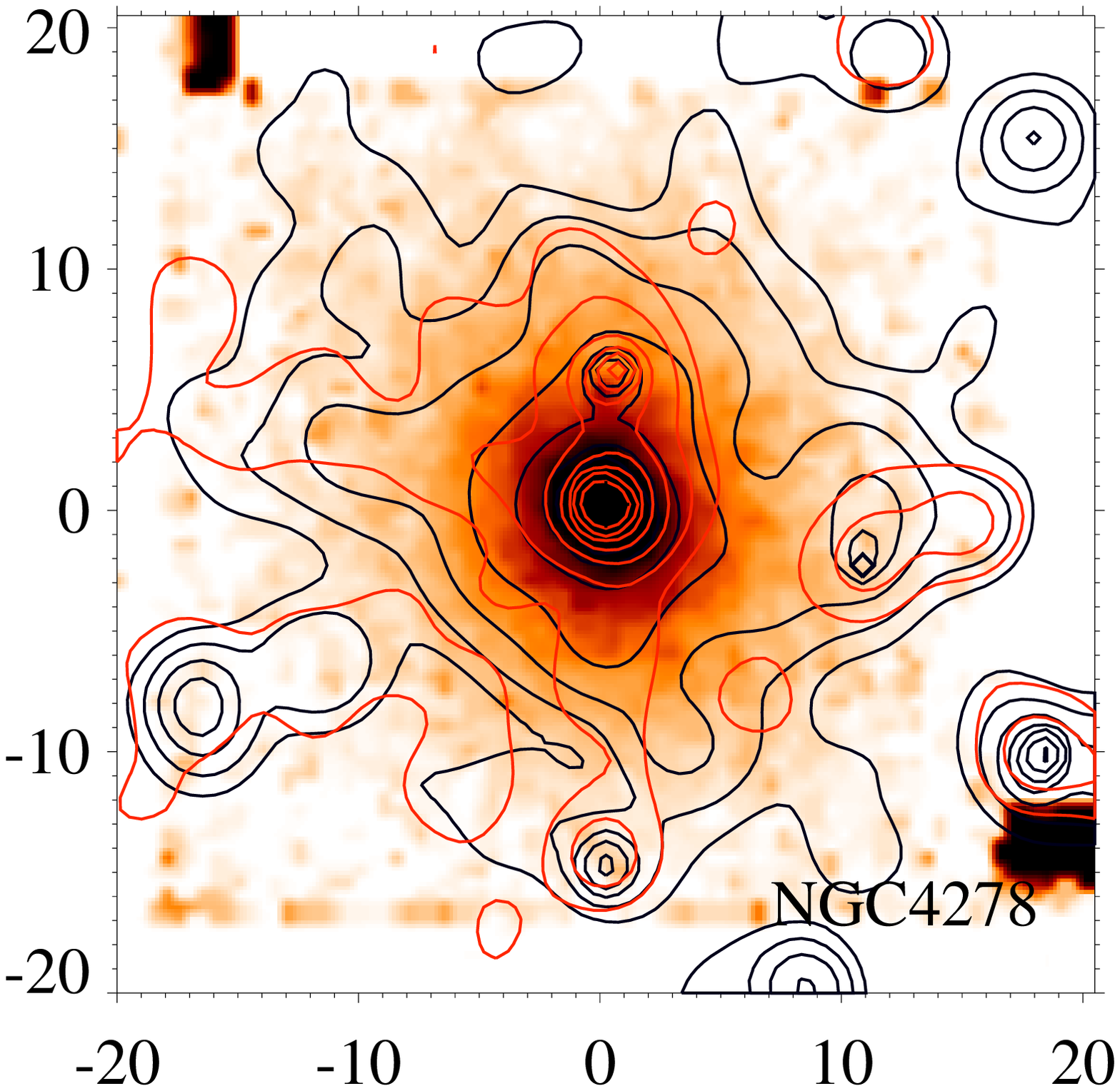}
\includegraphics[width=0.90\columnwidth]{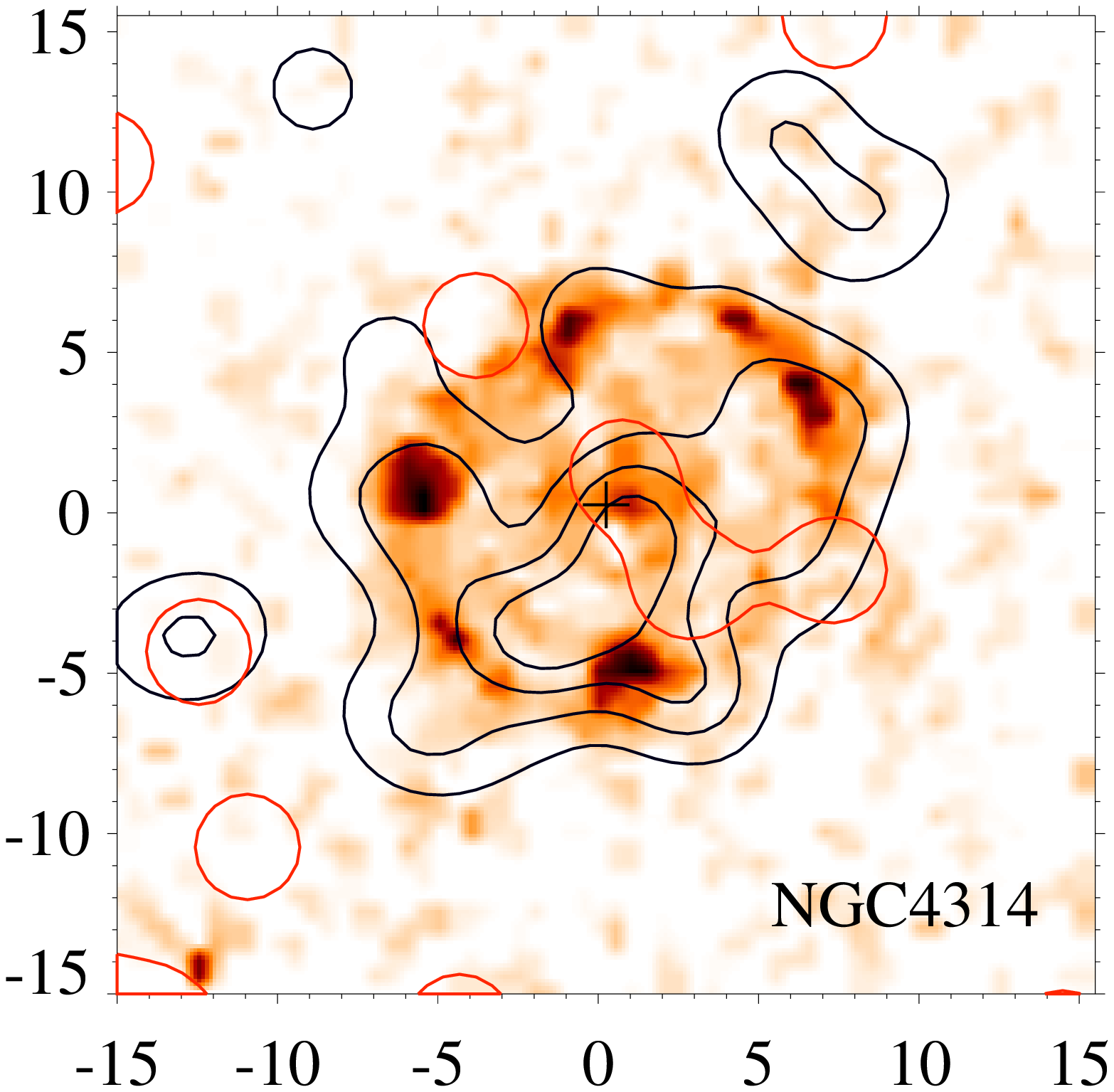}
\includegraphics[width=0.90\columnwidth]{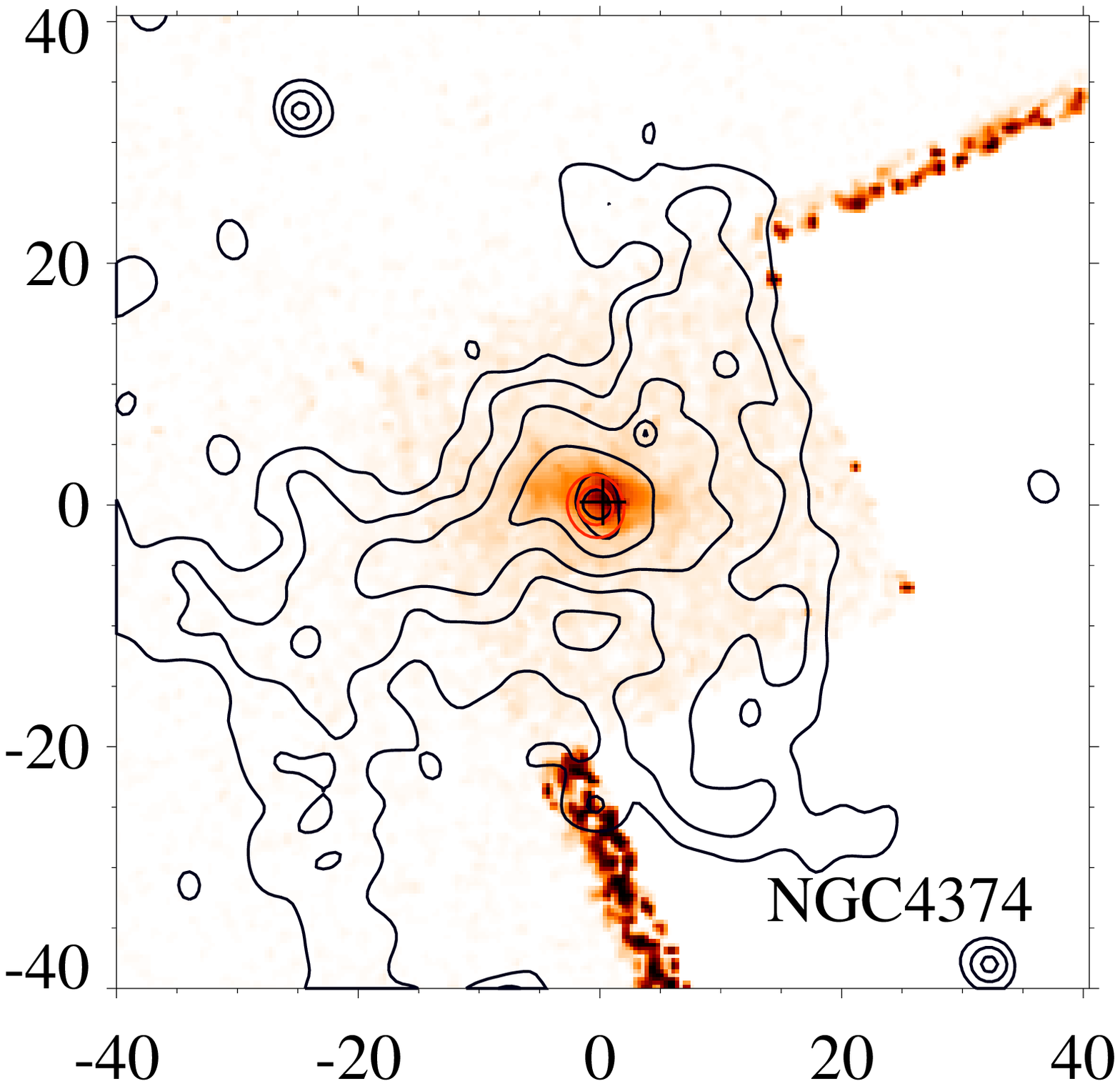}
\includegraphics[width=0.90\columnwidth]{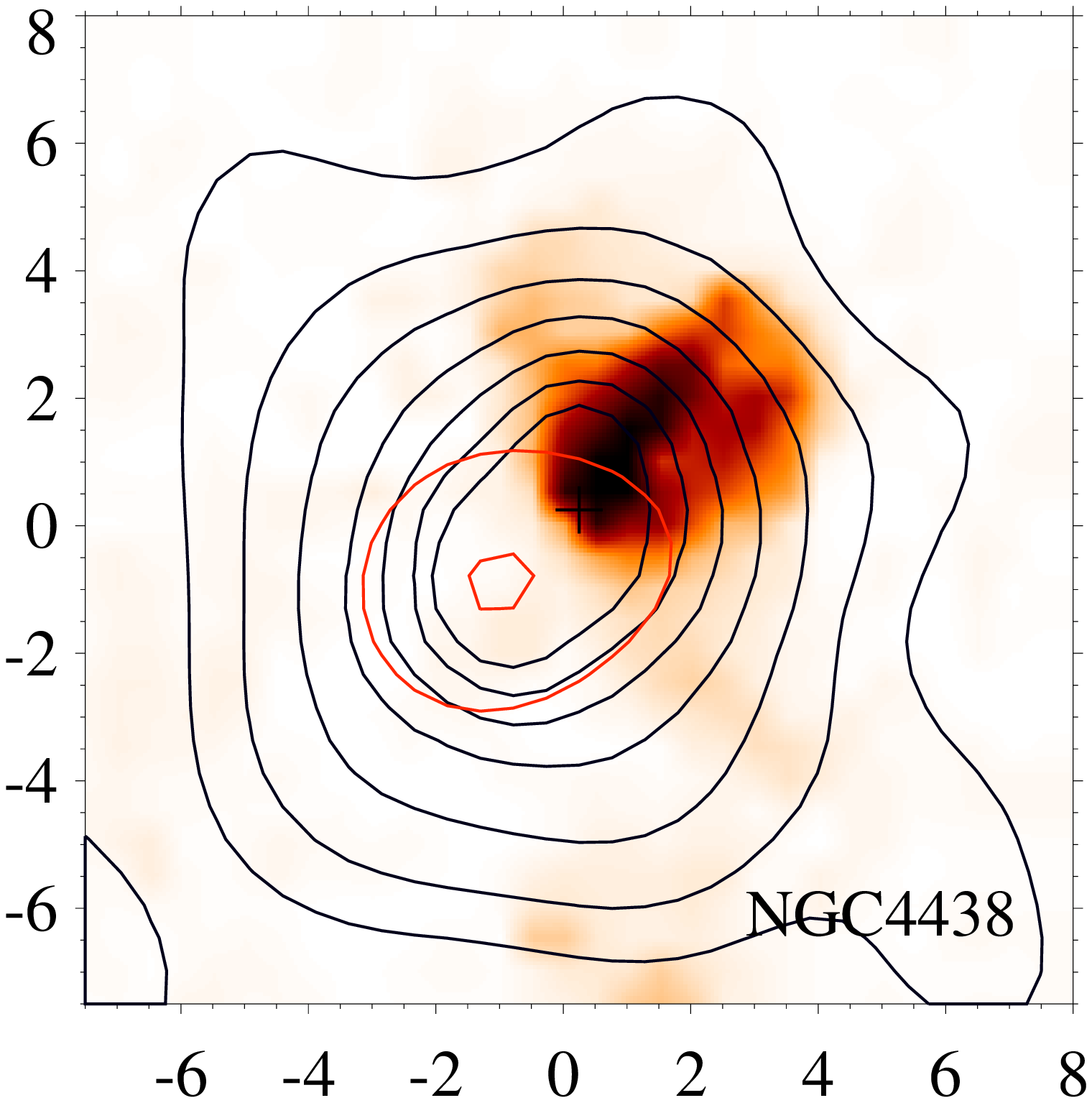}
\includegraphics[width=0.90\columnwidth]{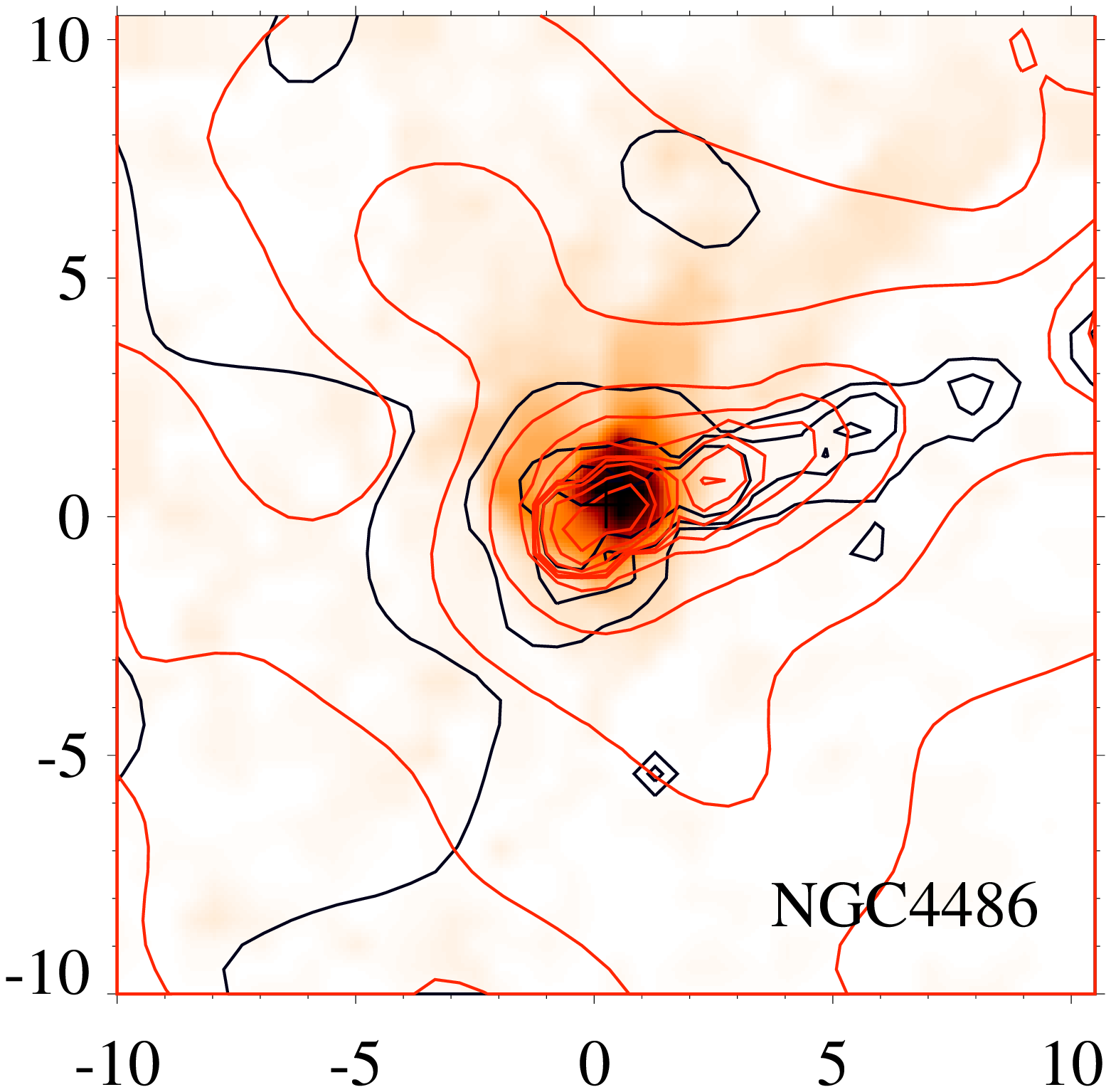}
\includegraphics[width=0.90\columnwidth]{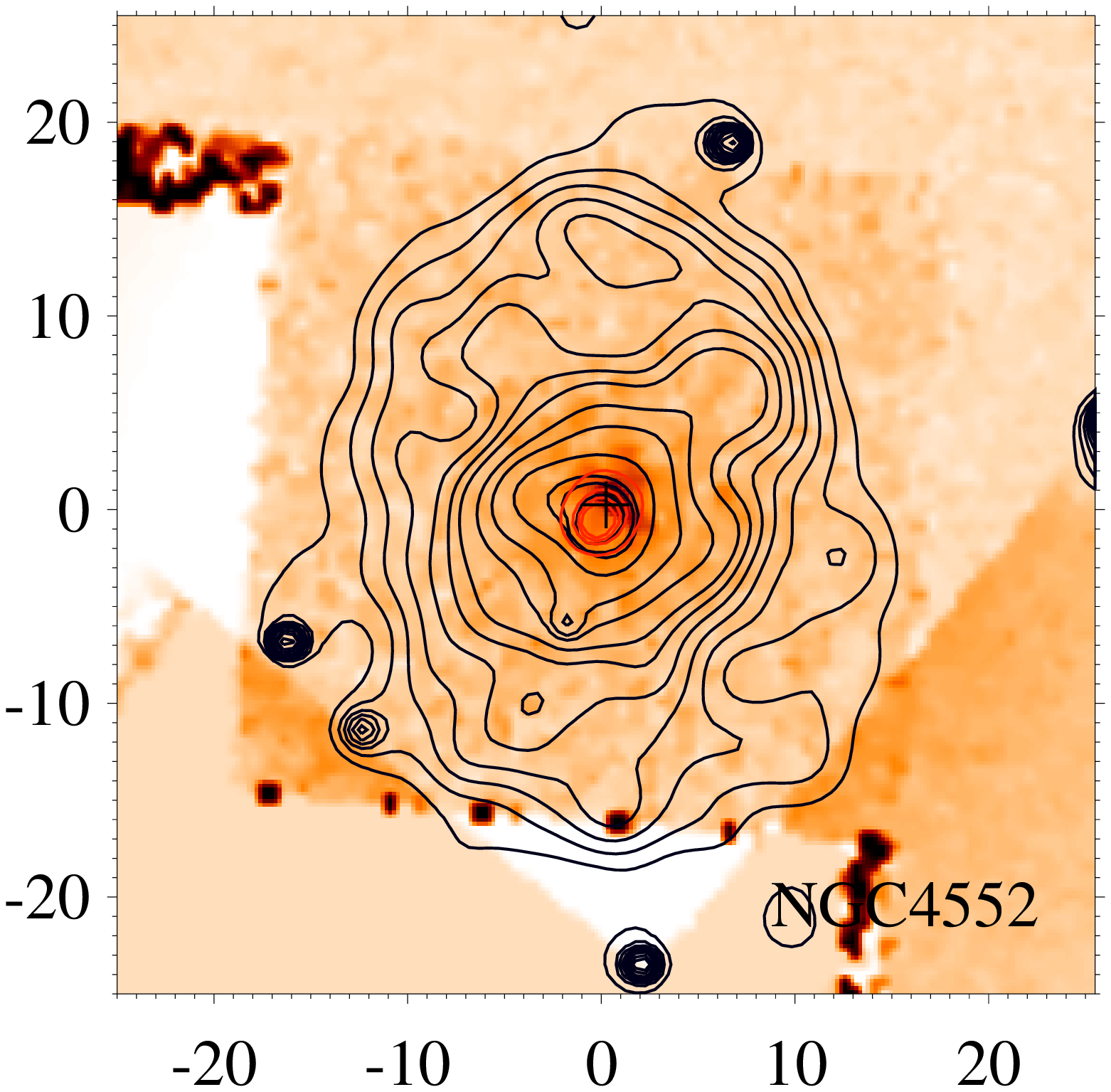}
\addtocounter{figure}{-1}
 \caption{Continued.}
         \label{}
  \end{figure*}

 \begin{figure*}
  \centering
\includegraphics[width=0.90\columnwidth]{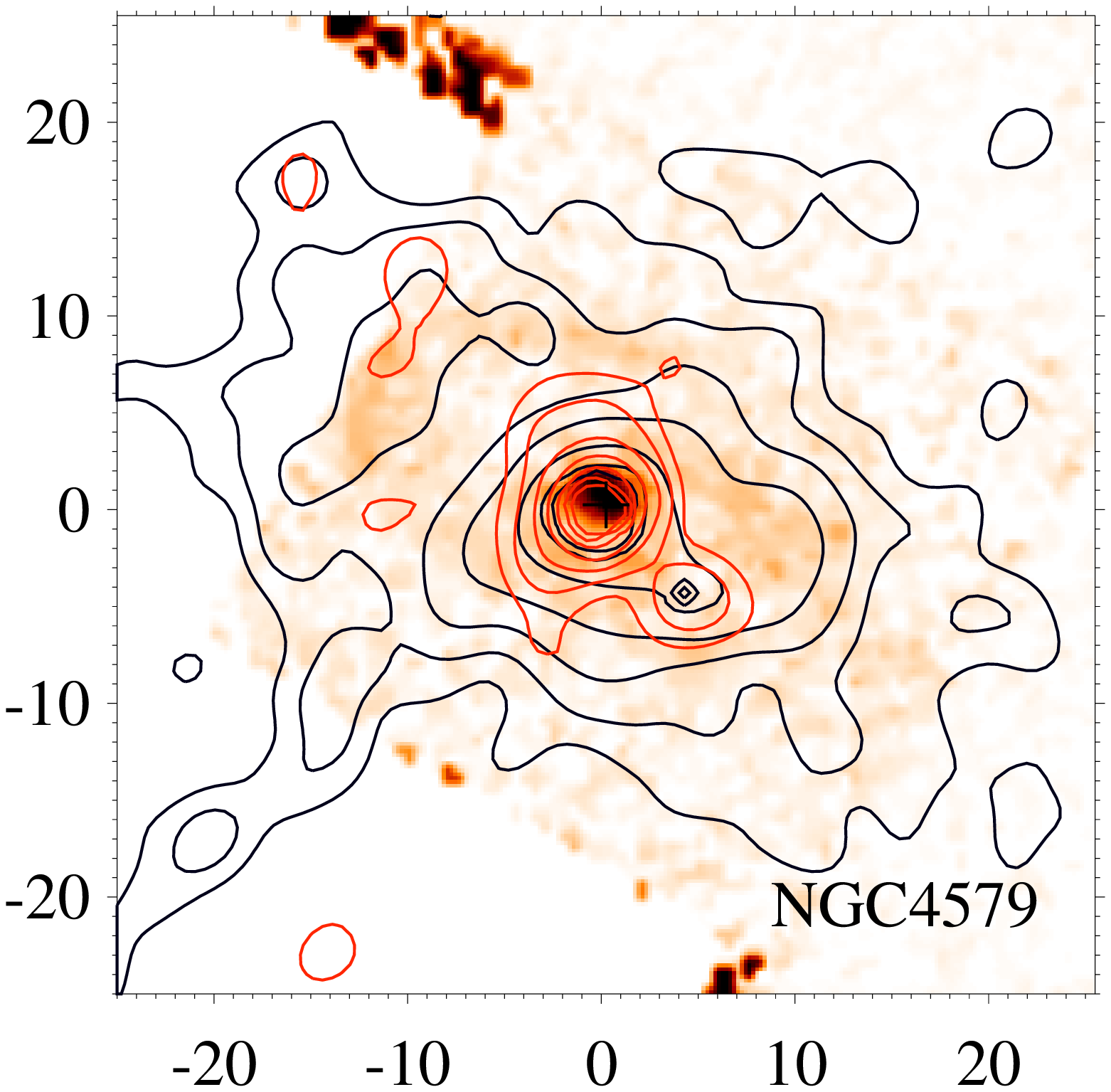}
\includegraphics[width=0.90\columnwidth]{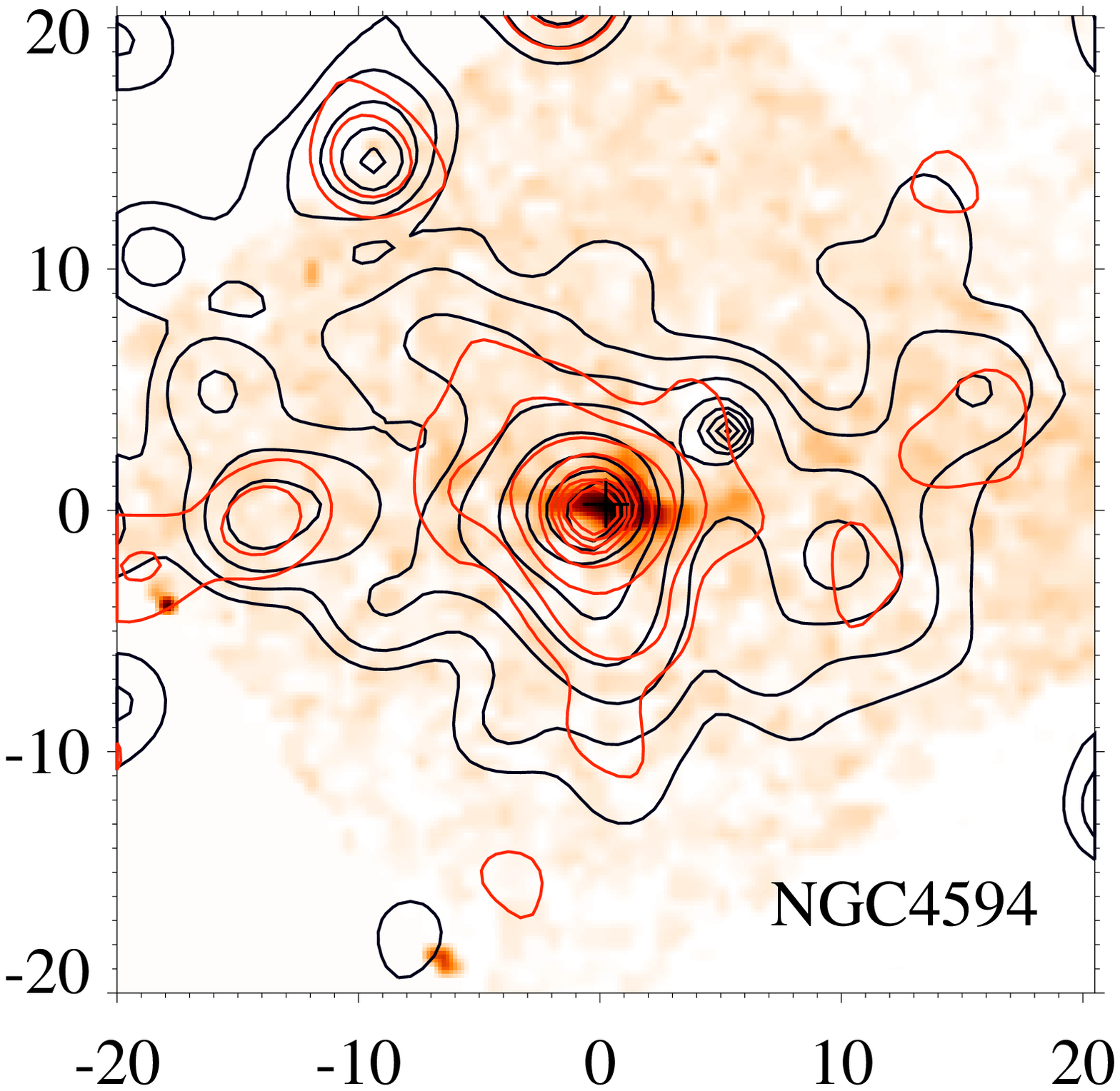}
\includegraphics[width=0.90\columnwidth]{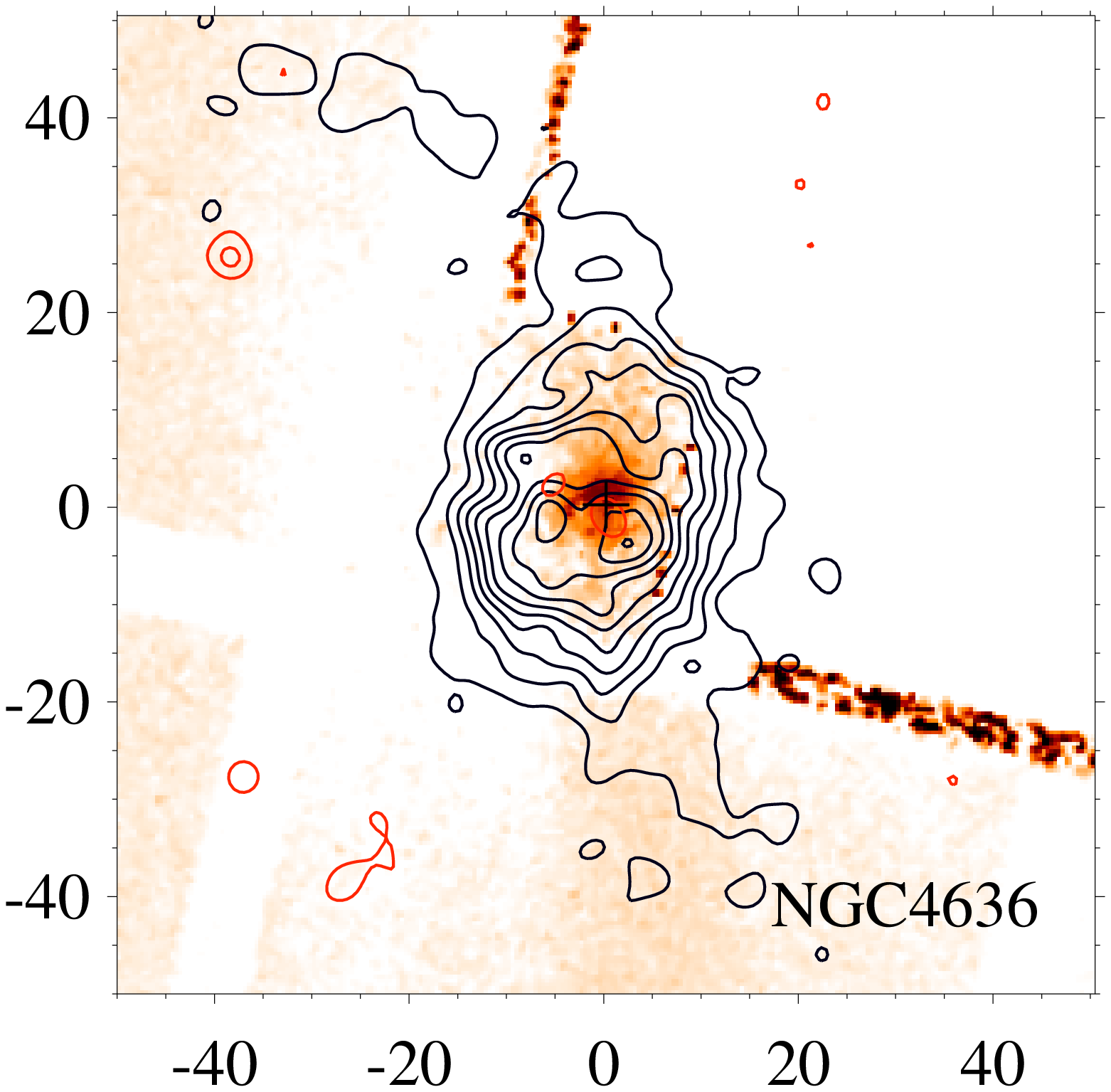}
\includegraphics[width=0.90\columnwidth]{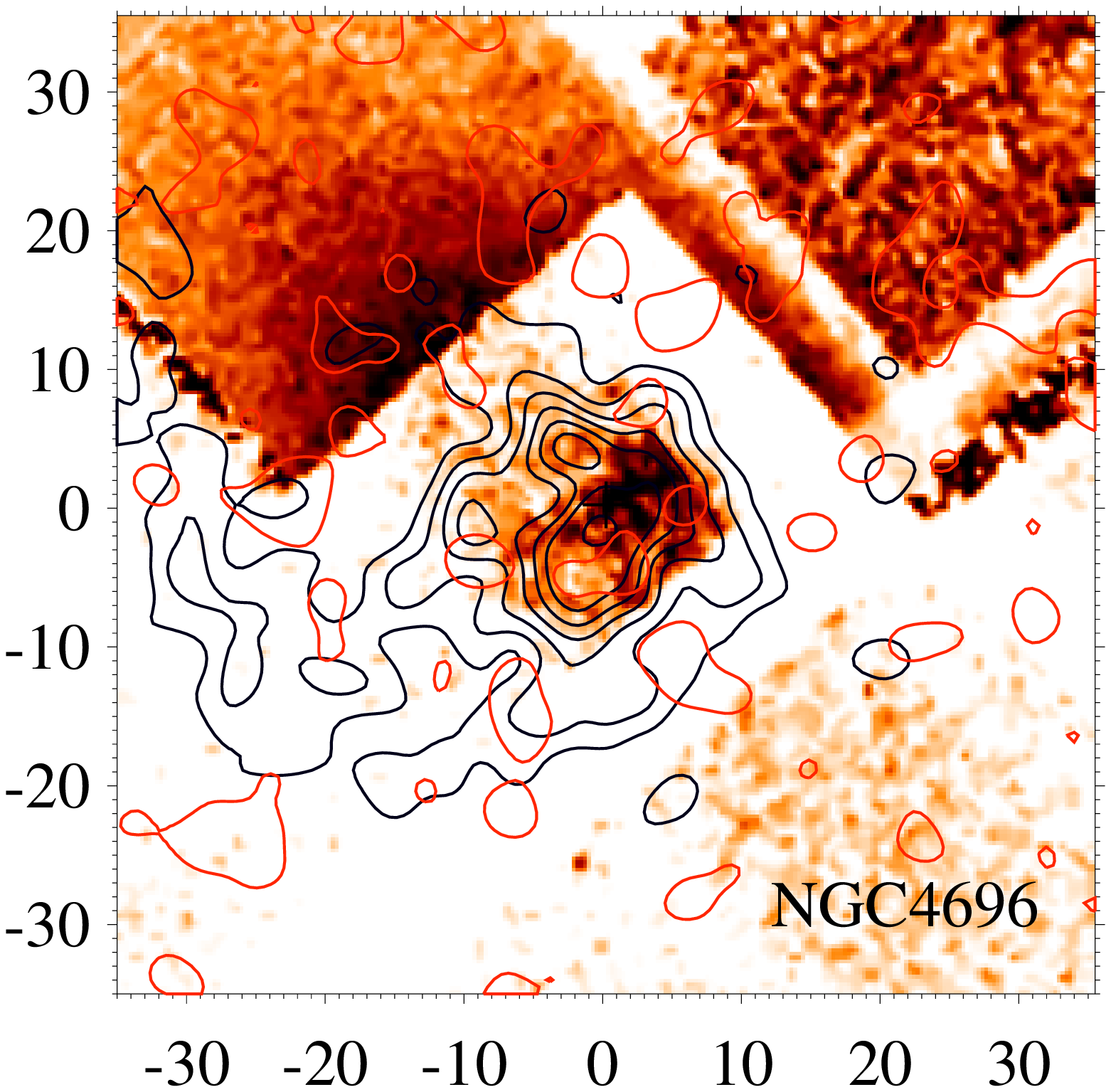}
\includegraphics[width=0.90\columnwidth]{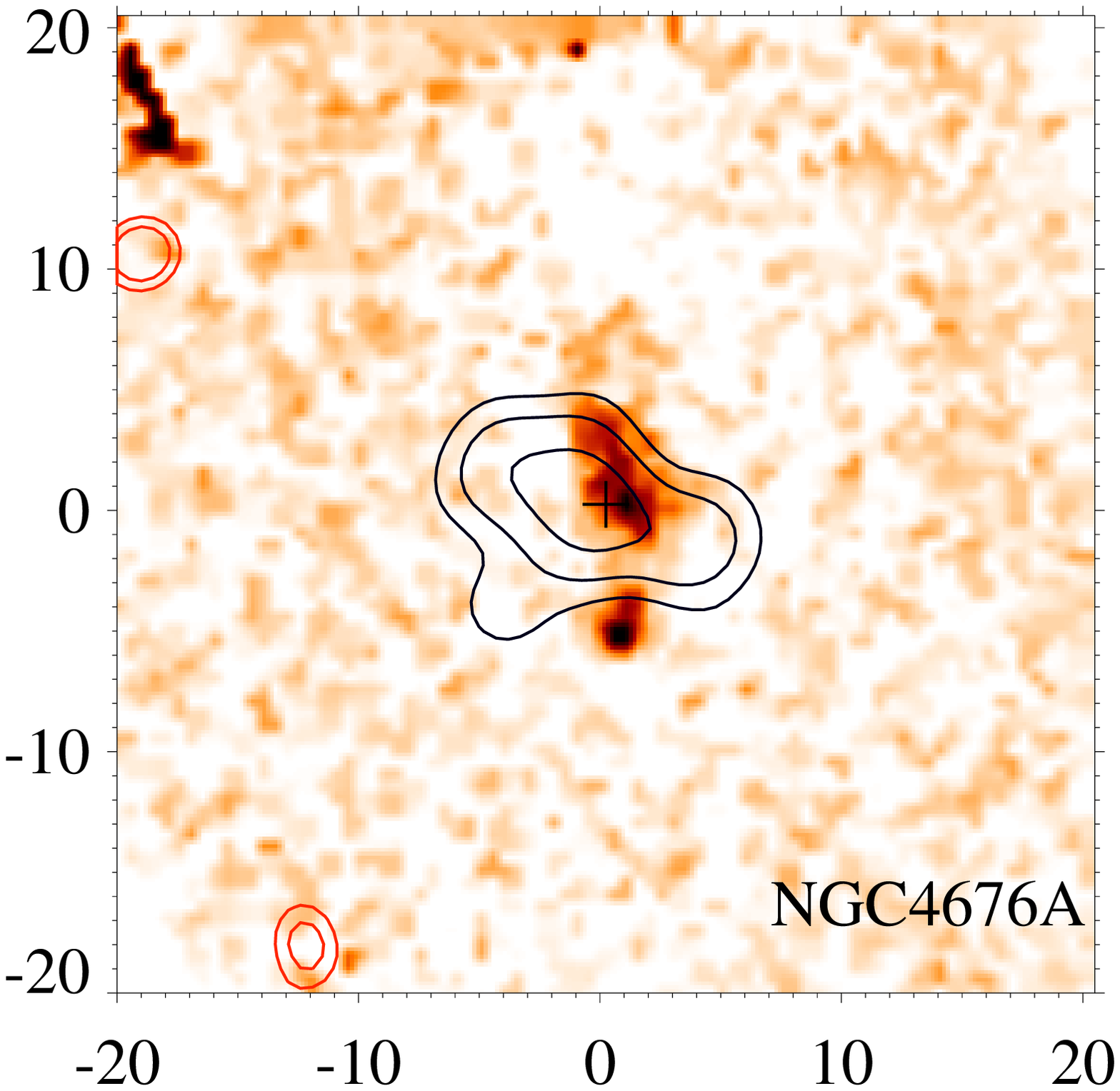}
\includegraphics[width=0.90\columnwidth]{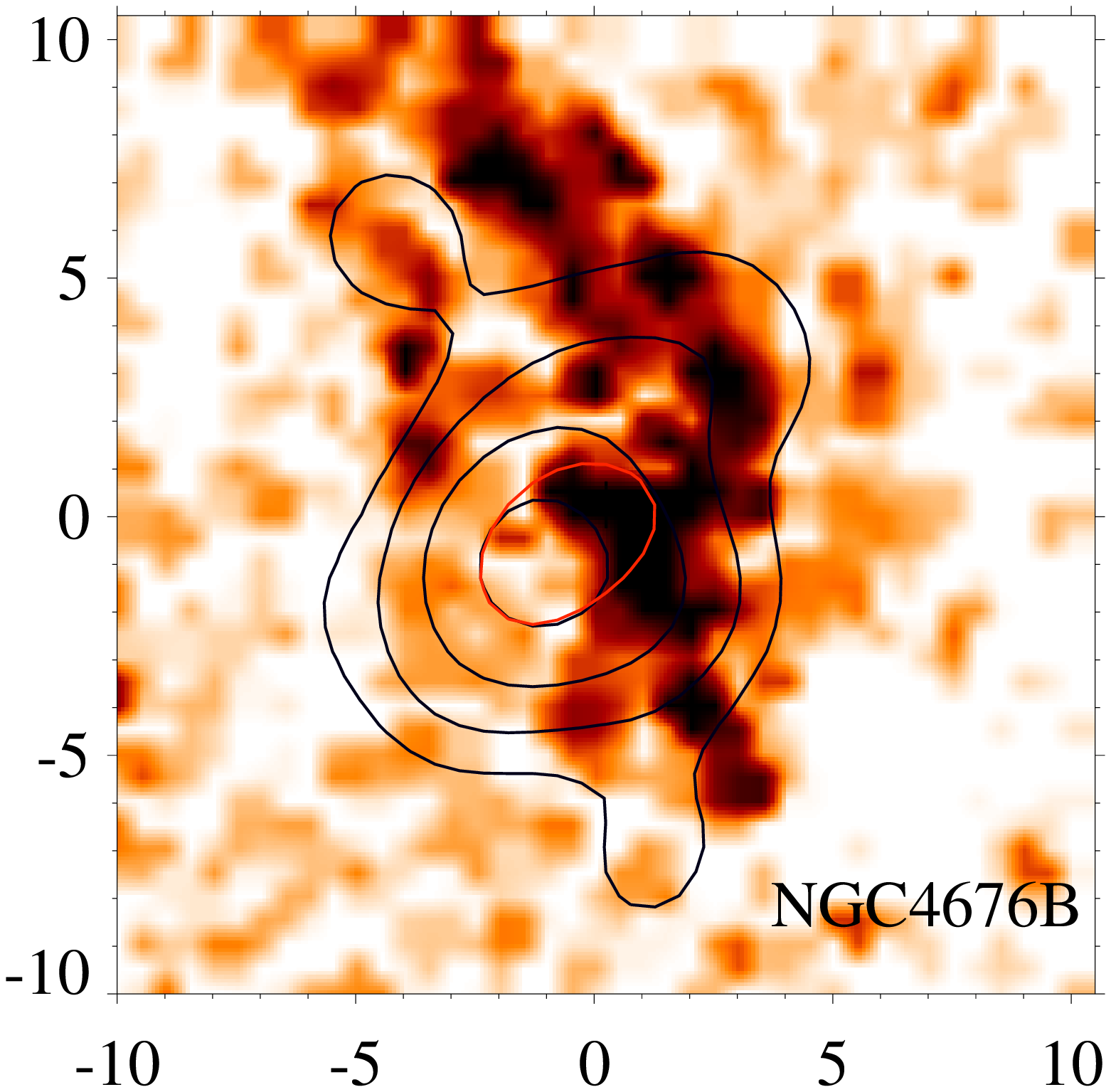}
\addtocounter{figure}{-1}
 \caption{Continued.}
         \label{}
  \end{figure*}

 \begin{figure*}
  \centering
\includegraphics[width=0.90\columnwidth]{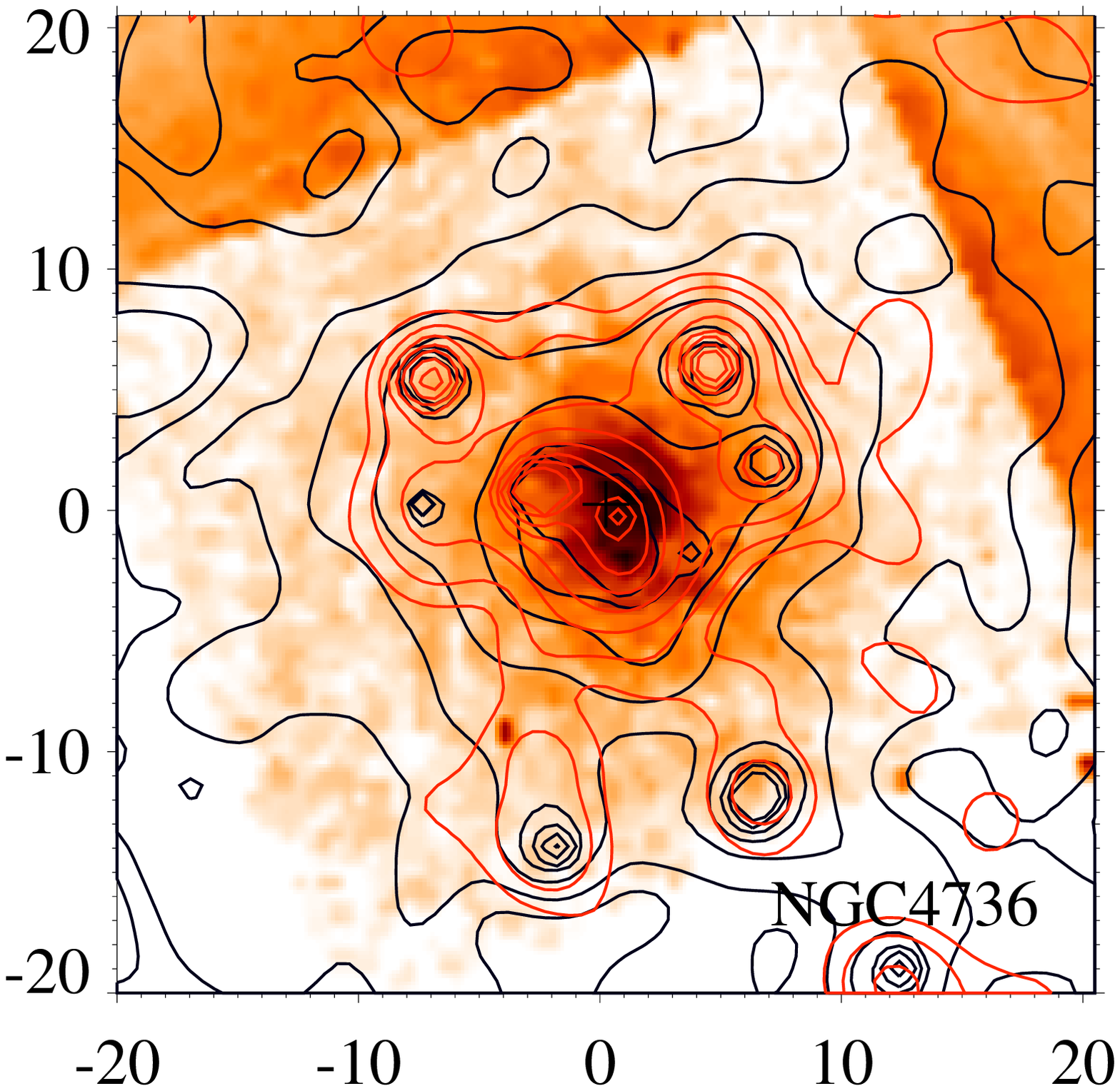}
\includegraphics[width=0.90\columnwidth]{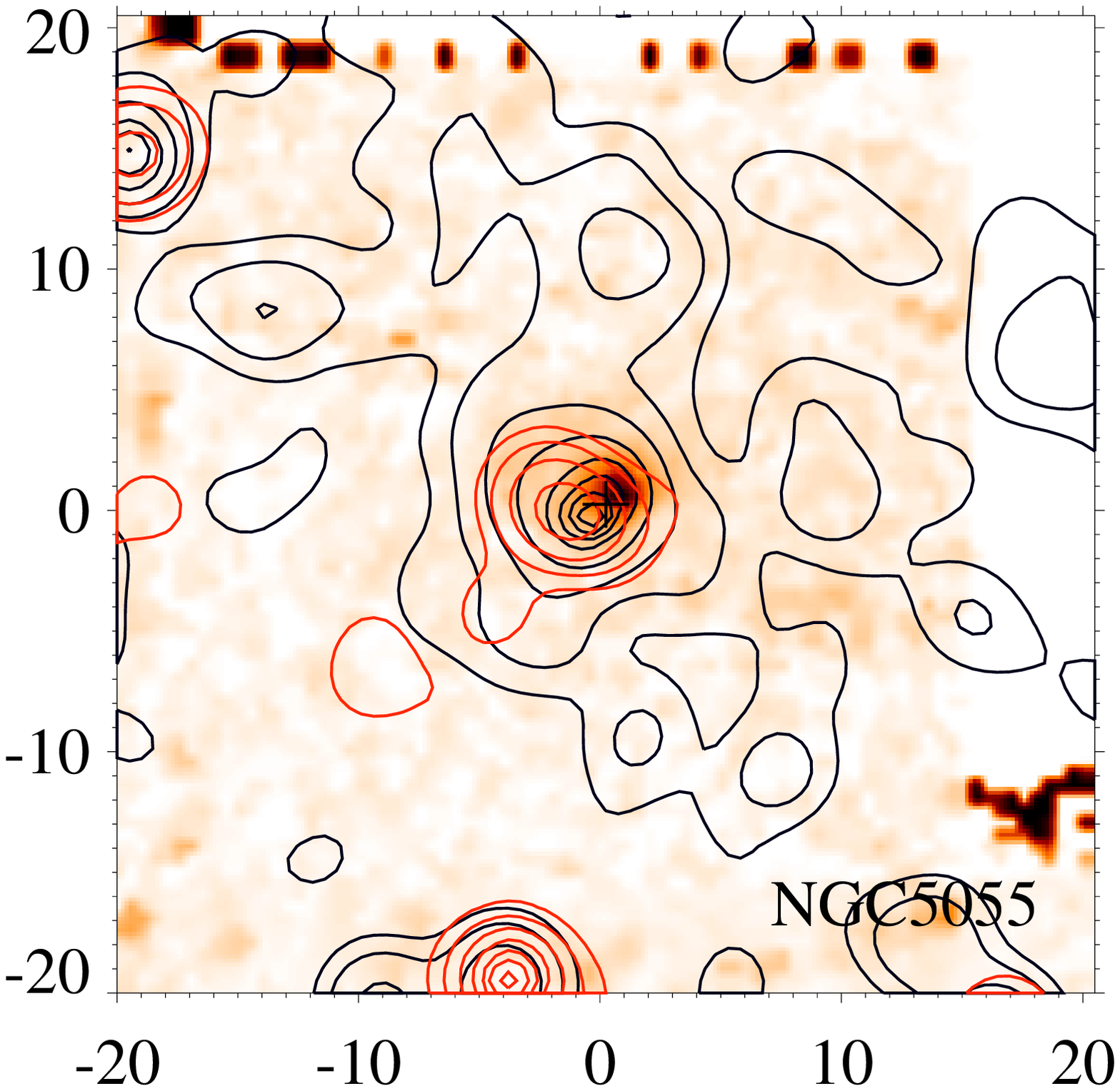}
\includegraphics[width=0.90\columnwidth]{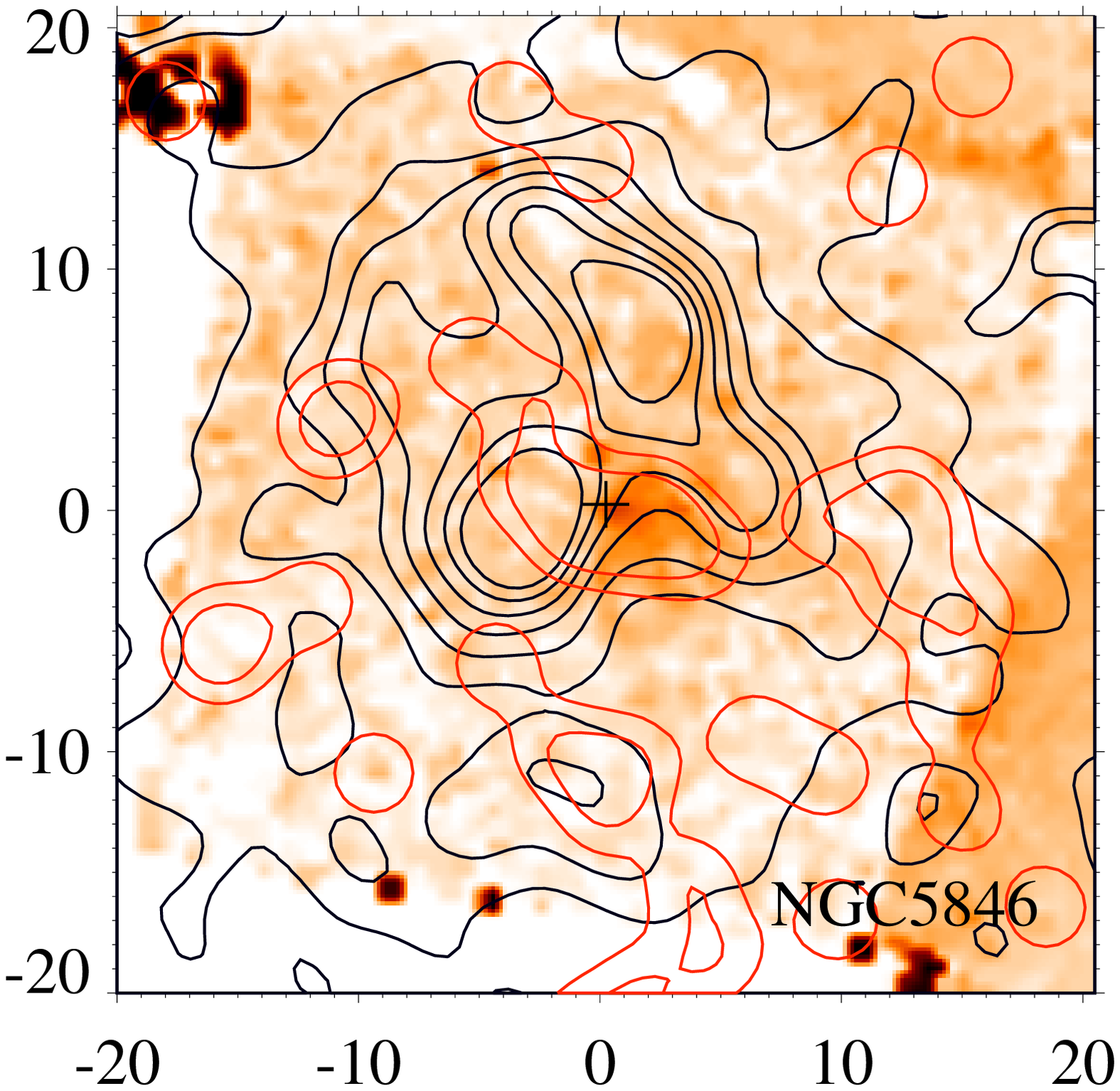}
\includegraphics[width=0.90\columnwidth]{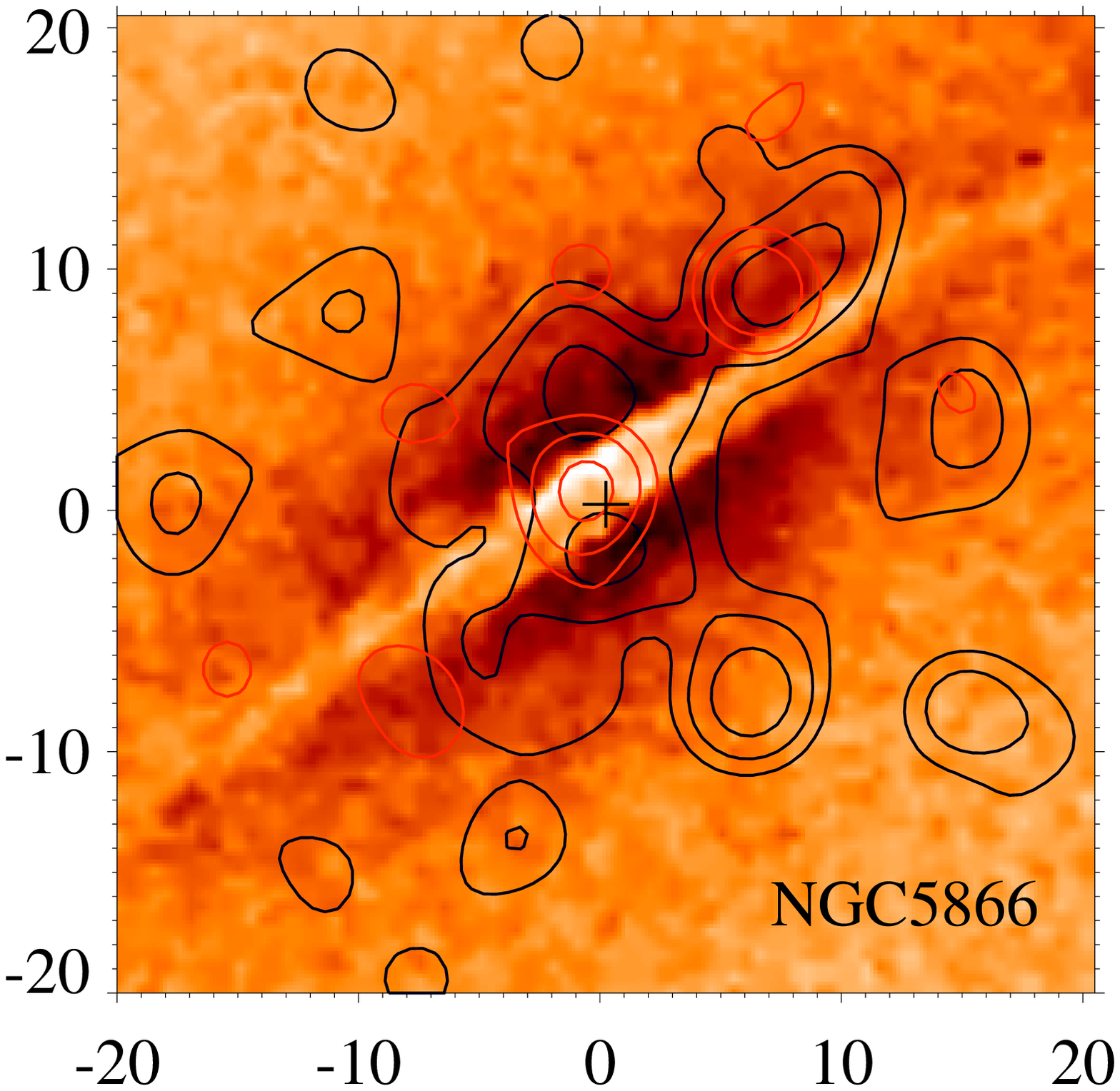}
\addtocounter{figure}{-1}
 \caption{Continued.}
         \label{softXvsHST}
  \end{figure*}

\end{document}